\documentclass[11pt,a4paper]{article}
\usepackage{jstyle}
\usepackage{hyperref}
\usepackage{slashed,framed}
\usepackage{empheq}
\usetikzlibrary{calc,decorations.markings}
\author[a]{Jin-Beom BAE}
\author[b,c]{\quad Euihun JOUNG}
\author[d]{\quad Shailesh LAL}

\affiliation[a]{Scranton Honors Program, Ewha Womans University, Seoul 120-750, Korea}
\affiliation[b]{School of Physics and Astronomy, Seoul National University, Seoul 151-747, Korea}
\affiliation[c]{Gauge, Gravity \& Strings, Center for Theoretical Physics of the Universe, Institute for Basic Sciences, Daejeon 34047, Korea }
\affiliation[d]{LPTHE -- UMR 7589, UPMC Paris 06, Sorbonne Universit{\'e}s,  Paris 75005, France}

\emailAdd{kastalean4@gmail.com}
\emailAdd{euihun.joung@snu.ac.kr}
\emailAdd{shailesh@lpthe.jussieu.fr}

\begin{document}

\date{\currenttime}

\title{\centering
\LARGE{One-loop Test of \\Free $SU(N)$ Adjoint Model Holography}}

\abstract{We consider the holographic duality where
the CFT side is given by $SU(N)$ adjoint free scalar field theory.
Compared to the vector models, the set of single trace operators is immensely extended
so that the corresponding AdS theory also contains infinitely many massive higher spin fields 
on top of the massless ones. 
We compute the one-loop vacuum energy of these AdS fields to test this duality at the subleading order in large $N$ expansion.
The determination of the bulk vacuum energy requires a proper scheme to sum up the infinitely many contributions.
For that, we develop a new method and apply it first to calculate the 
vacuum energies for the first few `Regge trajectories' in AdS$_4$ and AdS$_5$\,.
In considering the full vacuum energy
of AdS theory dual to a matrix model CFT,
we find that there exist
more than one available prescriptions
for the one-loop vacuum energy.
Taking a particular prescription,
we determine the full vacuum energy of the AdS$_5$ theory, 
whereas the AdS$_4$ calculation still remains technically prohibitive.
This result shows that the full vacuum energy 
of the AdS$_5$ theory
coincides with minus of the free energy of a single scalar field on the boundary.
This is analogous to the $O(N)$ vector model case,
 hence suggests an interpretation of
the \emph{positive} shift of the bulk coupling constant, 
i.e. from $N^2-1$ to $N^2$\,.
}

\maketitle

\section{Introduction}
Recently, there has been significant progress in the vectorial AdS/CFT correspondence \cite{Klebanov:2002ja, Sezgin:2002rt}. It relates free/critical CFTs in the $O(N)$ (or $U(N)$) vector multiplet to the Vasiliev's higher spin theory \cite{Vasiliev:1990en,Vasiliev:2003ev} with a certain boundary conditions. An important feature of this duality is the precise one-to-one correspondence between the spectrum of `light' conformal primaries\footnote{By `light' we mean that the conformal dimension of the primary does not scale with $N$.} on the CFT side and the spectrum of quadratic fluctuations near the AdS vacuum. In fact, the match of the spectrum --- shown by Flato and Fronsdal \cite{Flato:1978qz} --- predates the AdS/CFT conjecture and even Vasiliev's theory. It states that 
the tensor product of two singleton representations of $so(2,3)$
--- the scalar one, Rac or the spinor one, Di --- can be decomposed into the infinite sum of massless higher spin representations:
\ba
	{\rm Rac}\otimes{\rm Rac}
	=\bigoplus_{s=0}^{\infty} \cD(s+1,s)\,,
	\qquad
	{\rm Di}\otimes{\rm Di}
	=\cD(2,0)\oplus
	\bigoplus_{s=1}^{\infty} \cD(s+1,s)\,,	
\ea
where $\cD(\D,s)$ is the representation with spin $s$ and the conformal dimension $\D$. 
This mathematical theorem can be translated into the AdS/CFT language as:
all bilinear scalar operators (the tensor product of Rac) in free conformal scalar/spinor fields in three dimensions have an one-to-one correspondence with the massless higher spin fields in the bulk of AdS$_{4}$ 
(the representations of $\cD(s+1,s)$). The fact that the CFT fields are 
in the vectorial representation --- not in the adjoint one ---
singles out only bilinear operators as the single trace operators.
The contributions of higher trace
operators are suppressed in the large $N$ limit, hence we are left with a minimal set of operators in the spectrum.

A trivial but important property of this duality is that the boundary CFTs do not have any $1/N$ subleading contributions as they are free theories. An immediate implication of this property towards the bulk physics
is the absence of any quantum corrections. This is a remarkable feature because it necessitates a precise cancellation of infinitely many loop diagrams in the bulk. This aspect has been examined in the series of the papers \cite{Giombi:2013fka, Giombi:2014iua, Giombi:2014yra} and \cite{Beccaria:2014xda,
Beccaria:2014zma,Beccaria:2014qea,Beccaria:2015vaa,Beccaria:2016tqy}\footnote{See also \cite{Tseytlin:2013jya,Tseytlin:2013fca,Beccaria:2014jxa,Joung:2015eny} for related discussions on the conformal higher spin theory.} where the authors considered the simplest example, 
the vanishment of the  one-loop vacuum energy.\footnote{By `vacuum energy' we mean the log of the partition function about Euclidean AdS$_{d+1}$. We note that this terminology is different from that adopted in \cite{Giombi:2014yra}, where the same object was
referred to as `$S^d$ partition function'.}
 Since the vacuum energy in AdS$_{d+1}$ ought to be dual to the CFT zero-point function, they can only depend on the size of the radius of the boundary $S^d$\,. The summation of the vacuum energies over all field contents results in an infinite series: 
\be
	\sum_{s=0}^{\infty}\ \ 
	\parbox{63pt}{
	\begin{tikzpicture}
	\draw [blue, semithick] (0,0) circle [radius=1];
	\draw [semithick] (0,0) circle [radius=0.4];
	\node at (0,-0.55) {$\st s$};
	\end{tikzpicture}}
	=\sum_{s=0}^{\infty}\,\G^{\sst (1)}_{s}(z)=\G^{\sst (1)}(z)\,,
\ee
where $\G^{\sst (1)}_{s}(z)$ is the UV regularized (with a regulator $z$) vacuum energy  with the massless spin $s$ field in the loop. Two different methods have been considered to analyze this series. In the first method, the summation over $s$ is carried out before getting the function $\G^{\sst (1)}(z)$. The resulting vacuum energy is free from UV divergence and vanishes as $z\to 0$, for the minimal Vasiliev theory.
This method does not require an additional  regularization scheme and it is used for even $d$ cases where $\G^{\sst (1)}_{s}(z)$ have relatively simple forms. 
In odd $d$, however, the expression of $\G^{\sst (1)}_{s}(z)$ is more involved such that we cannot identify $\G^{\sst (1)}(z)$ with preceding method. This 
necessitates another approach.
In the second method, we take only the finite part $\G_{s}^{\sst (1)\, \rm ren}$ of $\G_{s}^{\sst (1)}(z)$ (and neglect the divergent part)
to end up with the series $\sum_{s=0}^{\infty}\G_{s}^{\sst (1)\,\rm ren}$\,.
This series is divergent so requires a new regularization in order to show that it indeed vanishes.

\bigskip

Motivated by these developments, we study the quantum property of the AdS theory
dual to the free scalar CFT in $SU(N)$ adjoint representation \cite{TseytlinHS3},
\be
	S_{\rm\sst CFT}[\bm\phi]
	=\int d^{d}x\,\tr\left[\bm\phi^\dagger\,\Box\,\bm\phi\right],
	\label{free matrix}
\ee
 by focusing on its vacuum energy. 
 There are several reasons which lead us to do so. Firstly, in contrast to the vector models, the set of single trace operators includes not only bilinear but also operators multi-linear in the field $\bm\phi$. This greatly extends the field content of the dual theory as compared to Vasiliev's theory. Standard AdS/CFT considerations lead us to expect that the holographic dual of such theory is the  Vasiliev higher spin theory coupled to 
 infinitely many massive higher spin fields.
 The whole spectrum organizes itself into
 infinitely many `Regge trajectories',\footnote{By Regge trajectory, we mean the 
 set of AdS fields dual to the CFT operators
 made by a fixed number of conformal fields:
  the $n$-th trajectory is dual to
 the CFT operators involving $n+1$ conformal fields.}
 each of which forms
 a `matter' multiplet of the higher spin algebra.\footnote{See Appendix C of \cite{Skvortsov:2015pea} 
 for a recent discussion.}
 Secondly, the putative holographic duality would closely mimic the dualities involving string theory in AdS in many ways. In particular, the theory already has interesting thermodynamics, exhibiting a Hagedorn phase transition \cite{Sundborg:1999ue,Aharony:2003sx} much like string theory \cite{Atick:1988si}. Thirdly, as a consequence of the usual AdS/CFT dictionary, we expect that the free CFT limit of stringy AdS/CFT dualities
 corresponds to taking the tensionless limit of string theory in AdS \cite{HaggiMani:2000ru,Sundborg:2000wp,wittenHS,Tseytlin:2002gz,Karch:2002vn,Gopakumar:2003ns,Bonelli:2003zu,Gopakumar:2004qb,Gopakumar:2004ys,Gopakumar:2005fx,Aharony:2006th,Yaakov:2006ce,Aharony:2007fs}.

Further, since we are working with a free CFT, the determination of the spectrum in closed form is available, even if it is a complicated task. As in the Flato-Fronsdal theorem which dictates the spectrum of the vector models, the spectrum of $SU(N)$ adjoint model can be identified by decomposing the multiple tensor products of singletons into irreducible representations:
\be
	\bigoplus_{n=2}^{\infty}\,{\rm Rac}^{\otimes n}=
	\bigoplus_{\D,s}\,N_{\D,s}\,\cD(\D,s)\,,
	\label{matrix decomp}
\ee
where $N_{\D,s}$ is the multiplicity of the representation $\cD(\D,s)$
and $n$ is the number of conformal fields. While doing this decomposition, the tensor product should be properly projected for the consistency with the cyclic invariance of trace \cite{Polyakov:2001af,Bianchi:2003wx,Beisert:2003te,Beisert:2004di,Spradlin:2004pp,Bianchi:2005ze}.

We now present a brief overview of our strategy. Firstly, we determine the operator spectra of the CFT which will be identified with the spectrum of the bulk theory. To determine the spectrum, we mainly take the most standard way of character analysis \cite{Barabanschikov:2005ri,Newton:2008au} but will also present the oscillator analysis for the $d=3$ case. These analyses give us information about the operator spectrum so that we can calculate in principle the corresponding AdS vacuum energies, knowing the one-loop contribution from each bulk field \cite{Camporesi:1990wm,Camporesi:1993mz,Camporesi:1994ga,Camporesi,Camporesi:1995fb}. However the difficulties arise in the summation of the vacuum energies from different fields. This is due to both the increasing complexity of the spectrum as higher and higher Regge trajectories are included, as well as the careful regularization of many formally divergent sums, as was already encountered in the vector model case \cite{Giombi:2013fka, Giombi:2014iua, Giombi:2014yra}. In order to surpass this problem, we introduce a new technique which enables us to access the resummed  vacuum energy directly from the character bypassing the steps of decomposition and resummation. This is realized in terms of a functional $\cF$ whose input is the character $\chi$ (or generalized partition function\footnote{By this we mean `refined' partition function $\tr\left(e^{-\beta H+ \sum_{i}\alpha_i J^i}\right)$ computed over the the \textit{one-particle} Hilbert space of the theory. Here $\beta$ is the temperature and $\alpha_i$ are chemical potentials for the Cartan
subalgebra of $so(d)\subset so(2,d)$.}) of the CFT and the output is the UV regularized AdS vacuum energy:
\be
	\sum_{\D,s}\, N_{\D,s}\
	\parbox{63pt}{
	\begin{tikzpicture}
	\draw [semithick, blue] (0,0) circle [radius=1];
	\draw [semithick] (0,0) circle [radius=0.4];
	\node at (0,-0.6) {$\st \D,s$};
	\end{tikzpicture}}
	=\cF[\chi](z)\,.
\ee
In this paper, 
we revisit the Vasiliev's theories
as test examples of applying the new method.
Then, we challenge the AdS theory dual 
to the $SU(N)$ adjoint matrix model.

\subsection*{Organization of the paper}

The paper is organized as follows. 
We begin with a review of unitary irreducible representations (UIRs) of the $d$--dimensional conformal algebra $so(2,d)$ in Section \ref{sec: spec}, including the introduction of singleton representations Di and Rac, construction of (reducible) representations by taking tensor products of singletons, as well as character formulae for the various UIRs. Based on the character of conformal algebra, we present decomposition rules of singleton tensor products, by using a generating function method. We also discuss an oscillator construction for arriving at these decomposition rules. 
Section \ref{sec: duality} contains a review of the heat kernel and zeta function formalism for computing one-loop effects in Euclidean Anti-de Sitter space, and general expectations from AdS/CFT duality for matching with free CFT answers.
In Section \ref{sec: AdS4}, we compute the spectral zeta function for AdS$_4$ using the results for the spectrum found in Section \ref{sec: spec}, and also by the new formalism alluded to above.
Section \ref{sec: AdS5} contains the extension of the above results to AdS$_5$ where we also discuss how mixed symmetry fields may be taken into account.
Section \ref{sec: conclusion} summarizes and concludes this paper and discusses related issues.
Appendices contain various additional details.

\section{Operator Spectrum of Free $SU(N)$ adjoint Model}
\label{sec: spec}

In this section, we will study the operator spectrum of free matrix models.
Many works have been devoted to this task 
\cite{Polyakov:2001af,Bianchi:2003wx,Beisert:2003te,Beisert:2004di,Spradlin:2004pp,Bianchi:2005ze,Barabanschikov:2005ri,Newton:2008au}.
Putting aside other interesting models, we only consider the simplest case of free scalar $SU(N)$ adjoint model. 
According to the standard scheme of AdS/CFT correspondence, in the large $N$ limit, single trace operators
are dual to the  \emph{single-particle} states (or fields) in the bulk theory.
Any single trace operator in a scalar CFT can be written as a linear combination of the operators,
\begin{equation}
\mbox{Tr}\Big[ 
\big(\partial_{\mu_{1}} \partial_{\mu_{2}} \cdots \partial_{\mu_{l}} 
\bm\phi \big)
\big(
\partial_{\nu_{1}} \partial_{\nu_{2}} \cdots \partial_{\nu_{m}} 
\bm\phi\big)
\cdots 
\big(
\partial_{\r_{1}} \partial_{\r_{2}} \cdots \partial_{\r_{n}} 
\bm\phi\big)\Big]\,.
\label{ST operators}
\end{equation}
These operators are in general reducible with respect to the conformal symmetry,
and can be decomposed into unitary irreducible representations (UIR).
Since the $d$-dimensional conformal symmetry 
and the isometry of AdS$_{d+1}$ are equivalent --- they are both $so(2,d)$ ---
the single trace operators carrying UIRs of the conformal symmetry 
are in one-to-one correspondence with the bulk fields carrying UIRs of the AdS isometry.
The decomposition does not mix
the operators \eqref{ST operators}
with different number of scalar fields in the trace,
hence can be performed for a given number of scalar fields.

The identification of all single trace operators 
reduces  to the decomposition of the operators \eqref{ST operators}
into UIRs for any number of scalar fields in the large $N$ limit.
Exact expressions of the decompositions ---
that is, the expression of UIR single trace operators in terms of $\bm\phi$ ---
involve rather complicated (anti)symmetrizations and contractions of indices.
However, for our purposes, it is sufficient  to find out the UIR labels of the 
resulting operators from the decompositions.
The latter task can be conveniently carried out 
relying on the representation theory of conformal symmetry $so(2,d)$\,.
The free scalar field carries a short UIR, called scalar singleton,
hence we are to analyze
the decomposition of multiple
tensor products of singletons into UIRs of $so(2,d)$\,.
The Lie algebra character is one of the most convenient tools for this analysis. 
In the following, we begin with a brief summary of the UIRs and 
the characters of  $so(2,d)$\,.
More detailed accounts and derivations can be found in \cite{Dolan:2005wy}.\footnote{A particularly accessible `physicist's account' for the UIRs of $so(2,d)$ is available in  \cite{Minwalla:1997ka}. The UIRs of $so(2,3)$ and $so(2,4)$ were first constructed in \cite{evans1967discrete} and \cite{mack1977all} respectively.
For the symmetries of the singleton representation,
see \cite{Bekaert:2009fg,Bekaert:2011js}.}

\subsection{Review: UIRs and Characters of $so(2,d)$}

The conformal algebra in $d$-dimension is isomorphic to $so(2,d)$.
The latter is generated by $M_{AB}$, whose commutation relations 
are given by 
\be
	[\,M_{AB}\,,\,M_{CD}\,]=
	i\left(\eta_{A[C}\,M_{B]D}
	-\eta_{B[C}\,M_{A]D}\right),
\ee
where the indices $A,B,\ldots$ run over $+,-,1,2,\ldots,d$ while \mt{a,b,\ldots=1,2,\ldots,d}\,. The non-vanishing components of the metric are
$\eta_{\pm\mp}=1$
and $\eta_{ab}=\delta_{ab}$\,. Its lowest weight (LW) representations $\mathcal V(\D,\bm\ell)$ are labeled by those of the $so(2)\oplus so(d)$ maximally compact subalgebra generated by $E=M_{+-}$ and $M_{ab}$\,. The $\D$ corresponds to the eigenvalue of the  generator $E$, and the $\bm\ell=(\ell_{1},\ldots,\ell_h)$ labels the irreducible representation of $so(d)$ with 
\be
	\ell_{1}\ge \cdots \ge \ell_{h-1} \ge 
	|\ell_{h}|\,;\qquad h=[\tfrac d2]\,;\qquad
	\ell_{h}\ge0
	\quad
	{\rm if}\quad d=2h+1\,,
\ee
where  $\ell_{i}$'s are either all integers, or all half-integers for a given representation.

The character of $\cV(\D,\bm\ell)$ is given by
\be
	 \chi_{\Delta,\bm\ell}(q,\bm x)=
	 \tr_{\cV(\D,\bm\ell)}\!\left[q^{E}\,x_{1}^{M_{12}}\cdots x_{h}{}^{M_{h(h+1)}}\right]
	=q^{\Delta}\,P_{d}(q,\bm x)\,\chi^{so(d)}_{\bm\ell}(\bm x)\,,
	\label{ch V}
\ee
where $\bm x=(x_{1},\ldots,x_{h})$ and $P_{d}(q,\bm x)$ is,
\be
P_{d}(q,\bm x)=\frac1{(1-q)^{d-2h}}
\prod_{i=1}^{h}\frac1{\left(1-q\,x_{i}\right)\left(1-q\,x_{i}^{-1}\right)}\,,
\label{Pd}
\ee
while $\chi^{so(d)}_{\bm\ell}(\bm x)$ is the character of $\bm\ell$ 
representation of $so(d)$\,. For even $d$, the character has a form
\be
\chi^{so(2h)}_{\bm\ell}(\bm x) = \frac{
\det\left[x_{i}{}^{k_{j}}+x_{i}{}^{-k_{j}}\right]+\det\left[x_{i}{}^{k_{j}}-x_{i}{}^{-k_{j}}\right]}
{2\,\D(\bm x)}\,,
\ee
and for odd $d$,
\be
\chi^{so(2h+1)}_{\bm\ell}(\bm x) =
\frac{\det\left[x_{i}{}^{k_{j}}-x_{i}{}^{-k_{j}}\right]}
{\D(\bm x)\,\prod_{i=1}^{h}\left(x_{i}{}^{\frac12}-x_{i}{}^{-\frac12}\right)}\,,
\ee
with $k_{i}=\ell_{i}+\frac d2-i$. The $\D(\bm x)$ is the Vandermonde determinant,
\be
	\D(\bm x)=\prod_{1\le i<j \le h}\left(x_{j}+x_{j}^{-1}-x_{i}-x_{i}^{-1}\right).
\ee

The irreducible representation $\cD(\D,\bm\ell)$ of $so(2,d)$ is the quotient of $\cV(\D,\bm\ell)$ by its maximal $so(2,d)$-invariant subspace $\cD(\D',\bm\ell')$ with a certain $\D'$ and $\bm\ell'$\,. The corresponding character of  $\cD(\D,\bm\ell)$ is then given by
\be
	 \chi_{\cD(\Delta,\bm\ell)}(q,\bm x)=\chi_{\Delta,\bm\ell}(q,\bm x)
	-\chi_{\cD(\Delta',\bm\ell')}(q,\bm x)\,.
	\label{ch D}
\ee
In the following, we summarize the UIRs $\cD(\D,\bm\ell)$ of $so(2,d)$
and their characters $\chi_{\cD(\D,\bm\ell)}$\,. 
For that, we need to define the number $p$ as
\be
	\ell_{1}=\ell_{2}=\cdots=|\ell_{p}|>\ell_{p+1}\,,
	\label{def p}
\ee
for a $so(d)$ representation $\bm\ell=(\ell_{1},\ldots,\ell_{h})$.

\subsubsection*{Long representations}

We consider first the representations $\cD(\D,\bm\ell)$
whose $so(d)$ part, $\bm\ell$,
satisfies
\be
	1\le p < \frac d2\,,
	\qquad
	\ell_1\ge d-2h\,,
	\label{semi-short}
\ee
which excludes the scalar $\bm\ell=(0,\ldots, 0)$ 
and the spinor $\bm\ell=(\frac12,\ldots,\pm\frac12)$ representations
in odd dimensions 
and  all the representations with $p=h$ in even $d$\,.
For the above class of representations, the 
unitarity bound is given 
in \cite{Metsaev:1995re} by
\be
	\D_{\bm\ell}= \ell_{1}+d-p-1\,.
	\label{D l}
\ee
Above the bound, the LW representation
does not develop any invariant subspace, hence
\be
	\cD(\D,\bm\ell)=\cV(\D,\bm\ell)\qquad [\,\D>\D_{\bm\ell}\,]\,.
\ee
These are \emph{long} representations.
Below the bound, \mt{\D<\D_{\bm\ell}}\,, the representation becomes 
non-unitary.

\subsubsection*{Semi-short representations}
If $\D$ \emph{saturates} the unitarity bound, i.e. \mt{\D=\D_{\bm\ell}}\,, the representation get shortened 
as
\be
	\cD(\D_{\bm\ell},\bm\ell)=
	\mathcal{V}(\D_{\bm\ell},\bm\ell)\,
	\big/\,\mathcal{D}\!\left(\D_{\bm\ell}+1,\bar{\bm\ell}\right),
\ee
where $\bar{\bm\ell}=(\bar\ell_{1},\ldots,\bar\ell_{h})$
is defined through $\bm\ell$ as
\be
	\bar\ell_{i}=\ell_{i}-\delta_{pi}\,.
	\label{bar ell}
\ee
These representations are often referred as to \emph{semi-short}
for the distinction from shorter representations, namely singletons,
which will be referred as to \emph{short}.
Notice that for $p>1$\,, we have $\D_{\bm\ell}+1=\D_{\bar{\bm\ell}}$\,, 
hence the invariant subspace corresponds again to a semi-short 
representation $\cD(\D_{\bar{\bm\ell}},\bar{\bm\ell})$\,.
Therefore, the quotient process should be repeated 
recursively until one reaches $p=1$\,.\footnote{This has been shown explicitly by means of an oscillator construction in \cite{Barabanschikov:2005ri} for the case of $d=4$.}
The characters of semi-short representations are given by 
\be
	\chi_{\cD(\D_{\bm\ell},\bm\ell)}(q,\bm x)=
	\chi_{\D_{\bm\ell},\bm\ell}(q,\bm x)
	-\chi_{\cD(\D_{\bm\ell}+1,\bar{\bm\ell})}(q,\bm x)\,,
\ee
which again can be recursively decomposed into $\chi_{\D_{\bm\ell},\bm\ell}$
with different $\bm\ell$'s. 
Both 
the semi-short and long representations
have the Gelfand-Kirillov (GK) dimension $d$ \cite{GK}:
they can be described by a function of $d$ continuous variables.
Hence they can be realized either as 
an operator on $d$-dimensional boundary without an on-shell condition
or as a field living on $AdS_{d+1}$\,.
More precisely,
in the bosonic case, $\cD(\D,\bm\ell)$ can be realized as a
boundary irreducible $so(d)$-tensor operator,\footnote{\label{fn}Here, $a_i(\ell_i)$ is a 
shorthand notation for a group of fully symmetric $\ell_i$ indices 
$a_{i|1}\,a_{i|2}\, \cdots\, a_{i|\ell_i}$.}
\be
	\cO_{\D}^{a_{1}(\ell_{1}),\ldots,
a_{h}(\ell_{h})}(x)\,,
\ee 
or as a  bulk irreducible $so(1,d)$-tensor field, 
\be
	\varphi_{\mu_{1}(\ell_{1}),\ldots,
\mu_{h}(\ell_{h})}(z,x)\,,
\ee
with the mass squared \cite{Metsaev:2003cu}, 
\be
    M_{\D,\bm\ell}^{2}=\frac{\D(\D-d)-\sum_{i=1}^h \ell_i}{R^2}\,,
	\label{mass}
\ee
where $R$ is the radius of AdS.
At the shortening point $\D=\D_{\bm\ell}$\,, 
the boundary  operator satisfies a conservation condition:
\be
	{\rm Y}_{\bar{\bm\ell}}\left[
	\partial_{a_{p|\ell_{p}}}
	\cO_{\D_{\bm\ell}}^{a_{1}(\ell_{1}),\ldots,a_{h}(\ell_{h})}(x)\right]=0\,,
\label{conserv}
\ee
whereas the bulk field admits a gauge symmetry
\be
	\delta\,\varphi_{\mu_{1}(\ell_{1}),\ldots,
\mu_{h}(\ell_{h})}(z,x)
={\rm Y}_{\bm\ell}\left[\partial_{\mu_{p|\ell_{p}}}
\varepsilon_{\mu_{1}(\bar\ell_{1}),\ldots,
\mu_{h}(\bar\ell_{h})}(z,x)\right].
\ee
Here ${\rm Y}_{\bm{\ell}}$ is the projection operator
to the irreducible Young diagram $\bm{\ell}$\,(see \cite{Metsaev:1995re,
Metsaev:2004ee,Metsaev:2014sfa} for the details).
Notice that the conservation condition corresponds to
the invariant subspace $\cD(\D_{\bar{\bm\ell}},\bar{\bm\ell})$\,.

Before moving to the short representation, let us consider
the example of the symmetric tensor representation 
$\bm\ell=(\ell,0,\ldots,0)=:(\ell,\bm0)$\,,
whose unitary bound is given by 
$\D_{(\ell,\bm 0)}=\ell+d-2$\,. 
The character is given simply by
\be
	\chi_{\cD(\D_{(\ell,\bm0)},(\ell,\bm0))}(q,\bm x)=
	\chi_{\D_{(\ell,\bm 0)},(\ell,\bm0)}(q,\bm x)
	-\chi_{\D_{(\ell,\bm 0)}+1,(\ell-1,\bm0)}(q,\bm x)\,,
\ee
since $p=1$ in this case.
This representation can be realized either as a conserved current $\cO^{a_{1}\cdots a_\ell}$ on the boundary or as a symmetric gauge field 
$\varphi_{\mu_{1}\cdots \mu_{\ell}}$ in the bulk.

\subsubsection*{Short representations: singletons}

The condition \eqref{semi-short}
leaves three exceptional cases,
where we get short representations,
instead of semi-short ones,
when $\D$ is on the boundary of unitarity.
The short representations have one less GK dimension,
that is $d-1$\,,
hence do
not admit a standard field theoretic realization
in the $d+1$ dimensional bulk
(see however the attempts \cite{Flato:1980we}).
More suitable realization of them is 
as boundary conformal field
operator subject to certain on-shell conditions.

\paragraph{Scalar singleton}
The first case is the scalar representation, 
$\bm\ell=(0,\ldots,0)=:\bm 0$\,, 
where the unitarity bound reads
\be
	 \D_{\bm 0}=\frac{d-2}2\,.
	 \label{D 0}
\ee
Above the bound $\D>\D_{\bm 0}$\,, we get a long representation.
On the border of the unitarity, $\D=\D_{\bm 0}$\,,
we have the scalar singleton,
\be
	\cD(\D_{\bm0},\bm 0)=
	\mathcal{V}(\D_{\bm0},\bm 0)\,\big/\,
	\mathcal{V}(\D_{\bm0}+2,\bm 0)\,.
\ee 
Its character is given by
\be
	\chi_{\cD(\D_{\bm0},\bm0)}(q,\bm x)=\chi_{\D_{\bm0},\bm0}(q,\bm x)
	-\chi_{\D_{\bm0}+2,\bm0}(q,\bm x)
	=q^{\frac{d-2}2}(1-q^{2})\,P_{d}(q,\bm x)\,.
\ee
The scalar singleton representation  can be realized as a conformal scalar $\phi$ on the boundary 
and the subspace $\cV(\D_{\bm0}+2,\bm 0)$ corresponds to the
LHS of the equation of motion 
for the conformal scalar, \mt{(\Box+\frac{d-2}{4(d-1)}\,\cR)\phi=0}\,.

\paragraph{Spinor singleton}

The second case is the spinor representation $\bm\ell=(\frac12,\ldots,\frac12)=:\bm{\frac12}$\,, where the unitarity requires
\be
	\D\ge \D_{\bm{\frac12}}=\frac{d-1}2\,.
	\label{D 1/2}
\ee
On the border, we get the spinor singleton,
\be
	\cD\left(\D_{\bm{\frac12}},\bm{\frac12}\right)=
	\mathcal{V}\left(\D_{\bm{\frac12}},\bm{\frac12}\right)\bigg/
	\mathcal{V}\left(\D_{\bm{\frac12}}+1,\bm{\frac12}\right).
\ee
It can be realized as a Dirac spinor $\psi$ on the boundary, 
and the subspace $\cV(\D_{\bm{\frac12}}+1,\bm{\frac12})$ corresponds to the equation of motion $\slashed\partial\,\psi=0$\,.
The character is given by
\be
	\chi_{\cD(\D_{\bm{\frac12}},\bm{\frac12})}(q,\bm x)
	=\chi_{\D_{\bm{\frac12}},\bm{\frac12}}(q,\bm x)
	-\chi_{\D_{\bm{\frac12}}+1,\bm{\frac12}}(q,\bm x)\nn
	=
	q^{\frac{d-1}2}(1-q)\,\chi^{so(d)}_{\bm{\frac12}}(\bm x)\,P_{d}(q,\bm x)\,,
\ee
with the $so(d)$ one,
\be
	\chi^{so(d)}_{\bm{\frac12}}(\bm x)
	=\prod_{i=1}^{h}\left(x_{i}{}^{\frac12}+x_{i}{}^{-\frac12}\right).
\ee

\paragraph{Higher spin singleton}

When $d$ is even,
the representations $\cD(\D,\bm\ell)$ with $\bm\ell=\bm s_{\pm}:=(s,\ldots,s,\pm s)$
also develop short representations on the unitarity bound:
\be
	\D_{\bm s_{\pm}}=s+\frac{d-2}2\,.
\ee
Since the above can be also written as $\D_{\bm s_{\pm}}=s+h-1$\,,
it corresponds in fact to particular cases of \eqref{D l}.
Moreover, for $\bm \ell=\bm 0$ and ${\bm{\frac12}}_+$,
it coincides with the bounds \eqref{D 0} and \eqref{D 1/2} for the scalar and spinor singletons.
Again, above the bound  the representations are long, while
on the border, we get the short representations,
\be
	\cD\left(\D_{\bm s_{\pm}},\bm s_{\pm}\right)=
	\mathcal{V}(\D_{\bm s_{\pm}},\bm s_{\pm})\,\big/\,
	\cD(\D_{\bm s_{\pm}}+1,\bar{\bm s}_{\pm})\,,
\ee
where $\bar{\bm s}_{\pm}$ is defined in \eqref{bar ell}
with \mt{(\bar{\bm s}_{-})_{h}=-(s-1)}\,.
For a more explicit expression of character,
we define
\be
	\bm e_{\pm,n}=(\underbrace{0,\ldots, 0}_{h-n}, 
	\underbrace{1, \ldots, 1,\pm 1}_{n})\,,
\ee
then we get
\be
	\chi_{\cD(\D_{\bm s_{\pm}},\bm s_{\pm})}(q,\bm x)
	=\sum_{n=0}^{h}\,(-1)^{n}\,
	\chi_{\D_{\bm s_{\pm}}\!+n\,,\,\bm s_{\pm}-\bm e_{\pm,n}}(q,\bm x)\,.
\ee
In the integer $s$ cases, higher spin singletons can be realized 
by boundary tensor fields
$\varphi^{a_{1}(s),\ldots,a_{h}(s)}$
and the subspace $\cD(\D_{\bm s_{\pm}}+1,\bar{\bm s}_{\pm})$ 
corresponds to the conservation condition \eqref{conserv}.
In terms of dual fields,\footnote{Here, $b_i[h]$ is a 
shorthand notation for a group of fully anti-symmetric $h$ indices 
$b_{i|1}\,b_{i|2}\,\cdots\, b_{i|h}$.}
\be
	\pi_{b_{1}[h],\ldots,b_{s}[h]}
	=\e_{b_{1}[h]\,a_{1|1}\, a_{2|1}\,\cdots\, a_{h|1}}
	\cdots\,\e_{b_{s}[h]\,a_{1|s}\,a_{2|s}\,\cdots\, a_{h|s}}\,
	\varphi^{a_{1}(s),\ldots,a_{h}(s)}\,,
\ee
the conservation condition get simplified as
\be
	\partial_{b_{1}}\pi_{b_{1}[h],\ldots,b_{s}[h]}=0\,.
\ee
This is the Bargmann-Wigner equation for massless higher spin field \cite{Bargmann:1948ck}.
The $\pm$ of $\bm s_{\pm}$ defines
different parity decompositions depending on the dimensionality. For even $h$ (that is $d=4m$),
they correspond to self-dual and anti-self-dual field,
whereas for odd $h$ (that is $d=4m+2$),
they correspond to chiral and anti-chiral field.
\subsubsection{Examples of lower dimensions}
Let us conclude the review of the UIRs and characters of $so(2,d)$
with the examples of lower dimensions $d=2,3,4$\,. 

\paragraph{$\bm{so(2,2)}$}

From \eqref{ch V}, the character of long representations are
\be
	\chi_{\D,\ell}(q,x)=q^{\D}\,P_{2}(q,x)\,\chi^{so(2)}_{\ell}(x)\,,
\ee
with $P_{2}$ and $so(2)$ character $\chi^{so(2)}$ given by
\be
P_{2}(q,x)=\frac1{(1-q\,x)(1-q\,x^{-1})}\,,
\qquad
\chi^{so(2)}_{\ell}(x) =x^{\ell}\,.
\ee
Since $so(2,2)\simeq so(1,2)\oplus so(1,2)$\,, 
the above character can be decomposed into that of $so(1,2)$ as
\be
	\chi_{\D,\ell}(q,x)=\chi^{so(1,2)}_{j}(z_{+})\,\chi^{so(1,2)}_{\bar j}(z_{-})\,,
\ee
where $\D,\ell$ are related to $j,\bar j$ as 
\be
	\D=j+\bar j\,,
	\qquad
	\ell=j-\bar j\,,
	\qquad
	q\,x=z_{+}\,,\qquad q\,x^{-1}=z_{-}\,,
\ee
and the character for $so(1,2)$
\be
	\chi^{so(1,2)}_{j}(z)=\frac{z^{j}}{1-z}\,.
\ee
Here, short representations correspond to the holomorphic
or anti-holomorphic ones,
\be
	\chi_{\cD(s,\pm s)}(q,x)=\chi_{s,\pm s}(q,x)
	-\chi_{s+1,\pm(s-1)}(q,x)
	=\chi^{so(1,2)}_{s}(z_{\pm})\,.
\ee
\paragraph{$\bm{so(2,3)}$}
In $d=3$\,, the character for long representations is given by
\be
	\chi_{\D,\ell}(q,x)=q^{\D}\,P_{3}(q,x)\,\chi^{so(3)}_{\ell}(x)\,,
	\label{so23 ch}
\ee
with  $P_{3}$ and $so(3)$ character given by
\be
P_{3}(q,x)=\frac1{(1-q)(1-q\,x)(1-q\,x^{-1})}\,,
\qquad
\chi^{so(3)}_{\ell}(x) =
\frac{x^{\ell+\frac12}-x^{-\ell-\frac12}}{x^{\frac12}-x^{-\frac12}}\,.
\ee
The only semi-short representation is
the symmetric tensor one, 
\be
	\chi_{\cD(\ell+1,\ell)}(q,x)=q^{\ell+1}\,P_{3}(q,x)
	\left(\chi^{so(3)}_{\ell}(x)-q\,
	\chi^{so(3)}_{\ell-1}(x)\right).
\ee
For short representations, we have scalar and spinor representations
corresponding to the characters,
\be
	\chi_{\rm Rac}(q,x)=\frac{q^{\frac12}\,(1+q)}{(1-q\,x)(1-q\,x^{-1})}\,,
	\qquad
	\chi_{\rm Di}(q,x)=\frac{q\,(x^{\frac12}+x^{-\frac12})}{(1-q\,x)(1-q\,x^{-1})}\,.
	\label{Di Rac}
\ee
These representations are 
often referred as Rac and Di following \cite{Flato:1978qz}.

\paragraph{$\bm{so(2,4)}$}

Finally, in $d=4$\,, the characters of long representations read 
\be
	\chi_{\D,(\ell_{1},\ell_{2})}(q,x_{1},x_{2})=q^{\D}
	\,P_{4}(q,x_{1},x_{2})\,\chi^{so(4)}_{(\ell_{1},\ell_{2})}(x_{1},x_{2})\,.
	\label{so24 ch}
\ee
Since $so(4)\simeq so(3)\oplus so(3)$\,,
the $so(4)$ character can be decomposed into
the $so(3)$ ones as
\be
	\chi^{so(4)}_{(\ell_{1},\ell_{2})}(x_{1},x_{2})
	=\chi^{so(3)}_{j_{+}}(x_{+})\,\chi^{so(3)}_{j_{-}}(x_{-})\,.
\ee
with
\be
	\ell_{1}=j_{+}+j_{-}\,,\qquad
	\ell_{2}=j_{+}-j_{-}\,,
	\qquad
	x_{1}\,x_{2}=x_{+}\,,
	\qquad
	x_{1}\,x_{2}^{-1}=x_{-}\,.
	\label{ell j}
\ee
We get the explicit form for $P_4$ from \eqref{Pd} as
\ba
	&&P_4(q,x_1,x_2)=
	\frac{1}{\left(1-q\,x_1\right)\left(1-q\,x_1^{-1}\right)
	\left(1-q\,x_2\right)\left(1-q\,x_2^{-1}\right)}\nn
	&&	=\,
	\frac{1}{
	\left(1-q\,{x_+}^{\!\frac12}\,{x_-}^{\!\frac12}\right)
	\left(1-q\,{x_+}^{\!\frac12}\,{x_-}^{\!\!-\frac12}\right)
	\left(1-q\,{x_+}^{\!\!-\frac12}\,{x_-}^{\!\frac12}\right)
	\left(1-q\,{x_+}^{\!\!-\frac12}\,{x_-}^{\!\!-\frac12}\right)}\,.
\ea
With these, the character of semi-short representations are given by
\ba
	&& \chi_{\cD(\ell_{1}+2,(\ell_{1},\ell_{2}))}(q,x_1,x_2)=
	q^{\ell_{1}+2}\, 
	P_{4}(q,x_{1},x_{2})\,\times\nn
	&&\qquad \times
	\left(\chi_{j_{+}}^{so(3)}(x_{+})\,\chi_{j_{-}}^{so(3)}(x_{-})-
	q\,\chi_{j_{+}-\frac12}^{so(3)}(x_{+})\,\chi_{j_{-}-\frac12}^{so(3)}(x_{-})\right).
\ea
while that of the short representations by
\ba
	&&\chi_{\cD(1,(0,0))}(q,x_1,x_2)
	=q\,(1-q^{2})\,P_{4}(q,x_1,x_2)\,,
	\label{char Rac5}\\
	&& \chi_{\cD(s+1,(s,\pm s))}(q,x_{1},x_{2})=
	q^{s+1}\,P_{4}(q,x_{1},x_{2})\,\times\nn
	&&\qquad \times
	\left(\chi_{s}^{so(3)}(x_{\pm})-q\,\chi^{so(3)}_{s-\frac12}(x_{\pm})\,\chi_{\frac12}^{so(3)}(x_{\mp})
	+q^{2}\,\chi_{s-1}^{so(3)}(x_{\pm})\right).
\ea
We shall use this form of the character in computing
corresponding one-loop diagrams.

\subsection{Decomposition of singleton tensor product}
\label{sec: Decomposition}

Any physical Hilbert space $\cH$ of conformal field theory in $d$ dimensions
carries a unitary representation of $so(2,d)$,
hence can be decomposed into the UIRs  as
\be
	\cH=\bigoplus_{\D,\bm\ell}N^{\cH}_{\cD(\D,\bm\ell)}\,\cD(\D,\bm\ell)\,,
	\label{decomp H}
\ee
where $N^{\cH}_{\cD(\D,\bm\ell)}$ are the multiplicities
of the UIR, $\cD(\D,\bm\ell)$ in $\cH$\,.
Via the state-operator and AdS/CFT correspondences,
each state carrying a UIR corresponds first to a CFT operator
then to a bulk field.
Since all UIRs of $so(2,d)$ are identified in the previous section, 
we can determine the bulk spectrum which corresponds to $\cH$ 
if the multiplicities $N^{\cH}_{\cD(\D,\bm\ell)}$ are determined. 
  
The typical way to extract the multiplicities from a given representation $\cH$
is the decomposition of its characters (or generalized partition function):
\begin{equation}
\chi_{\cH}(q,\bm x)= \tr_{\cH}\!\left[q^{E}\,
\,x_{1}^{M_{12}}\cdots x_{h}{}^{M_{h(h+1)}}\right]
=
\sum_{\Delta,\bm\ell} N^{\cH}_{\cD(\Delta,\bm\ell)}\,\chi_{\cD(\D,\bm\ell)}(q,\bm x)\,,
\end{equation}
where the character $\chi_{\cD(\D,\bm\ell)}$ 
 is given in \eqref{ch D} with \eqref{ch V}.
The standard partition function $Z_{\cH}(q)$
of the theory with Hilbert space $\cH$ is related to 
this character simply by
\be
	Z_{\cH}(q)=\chi_{\cH}(q,1,\ldots,1)\,,
\ee
but for the identification of the multiplicities, we need the full dependence
in $\bm x$\,. 
In principle we can use an orthogonality relation between characters
to extract the multiplicities \cite{Barabanschikov:2005ri}, but in practice it 
is not simple to evaluate the necessary integrals.
Instead, when the form of characters are simple enough, one can
use more plain functional properties.

In the following, we shall derive various decomposition formulas
by concentrating on  the $so(2,3)$ case.
Note however that this will not limit our analysis of bulk quantum effect 
to the $d=3$ case
because we shall adopt a new method later on
(hence, the rest of Section \ref{sec: Decomposition}
is not prerequisite for the following sections).

\subsubsection{Laurent expansion of character}

The key observation for the $so(2,d)$ character decomposition is that 
for a few lower dimensions the character take essentially a monomial form 
up to an overall function factor \cite{Newton:2008au}.
For $so(2,3)$, the character of $\cV(\D,\ell)$ satisfies
\be
	\frac{1-x^{-1}}{P_{3}(q,x)}\,\chi_{\D,\ell}(q,x)
	=q^{\D}\left(x^{\ell}-x^{-\ell-1}\right),
\ee
the multiplicities of $\cV(\D,\ell)$
can be obtained from Laurent expansion of 
the character as
\be
	N^{\cH}_{\cV(\Delta,\ell)}
	=\oint \frac{dx}{2\pi i\,x^{\ell+1}}\oint \frac{dq}{2\pi i\,q^{\D+1}}
	\frac{1-x^{-1}}{P_{3}(q,x)}\,\chi_{\cH}(q,x)\,.
\ee
Hence, with this formula, for any reducible representation $\cH$\,,
once its character $Z_{\cH}(q,x)$ is known, one can get the
decomposition formula,
\be
	\cH=\bigoplus_{\D,\ell}N^{\cH}_{\D,\ell}\,\cV(\D,\ell)\,,
\ee
in terms of the LW representation $\cV(\D,\ell)$\,.
Note that in this case, the multiplicity $N^{\cH}_{\cV(\D,\ell)}$ might
be negative integers.
Afterwards, the decomposition \eqref{decomp H} in terms of UIRs
can be obtained by recollecting $\cV(\D,\ell)$'s into
$\cD(\D,\ell)$'s,
then the multiplicities $N^{\cH}_{\cD(\D,\ell)}$ are non-negative integers. 

\subsubsection{Oscillator  representation}

For $d=3$ case, the oscillator representation of singletons
\cite{Fradkin:1986ka,Konshtein:1988yg} has proven crucial in describing higher spin algebra
as well as the Vasiliev's equations.
It is also useful in studying the Flato-Fronsdal theorem \cite{Flato:1978qz} and its extensions \cite{Vasiliev:2004cm}.
In this section, we show how the oscillator representation 
can be used to analyze the decomposition rule of the generic
singleton tensor products,
\be
	\cH={\rm Di}^{\otimes q}\otimes {\rm Rac}^{\otimes p-q}\,,
\ee
where Di and Rac refers (following the standard terminology \cite{Flato:1978qz}) respectively 
the spinor and scalar singleton representation:
${\rm Di}=\cD(1,\tfrac12)$
and ${\rm Rac}=\cD(\tfrac12,0)$\,.
This generalizes the Flato-Fronsdal theorem,
to the cases of generic powers, 
\be
	{\rm Di}^{\otimes q}\otimes {\rm Rac}^{\otimes p-q}
	=\bigoplus_{n=0}^{\infty}\bigoplus_{2s=0}^{\infty}
	N^{[q,p-q]}_{(n+2s,n)}\,\cD\Big(s+n+\frac{p}2,s\Big).
	\label{Di * Rac}
\ee
The explicit expression of  $N^{[q,p-q]}_{(m,n)}$ can be 
identified using characters, whereas 
the oscillator representation
provides a simple combinatoric account for 
the multiplicities: 
\begin{framed}
	\noindent
	The multiplicity $N^{[q,p-q]}_{(m,n)}$
	of the tensor-product decomposition \eqref{Di * Rac}
	is equal to the 
	number of components in\ \,
	\parbox{60 pt}{
	\begin{tikzpicture}
	\draw (0,0) rectangle (2,0.3);
	\draw (0,0) -- (0,-0.3) -- (1.6,-0.3) -- (1.6,0);
	\node at (1,0.15){$\st m$};
	\node at (0.8,-0.15){$\st n$};
	\end{tikzpicture}} representation of $O(p)$\,,
	which are odd (and even) under the reflection of 
	first $q$ directions (last $p-q$ directions).
	\label{O(p)}
\end{framed}

The singleton representations of $so(2,3)$ algebra 
are realized by two sets of oscillators:
\be
	[\,a\,, a^{\dagger}\,]=1=[\,b\,,b^{\dagger}\,]\,.
\ee
In terms of these oscillators, the generators of $so(2)\oplus so(3)$
are given by
\be
	E=\frac12\left(a^{\dagger}\,a+b^{\dagger}\,b+1\right)\,,
	\qquad 
	J_{3}=\frac12\left(a^{\dagger}\,a-b^{\dagger}\,b\right)\,,
	\qquad 
	J_{+}=a^{\dagger}\,b\,,\qquad J_{-}=b^{\dagger}\,a\,.
\ee
The lowering operators are given by
\be
	M^{-}{}_{-}=a^{2}\,,
	\qquad
	M^{-}{}_{+}=b^{2}\,,
	\qquad 
	M^{-}{}_{3}=a\,b\,,
	\label{lowering}
\ee
and raising operators as the complex conjugate of the above.
In this oscillator representation of $so(2,3)$, two singletons, Di and Rac, are given 
as follows:

\begin{itemize}
\item

Rac is the representation
whose lowest weight state is the Fock vacuum 
$|0\ra$
with \mt{a\,|0\ra=0=b\,|0\ra}\,.
Obviously the lowering operators \eqref{lowering} annihilate this state,
and generic states of the Rac representation
are constructed by acting raising operators on $|0\ra$,
\be
	(M^{+}{}_{+})^{m}\,(M^{+}{}_{-})^{n}\,(M^{+}{}_{3})^{\ell}\,|0\ra=
	(a^{\dagger})^{2m+\ell}\,(b^{\dagger})^{2n+\ell}\,|0\ra\,.
\ee
Hence, they have even number of oscillators.
By acting $E$ and $J_{3}$ on $|0\ra$\,, we can immediately see that 
$|0\ra$ defines $\cD(\tfrac12,0)$\,.

\item

Di is the representation whose lowest weight state is 
the doublet $a^{\dagger}\, |0\ra \oplus b^{\dagger}\,|0\ra$\,.
Since these states involve only one creation operator,
they are still annihilated by the lowering operators \eqref{lowering}. 
The generic states of Di have odd number of oscillators. 
By acting $E$ and $J_{3}$ on $a^{\dagger}|0\ra$\,
(the highest $J_{3}$ state), we find
the doublet vacuum defines $\cD(1,\tfrac12)$\,.

\end{itemize}

\paragraph{Higher spin algebra}

This oscillator representation makes clear that
the higher spin algebra is the maximal symmetry 
of singleton representations,
namely the endomorphism of Rac (or Di) \cite{Konshtein:1988yg}. 
Since the singletons are constructed by acting 
the creation operators $a^{\dagger}$ and $b^{\dagger}$ on the Fock vacuum (or
the doublet $a^{\dagger}\, |0\ra \oplus b^{\dagger}\,|0\ra$ for Di),
any operators even orders in oscillators,
\be
	(a^{\dagger})^{m}\,(b^{\dagger})^{n}
	\,a^{\,p}\,b^{\,q}\qquad [m+n+p+q\in 2\,\mathbb N]\,,
\ee
belongs to the endomorphism, so the higher spin symmetry. 
The precise relation to the usual oscillators of high spin algebra
reads
\be
	y_{1}=a+b^{\dagger}\,, \qquad y_{2}=i\,(a^{\dagger}-b)\,,
	\qquad \bar y_{\dot \a}=(y_{\a})^{\dagger}\,.
\ee

\paragraph{Tensor product} We now consider tensor products of
$p$ singleton representations.
They are realized by $p\times 2$ sets of oscillators:
\be
	[\,a_{i}\,,a^{\dagger}_{j}\,]=\delta_{ij}=
	[\,b_{i}\,,b^{\dagger}_{j}\,]
	\qquad [i=1,\ldots,p]\,.
\ee
A generator $T$ of $so(2,3)$ is represented by
\be
	T=T_{1}+\cdots+T_{p}\,,
\ee
where $T_{i}$ is the representation given only by the $i$-th oscillators.
For instance, we have
\be
	E=\frac12(N_{a}+N_{b}+p)\,,
	\qquad
	J_{3}=\frac12(N_{a}-N_{b})\,,
\ee
with
\be
	N_{a}=\sum_{i=1}^{p}\,a^{\dagger}_{i}\,a_{i}\,,
	\qquad
	N_{b}=\sum_{i=1}^{p}\,b^{\dagger}_{i}\,b_{i}\,,
	\qquad
	J_{+}=\sum_{i=1}^{p}\,a^{\dagger}_{i}\,b_{i}\,,\qquad
	M_{-}{}^{-}=\sum_{i=1}^{p}\,a_{i}^{2}\,.
\ee
We are looking for LW states in the $k$ singleton tensor product space.
Such LW states are not singlet under $so(3)$
so we focus only on the highest $J_{3}$ state among the $so(2,3)$ 
LW states. 
Such a state with $(E,J_{3})=(\D,s)$ is an eigenstate of the number operators
$(N_{a},N_{b})$\,:
\be
	n_{a}=\D+s-\frac p2\,,\qquad n_{b}=\D-s-\frac p2\,,
\ee
and can be expressed as
\be
	C_{n_{a},n_{b}}(\bm a^{\dagger},\bm b^{\dagger})\,
	|0\ra\,,
\ee
where $\bm a^{\dagger}=(a_{1}^{\dagger},\ldots,a_{p}^{\dagger})$ and 
$\bm b^{\dagger}=(b_{1}^{\dagger},\ldots,b_{p}^{\dagger})$ are $p$-dimensional vectors.
The function $C_{n_{a},n_{b}}(\bm x,\bm y)$ satisfies
\be
	C_{n_{a},n_{b}}(\l_{a}\,\bm x,\l_{b}\,\bm y)
	=\l_{a}^{n_{a}}\,\l_{b}^{n_{b}}\,C_{n_{a},n_{b}}(\bm x,\bm y)
	\qquad [\l_{a},\l_{b}\in \mathbb{R}_{>0}]\,.
\ee
Then, in terms of this function, the vanishing $M^{-}{}_{-}$ 
and $J_{+}$ conditions read
\be
	\partial_{\bm x}^{2}\,C_{n_{a},n_{b}}(\bm x,\bm y)=0\,,
	\qquad
	\bm x\cdot\partial_{\bm y}
	C_{n_{a},n_{b}}(\bm x,\bm y)=0\,.
	\label{M condition}
\ee
This defines actually two-row $O(k)$ Young diagram 
where the length of first and second rows are $n_{a}$ and $n_{b}$, respectively.
More precisely,
\be
	C_{n_{a},n_{b}}(\bm x,\bm y)=
	C_{i_{1}\cdots i_{n_{a}},j_{1}\cdots j_{n_{b}}}
	\frac{x_{i_{1}}\cdots x_{i_{n_{a}}}}{n_{a}!}
	\frac{y_{j_{1}}\cdots y_{j_{n_{b}}}}{n_{b}!}\,.
\ee
and 
\be
	C_{i_{1}\cdots i_{n_{a}},j_{1}\cdots j_{n_{b}}}
	\sim\,
	\parbox{115 pt}{
	\begin{tikzpicture}
	\draw (0,0) rectangle (4,0.4);
	\draw (0,0) -- (0,-0.4) -- (3,-0.4) -- (3,0);
	\node at (2,0.2){$ n_{a}$};
	\node at (1.5,-0.2){$n_{b}$};
	\end{tikzpicture}}\ .
\ee
Hence, the number of LW states with fixed $n_{a}$ and $n_{b}$ 
correspond to the dimensions of $(n_{a},n_{b})$ Young diagrams, 
${\rm dim}\left(\p^{\sst O(p)}_{(n_{a},n_{b})}\right)$\,:
\ba
	&&{\rm dim}\left(\p^{\sst O(k)}_{(n_{a},n_{b})}\right)
	 =\\
	&&=\,
	 \frac{(n_a-n_b+1)}{(n_a+1)!\,n_b!} \frac{(k+n_a-4)!}{(k-2)!}
	 (k+2n_a-2)(k+n_a+n_b-3) 
	 \frac{(k+n_b-5)!}{(k-4)!}
	 (k+2n_b-4)\,.\nonumber
	\label{young_dim}
\ea

So far, we did not care where belong the LW states we found.
Therefore, they correspond to tensor products of
Di\,$\oplus$\,Rac,
\be
	({\rm Di}\oplus{\rm Rac})^{\otimes p}
	=\bigoplus_{n=0}^{\infty}\bigoplus_{2s=0}^{\infty}
	{\rm dim}\left(\p^{\sst O(p)}_{(n+2s,n)}\right)\,
	\cD\Big(s+n+\frac p2,s\Big)\,. 
	\label{DiRac_p}
\ee
Now, let us consider tensor products of Di's and Rac's.
The Fock space of Di and Rac are constructed by odd and even numbers of oscillators, respectively. Hence, 
Di and Rac carry the alternating ($-1$) and the trivial ($+1$) representations
of  $\mathbb Z_{2}=\{1,\s\}$ generated by the oscillator sign flip operation
$\s\,:\,\left(a^{\dagger},b^{\dagger}\right)\mapsto\,\left(-a^{\dagger},-b^{\dagger}\right) $.
When considering the tensor product of $k$ singletons, 
the group $\mathbb Z_{2}$ extends to $\mathbb Z_{2}^{\otimes p}$
which is the  reflection subgroup of $O(p)$ and
generated by 
\be
	\s_{i}\,:\quad (a^{\dagger}_{j}, b^{\dagger}_{j})\ \ \longmapsto \ \ 
	(-1)^{\delta_{ij}}\,(a^{\dagger}_{j}, b^{\dagger}_{j})\,.
\ee
Renaming Rac=S$^{(+1)}$ and Di=S$^{(-1)}$\,, 
the tensor product
$\rm{S}^{(\e_{1})}\otimes \rm{S}^{(\e_{2})}\otimes \cdots\otimes 
\rm{S}^{(\e_{p})}$
carries the $(\e_{1},\ldots,\e_{p})$ representation of $\mathbb Z_{2}^{\otimes p}$\,.
We consider the branching $O(p) \downarrow \mathbb Z_{2}^{\otimes p}$ of 
the $(n_{a},n_{b})$ Young Diagram representation:
\be
	\p^{\sst O(p)}_{(n_{a},n_{b})}\,\Big|_{\mathbb Z_{2}^{\otimes p}}=
	\bigoplus_{\e_{1}=\pm1}\cdots \bigoplus_{\e_{p}=\pm1}N^{(\e_{1},\ldots,\e_{p})}_{(n_{a},n_{b})}\,
	\pi^{\sst \mathbb Z_{2}^{\otimes p}}_{(\e_{1},\ldots,\e_{p})}\,,
	\label{branching}
\ee
where $N^{(\e_{1},\ldots,\e_{p})}_{(n_{a},n_{b})}
=N^{[q,p-q]}_{(n_{a},n_{b})}$ are the multiplicities ($q$ is 
the number of Di's)
and it also gives the multiplicities of the tensor product decompositions.
Therefore, the tensor product rule for $\rm{S}^{(\pm1)}$ can be written as
\eqref{Di * Rac}.
The explicit expression for $N^{[q,p-q]}_{(n+2s,n)}$
can be found using the combinatorics 
and its generating function 
turns out to coincide with the $so(2,3)$ character. 
See Appendix \ref{app: oscil} for the details.

\subsection{Single trace operators}
\label{sec: single trace}

So far, in considering tensor products of singletons, we have not 
taken any permutation symmetry of singletons into account.
Let us denote such tensor-product spaces as
\be
	T^{\sst (n)}(V)=V_1\otimes \cdots \otimes V_n\,,
\ee
where $V_i$ is the space of singleton representation and the subscript $i$ 
is introduced to distinguish different copies of $V$.
Notice however that the single trace operators \eqref{ST operators}
are invariant under cyclic permutations of $\partial^k\bm\phi$'s
 due to the cyclicity of trace.
This means that the space of the $n$-th order single trace operators 
corresponds not to $T^{\sst (n)}(V)$ but to 
the subspace $T_{\rm cyc}^{\sst (n)}(V)\subset T^{\sst (n)}(V)$,
\be
	T^{\sst (n)}_{\rm cyc}(V)
	= \bigoplus_{\p\in C_n} V_{\p(1)}\otimes \cdots \otimes V_{\p(n)}
	= \bigoplus_{i=1}^{n} V_{i}\otimes V_{i+1}
	\otimes \cdots \otimes V_{n}\otimes V_{1}\otimes
	\cdots \otimes V_{i-1}\,,
\ee
 which is invariant under the
actions of the cyclic group $C_n$. Therefore, when single trace operators admit symmetries such as cyclic permutations,
we have to decompose the properly symmetrized tensor product   of singletons 
into the UIRs of $so(2,d)$\,.
Depending on the symmetry of the space, single trace operators have different symmetries,
hence different tensor product should be used.
In this paper, we focus on  the CFT 
where the scalar field $\bm\phi$ takes value in the adjoint representation of $SU(N)$\,: under the action of an $SU(N)$ element $\bm a$\,, 
the field transforms as 
\be
	\bm\phi\ \to\ \bm a\,\bm\phi\,\bm a^{-1}\,.
	\label{adj}
\ee
In this case, the matrix $\bm\phi$ does not admit any particular symmetry
and its single trace operators, invariant under \eqref{adj}, only admit the cyclic symmetries.
The character of the cyclic tensor-product space $T^{\sst (n)}_{\rm cyc}(V)$ is no more $[\chi_{V}(g)]^{n}$ but 
given by 
\cite{Polyakov:2001af,Bianchi:2003wx,Beisert:2003te,Beisert:2004di,Spradlin:2004pp,Bianchi:2005ze,Barabanschikov:2005ri,Newton:2008au}
\be
		\chi_{{\rm cyc}^n}(g)=\frac1{n}\sum_{k|n}
	\varphi(k)\left(\chi_{V}(g^{k})\right)^{\frac nk}
	\qquad \big[\,g^{k}=(q^{k},x_{1}^{k},\ldots, x_{h}^{k})\,\big],
	\label{n adj}
\ee
where $k|n$ means the $k\in\{1,\ldots,n\}$ which divides $n$ 
and $\varphi(k)$ is the Euler totient function
which counts the number of relative primes of $k$ in $\{1,\ldots,k\}$\,.
The derivation of the above character is a result of the Polya's enumeration theorem.
Hence, the character for the entire space of singlet trace operators 
is given by the sum of \eqref{n adj} over $n\ge2$
(the \mt{n=1} contribution drops out due to  \mt{\tr(\bm\phi)=0}).
By changing the summation as $\sum_{n=1}^{\infty}\sum_{k|n}=\sum_{k=1}^{\infty}\sum_{m=1}^{\infty}$
with $n=m\,k$\,, we can perform the summation over $m$ and get
\be
	\chi_{\rm adj}(g)=\sum_{n=2}^\infty \chi_{{\rm cyc}^n}(g)=-\chi_{V}(g)
-\sum_{k=1}^{\infty}\,\frac{\varphi(k)}{k}\,\log\!\left(1-\chi_{V}(g^{k})\right).
\label{cyc}
\ee
By decomposing this character into the UIRs, we can 
identify the spectrum of all single trace operators in the free scalar $SU(N)$ model.
We remark here that the above formula relies on the infinite summation over $m$ giving the $\log$ function. Hence, it turns out that the partition function develops singularities at finite non-zero values of $\beta$ in contrast to the vector models \cite{Shenker:2011zf}. 
When the CFT$_d$ is placed on $S^1\times S^{d-1}$, $\beta$ has the physical interpretation of inverse temperature and this divergence is related to the Hagedorn phase transition \cite{Sundborg:1999ue,Aharony:2003sx}. In our case $\beta$ is just a parameter which is useful for counting conformal dimensions of primaries. Nonetheless, we will see that these singularities play an important role in our vacuum energy computations and indeed introduce new ambiguities and subtleties not presented in the vector model CFT holography.

\subsection{Explicit Examples of Decompositions}
\label{sec: decomp ex}

Let us conclude this section with a few examples of singleton tensor-product decompositions. We consider various bosonic tensor-products of Rac's and Di's up to order four. About the single trace operators, we confine ourselves to the scalar field cases, that involve only Rac's.
In $O(p)$ Young diagrams, the number of boxes cannot exceed $p$ for the first two columns. Therefore, the decomposition rules for two and three singleton tensor-products are particularly simple.

\subsubsection*{Two singletons}

For the tensor-products of two singletons, we recover the well known result of Flato and Fronsdal:
\ba
	{\rm Rac}^{\otimes 2}
	=\bigoplus_{s=0}^{\infty} \cD(s+1,s)\,,
	\qquad
	{\rm Di}^{\otimes 2}
	=\cD(2,0)\oplus
	\bigoplus_{s=1}^{\infty} \cD(s+1,s)\,.
\ea
Let us now consider the  adjoint model.
Its operator spectrum makes use of the cyclic tensor-product,
\be
	T^{\sst (2)}_{\rm cyc}({\rm Rac})=
	\bigoplus_{n=0}^{\infty} \cD(2n+1,2n)\,,
\ee
which actually coincides with the result of the $O(N)$-vector model.

\subsubsection*{Three singletons}

For the tensor-products of three singletons, we obtain
\ba
	{\rm Rac}^{\otimes 3}\eq
	\bigoplus_{s=0}^{\infty}\,
	(s+1)\left[
	\cD(s+\tfrac{3}2,s)
	\oplus 
	\cD(s+\tfrac{7}2,s+1)\right],\\
	{\rm Di}^{\otimes2}\otimes{\rm Rac}
	\eq
	\bigoplus_{s=0}^{\infty}\,
	(s+1)\left[
	\cD(s+\tfrac{5}2,s)
	\oplus 
	\cD(s+\tfrac{5}2,s+1)\right]. \label{rac3_decom}
\ea
In the adjoint models, the spectrum of single trace operators is given by
\be
	T^{\sst (3)}_{\rm cyc}({\rm Rac})=
	\bigoplus_{s=0}^{\infty}
	\left(s+1+2\,[-\tfrac s3]\right)
	\left[
	\cD(s+\tfrac{3}2,s)
	\oplus 
	\cD(s+\tfrac{7}2,s+1)\right],
\ee
where $[x]$ is the biggest integer not greater than $x$\,.

\subsubsection*{Four singletons}

 For the tensor-products of four singletons, we obtain
\ba
	{\rm Rac}^{\otimes 4} \eq \bigoplus_{s=0}^{\infty}
	\frac{(1+s)(2+s)}2\,
	\cD(s+2,s) \nn
	&&\,\oplus \bigoplus_{s=0}^{\infty}\bigoplus_{n=1}^{\infty}
	\frac{(2n+2s+1) (2s+1)+3 (-1)^n}4\,
	\cD(s+n+2,s)\,, \label{rac4_decom}
\ea
\ba
	{\rm Di}^{\otimes2}\otimes{\rm Rac}^{\otimes 2}\eq \bigoplus_{s=0}^{\infty}
	\frac{(1+s)(2+s)}2\,
	\cD(s+2,s) \nn
	&&\,\oplus \bigoplus_{s=0}^{\infty}\bigoplus_{n=1}^{\infty}
	\frac{(2n+2s+1) (2s+1)- (-1)^n}4\,
	\cD(s+n+2,s)\,,
\ea
\ba
	{\rm Di}^{\otimes4}\eq  \bigoplus_{s=0}^{\infty}\, 
	\frac{s(s-1)}2\,
	\cD(s+2,s)\nn
	&&\,\oplus \bigoplus_{s=0}^{\infty}\bigoplus_{n=1}^{\infty}
	\frac{(2n+2s+1) (2s+1)+3 (-1)^n}4\,
	\cD(s+n+2,s)\,.
\ea
We can also obtain  the spectrum of single trace operators for  the $SU(N)$ adjoint model,
but the formula become too lengthy and does not seem to be illuminating. 

It is worth to note that the spectra with four fields
contain the operators $\cD(s+2,s-1)$ of twist $3=(s+2)-(s-1)$
which can be interpreted
as the one dual to the higher spin Goldstone modes \cite{Bianchi:2003wx,Beisert:2003te,Beisert:2004di,Bianchi:2005ze}. 
In fact, only the order four and six can give this contributions
because the minimum twist $\t=\D-s$ of the spectrum
is larger than $n/2$ where $n$ is the number of the conformal fields.
The massless higher spin fields 
from the order two may acquire masses after combining 
with these modes from the order four or six.
The fact that the Goldstone modes arises at these orders 
is the particularity of three dimensional scalar models where the conformal weight of 
scalar field is $1/2$ hence requires two or four more orders to give the Goldstone ones.
In four dimensions where scalar field has weight one, 
only the order three operators can give the Goldstone modes.
In particular, the operator dual to the scalar Goldstone
can trigger a marginal deformation.

As one can see from the above results, the analytic formulas for the
multiplicities of single trace operators become highly non-trivial as 
the number of Rac increases:
the order four cyclic result would not fit in a single page.
Consequently, the summation of physical quantities over such spectrum
becomes practically intractable apart from a first few powers of Rac.
In the following sections, we nevertheless makes use of this decomposition
for some concrete calculations, but eventually proceed in a new approach.

\section{AdS/CFT and Bulk Vacuum Energy}
\label{sec: duality}

In this paper, we aim to study the AdS theory which is dual to free scalar CFT
in adjoint representation of $SU(N)$ by computing their one-loop vacuum energy.
Before entering to the analysis of AdS side, let us remind the general picture behind 
this correspondence.

\subsection{Holography for Free Matrix Model CFTs}

Let us first consider  the CFT side which is
described by a scalar action $S_{\rm\sst CFT}[\bm\phi]$
where $\bm\phi$ takes value in a matrix space of dimension $\mathsf N$\,.
For a $SU(N)$ adjoint model, $\mathsf N=N^2-1$\,. Let $\{\cO^I_{\D,\bm\ell}\}$ 
denote the full set of single trace primary operators: $\D$ and $\bm\ell$ are the labels for the 
conformal weight and spin,\footnote{In more than four spacetime dimensions, $\bm\ell$ indicates the set of all quantum numbers required to specify the spin of the field.}
and  $I$ is for the multiplicity.
The connected correlators of single trace operators take the form,
\be
	\big\la\, \cO^{I_1}_{\D_1,\bm\ell_1}(x_1)\,\cdots\,
	\cO^{I_n}_{\D_n,\bm\ell_n}(x_n)\,\big\ra_{\rm con}
	=\sum_k C^{\{I_i\}}_{\{\D_i,\bm\ell_i\}\, k}\,
	G^k_{\{\D_i,\bm\ell_i\}}(\{x_i\})\,,
\ee
where the functions $G^k_{\{\D_i,\bm\ell_i\}}(\{x_i\})$ are 
the  model independent tensor structures (labeled by $k$) allowed by the conformal symmetry,
whereas
the coefficients $C^{\{I_i\}}_{\{\D_i,\bm\ell_i\}\, k}$ 
encode the particularity of the model\footnote{See \cite{Maldacena:2011jn,Maldacena:2012sf,Stanev:2012nq,Stanev:2013qra} for explicit expressions for these tensor structures for three and four-point functions in CFT$_3$ and CFT$_4$.}.
In the convention that the operators are not normalized by $\mathsf N$\,,
the coefficients $C$ (with the labels ${}^{\{I_i\}}_{\{\D_i,\bm\ell_i\}\, k}$ suppressed)
admit a $1/{\mathsf N}$ expansion: 
\be
	C
	=\mathsf N\,C^{(0)}
	+C^{(1)}
	+\frac1{\mathsf N}\,C^{(2)}+\cdots.
	\label{1/N}
\ee
A particular property of free CFTs
is that the $1/{\mathsf N}$ expansion becomes
exact with the leading term alone: $C^{(n\ge1)}=0$\,. This property  is a triviality from the CFT point of view,
but it imposes a highly non-trivial requirement to the dual theory in AdS.
Before moving to  the AdS side, let us rephrase this property
in terms of the  generating function $F_{\rm\sst CFT}[h]$ of connected correlators.
The latter admits the path-integral representation,
\be
	\exp\left(-F_{\sst\rm CFT}[h]\right)
	=\int D\bm\phi\,
	\exp\bigg(-S_{\rm\sst CFT}[\bm\phi]+\int d^dx\,\sum_{\D,\bm\ell,I}
	\big\la\, h^I_{\D,\bm\ell}\,\big|\,\cO^I_{\D,\bm\ell}\,\big\ra\bigg)\,,
\ee
where $h^I_{\D,\bm\ell}$ are the sources for single trace operators
and $\la\cdot|\cdot\ra$ means the index contraction.
Again, $F_{\sst\rm CFT}$ of a generic matrix-model CFT admits a $1/{\mathsf N}$ expansion:
\be
	F_{\rm\sst CFT}
	=\mathsf N\,F_{\rm\sst CFT}^{(0)}
	+F_{\rm\sst CFT}^{(1)}
	+\frac1{\mathsf N}\,F_{\rm\sst CFT}^{(2)}+\cdots,
\ee
but that of free CFTs have vanishing $F_{\rm\sst CFT}^{(n\ge1)}$.

The AdS theory has the field contents $\{\varphi^I_{\D,\bm\ell}\}$ (with the masses \eqref{mass}) which are in one-to-one correspondence with the single trace operators. 
The conjecture states that the CFT physical quantity $F_{\rm\sst CFT}[h]$ coincides with 
the AdS one $\G_{\rm\sst AdS}[h]$ given by
\be
	\exp\left(-\G_{\rm\sst AdS}[h]\right)
	=\prod_{\D,\bm\ell, I}\int_{\varphi|^{\phantom{g}}_{\partial\rm AdS}=h}
	 \cD\varphi^I_{\D,\bm\ell}\,
	\exp\left(-\frac1{G}\,S_{\rm\sst AdS}[\varphi]\right),
\ee
where the subscript \mt{\varphi|^{\phantom{g}}_{\partial\rm AdS}=h} of the path-integral means that the fields are subject to the Dirichlet-like boundary condition,
\be
	\varphi^I_{\D,\bm\ell}\sim z^\D\,h^I_{\D,\bm\ell}
	\qquad {\rm as}\qquad z\to 0\,,
	\label{bc}
\ee
where $z$ is the radial variable in the Poincar\'e coordinate of AdS
whose boundary is located at $z=0$.
By denoting the unique classical solution  smooth in the interior of (Euclidean) AdS and satisfying \eqref{bc} as $\varphi^I_{\D,\bm\ell}(h)$\,,
we can split the fields into the classical background and the quantum fluctuation parts as
\mt{\varphi^I_{\D,\bm\ell}=\varphi^I_{\D,\bm\ell}(h)+\pi^I_{\D,\bm\ell}}\,.
Then, the $G$ expansion of $\G_{\rm\sst AdS}$,
\be
	\G_{\rm\sst AdS}[h]=
	\frac1G\,\G^{\sst (0)}_{\rm\sst AdS}[h]
	+\G^{\sst (1)}_{\rm\sst AdS}[h]+G\,\G^{\sst (2)}_{\rm\sst AdS}[h]+\cdots,
\ee
admits the diagrammatic interpretation that 
the 1PI scattering amplitudes for $\pi^I_{\D,\bm\ell}$
are given by
\be
	\frac{\d}{\d h_1}\cdots \frac{\d}{\d h_n}\G^{\sst (\ell)}[h]\,\Big|_{h=0}
	=\	\parbox{100pt}{
	\begin{tikzpicture}
	\draw [blue, semithick] (0,0) circle [radius=1.4];
	\draw [fill=lightgray,lightgray] (0,0) circle [radius=0.65];
	\node at (0,0) {\footnotesize $\ell$ loops};
	 \foreach \n in {0,1,2,...,8}
      \draw[semithick] (180-30*\n:0.65) -- (180-30*\n:1.4);
      \foreach \n in {1,...,8}
      \node at (180-30*\n: 1.55) {\tiny \n};
      \node at (180: 1.55) {$\st n$};
        \foreach \n in {1,...,5}
      \node at (180+20*\n: 1) {$\cdot$};
	\end{tikzpicture}}.
	\label{l n witten}
\ee
The classical actions $S_{\rm\sst AdS}[\varphi]$ of the AdS theories dual to free scalar matrix models are not known, but 
the correspondence gives us various information about them:
\begin{itemize}

\item 
The massless sector of $S_{\rm\sst AdS}$ coincides with the Vasiliev's theory,
because free matrix models always have
infinitely many conserved currents operators 
which are bilinear in $\bm\phi$
and their correlators remains the same as the vector model case.
\item
The massive higher spin fields, which are the complement to the Vasiliev's spectrum in $S_{\rm\sst AdS}$\,, behave as matter sectors of the Vasiliev's higher spin gauge theory.
Moreover, they consist of an infinite number of multiplets of Vasiliev's higher spin algebra:
singleton can be regarded as the fundamental representation of Vasiliev algebra,
and any its tensor products provide faithful representations of 
the algebra.
\item
Since free CFTs do not have any coupling constant, 
the only parameter that $S_{\rm\sst AdS}$ involves is the cosmological constant  or the radius $R$ of AdS.
This in turn combines with the gravitational constant $G$ to form a dimensionless constant,
\be	
	g:=R^{1-d}\,G\,.
	\label{coup g}
\ee
The latter should be related to the dimensionless parameter $\mathsf N$
of the CFT.
\item
As we discussed above, the correlators of free CFTs have the same  tensor structure for any value of $\mathsf N$\,. This implies that higher order loop corrections of Witten diagram at AdS dual theory should have the same tensor structure. This implies that the 
AdS dual theory cannot have different tensor structures for its Witten diagrams in different loop orders. 
Hence, they should be all proportional to the leading-order one,
\be
	 G^{\ell}\,\G^{\sst (\ell)}_{\rm\sst AdS}[h]=n_\ell\,g^\ell\,\G^{\sst (0)}_{\rm\sst AdS}[h]\,,
\ee
where 
$n_\ell$ are dimensionless constants.
Furthermore, we have
\be
	\G^{\sst (0)}_{\rm\sst AdS}[h]=S_{\rm\sst AdS}[\varphi(h)]\,.
\ee
Hence, for a  free CFT holography, we should have 
\be
	\mathsf{N}\,F^{\sst (0)}_{\rm\sst CFT}[h]
	=n(g)\,R^{1-d}\,S_{\rm\sst AdS}[\varphi(h)]\,,
	\qquad
	n(g)=\frac1{g}\left(1+\sum_{\ell=1}^\infty n_\ell\,g^\ell\right).
	\label{prop}
\ee
 \end{itemize}
By expanding the action $S_{\rm\sst AdS}$ around AdS background in
the power of fields as
\be
	S_{\rm\sst AdS}[\varphi]
	=S_{\sst 0}+
	S_{\sst 2}[\varphi]+S_{\sst 3}[\varphi]+\cdots,
	\label{S expansion}
\ee
each interaction terms can be constructed in principle by 
comparing the corresponding Witten diagrams with the CFT correlators. For the quadratic and cubic terms $S_{\sst 2}$ and  $S_{\sst 3}$\,,
it is sufficient to attach the boundary-to-bulk propagators to the vertices.
In \cite{Bekaert:2010hk,Joung:2011ww,Joung:2012rv,Joung:2012hz}, all the cubic interactions for massive and massless symmetric higher spin fields have been constructed for their transverse and traceless pieces, which are enough for the on-shell calculations.\footnote{Recently,
the AdS cubic vertex which gives the 3-pt functions of the free scalar CFT
has been determined in \cite{Sleight:2016dba}. This vertex ought to 
be the metric-like form of the cubic interaction of Vasiliev's theory.}
The quartic term $S_{\sst 4}$ is more subtle as it requires to subtract infinitely many exchange diagrams,
and a certain non-locality of interaction may start to appear from this order.
In \cite{Bekaert:2014cea,Bekaert:2015tva}, the scalar quartic interaction of Vasiliev's theory has been identified in this way and it has been shown
that its form is indeed a non-local one.\footnote{Once one accepts that a certain non-locality cannot be avoided, then one needs to distinguish a \emph{good} non-locality from the \emph{bad} ones whose introduction to classical actions spoils the predictability.
See \cite{Bekaert:2016ezc} for a recent discussion.}

\subsection{Bulk Vacuum Energy and Zeta Function}

 If the classical action $S_{\rm\sst AdS}$ can be \emph{constructed} from the 
holographic correspondence, then the conjecture becomes tautological in a sense at least at the classical level.
A non-trivial test of the conjecture would be to calculate $n$-point functions starting from a given form of classical theory. The calculation of 3-pt correlators from Vasiliev's theory corresponds to this case:
the task has been carried out in \cite{Giombi:2009wh} showing an agreement with the free Vector models.
See \cite{Giombi:2010vg,Colombo:2010fu,Colombo:2012jx,Didenko:2012tv,Didenko:2013bj} for further examinations based on a different technique.\footnote{For a certain type of 3pt correlators, a straightforward perturbative calculation involves 
some subtleties (noticed first in \cite{Giombi:2009wh})
as reported in \cite{Boulanger:2015ova,Skvortsov:2015lja}
in relation with the allowed class of field redefinitions (see also \cite{Vasiliev:2015wma,Didenko:2015cwv}). Hence, it would be fair to say that there still remains several issues to understand about the holography  of Vasiliev's theory.}
On the contrary to the Vasiliev's theories, which are dual to the free vector models,
the AdS theories dual to free matrix models are not known.
The best guess is that it can be again described by a Vasiliev type equations extended by 
a certain matter sector.
Anyway, as the classical theory is not known, 
 the only available test of the correspondence would be the property \eqref{prop}:
 the quantum effects should be proportional to the semi-classical ones.

In the following,
we consider the simplest case of the quantum effect: the one-loop ($\ell=1$)
diagram without any leg ($n=0$) in \eqref{l n witten}, namely the vacuum energy.
In principle, the vacuum energy cannot provide a rigorous test for \eqref{prop}
since it only returns a number rather than a tensor structure.\footnote{Strictly speaking, for the test, we need to consider the 3-pt
correlators where
different tensor structures start to enter. The  2-pt correlator is unique 
once the masses of fields are tuned by the conformal dimensions of the single trace operators.}
However, since this number is typically a rather special rational or transcendental number,
it may allow us to guess what the whole picture should look like. 
The vacuum energies of higher spin theories have been calculated 
for the vector model dualities \cite{Giombi:2013fka, Giombi:2014iua, Giombi:2014yra} as well as some related extensions
\cite{Beccaria:2014xda,
Beccaria:2014zma,Beccaria:2014qea,Beccaria:2015vaa,Beccaria:2016tqy,Tseytlin:2013jya,Tseytlin:2013fca,Beccaria:2014jxa,Joung:2015eny}. As we shall comment later,
it has lead to an interesting guess on the relation between $\mathsf N$ and $g$\,.

The vacuum energy, $\G^{\sst (1)}_{\rm\sst AdS}:=\G^{\sst (1)}_{\rm\sst AdS}[0]$ is given by
\be
	\exp\left(-\G^{\sst (1)}_{\rm\sst AdS}\right)
	=\prod_{\D,\bm\ell, I}\int
	 \cD\pi^I_{\D,\bm\ell}\,
	\exp\left(-\frac1{G}\,S_{\sst 2}[\pi]\right),
\ee
where the quadratic action $S_{\sst 2}$ \eqref{S expansion} simply reduces to the sum
of the quadratic actions for the fluctuation fields $\pi^I_{\D,\bm\ell}$\,:
\be
	S_{\sst 2}[\pi]
	=\sum_{\D,\bm\ell, I}\,S_{\D,\bm\ell}[\p^I]\,.
\ee
So far, we were using shorthand notations for $\pi^I_{\D,\bm\ell}$ 
which are actually AdS tensor fields $\pi^I_{\mu(\ell_1),\ldots,\mu(\ell_h)}$
carrying in general a mixed symmetry representation $\bm\ell$ under the Lorentz symmetry. 
The precise form of the action $S_{\D,\bm\ell}$ describing the $\cD(\D,\bm\ell)$ representation
is non-trivial and generically requires to introduce traces and possibly other set of auxiliary fields.
See \cite{Alkalaev:2006hq,Alkalaev:2005kw,Alkalaev:2006rw, Reshetnyak:2010ga,Alkalaev:2010af,Burdik:2011cb,Campoleoni:2012th} and references therein for the construction of such classical actions.
For our purpose --- which is to evaluate 
the vacuum energy diagrams ---
we need anyway to reduce the action to the traceless and transverse gauge 
where the form of the action is simplified to
\be
	S_{\D,\bm\ell}[\pi_{\rm\sst TT}]=
	\frac12\,\int d^{d+1}x\sqrt{g_{\rm\sst AdS}}\ 
	\pi_{\rm\sst TT}^{\mu(\ell_1),\ldots,\mu(\ell_h)}\,
	\mathscr{D}_{\D,\bm\ell}\,\pi_{{\rm\sst TT}\,\mu(\ell_1),\ldots,\mu(\ell_h)}\,,
\ee
where the differential operator $\mathscr{D}_{\D,\bm\ell} = \Box - M^2_{\D,\bm\ell}$
is defined with the mass term \eqref{mass}.
Here, we assumed that the corresponding representations are long ones.
In case of short representations, the path-integral should be supplemented
by the proper Jacobian (or ghost contribution) ---
which eventually amounts to extending the field content (labeled by $I$)
to include the ghost fields with negative counting.
The Laplacian operator $\Box$ depends on the tensor field that it acts on, specified by $\bm\ell$\,. 
Since the mass term will only shift the eigenvalues of $\mathscr{D}_{\D,\bm\ell}$ from those of $\Box$ 
by an additive constant, we will mostly concern ourselves with the spectral problem for the Laplacian. The path-integral can be formally evaluated to give
\be
	\G^{\sst (1)}_{\rm\sst AdS}=\sum_{\D,\bm\ell, I}
	-\frac12\,\ln\det\mathscr{D}_{\D,\bm\ell}
	=\sum_{\D,\bm\ell}N^{\cH}_{\D,\bm\ell}\,\G^{\sst (1)}_{\D,\bm\ell}\,,
\ee
where $N^{\cH}_{\D,\bm\ell}=\sum_{I}$ is the multiplicity of the fields $\cV(\D,\bm\ell)$ including the ghost contributions
and $\G_{\D,\bm\ell}$ is the vacuum energy of the field corresponding to $\cV(\D,\bm\ell)$\,:
\be
	\G^{\sst (1)}_{\D,\bm\ell}=-\frac12\,
	\ln\det\mathscr{D}_{\D,\bm\ell}=
	-\frac12\int_0^\infty \frac{dt}{t}\, 
	\tr\left[e^{-t\,\mathscr{D}_{\D,\bm\ell}}\right].
	\label{hk}
\ee
After the last equality,
we have used the standard representation of the functional determinant 
assuming that $\mathscr{D}_{\D,\bm\ell}$ has positive definite eigenvalues. The expression
contains the traced heat kernel, 
\begin{equation}
	K_{\D,\bm\ell}\left(t\right):=\tr \left[e^{-t\,\mathscr{D}_{\D,\bm\ell}}\right] = 
	 \int d^{d+1}{x}\sqrt{g_{\sst\rm AdS}}\  K_{\D,\bm\ell}{}^{\bm\mu}
	 {}_{\bm\mu}\left(x,x;t\right),
\end{equation}
where $\bm\mu$ is the shorthand notation for the indices ${\mu(\ell_1),\ldots,\mu(\ell_h)}$ of 
$\chi^{{\mu(\ell_1),\ldots,\mu(\ell_h)}}$
(which is again a shorthand notation (see footnote \ref{fn})).
Given eigenfunctions $\psi_{n}^{\bm\mu}(x)$ belonging to an eigenvalue $E_n$ of the Laplacian $\Box$\,,
the heat kernel  $K_{\D,\bm\ell}{}^{\bm\mu\bm\nu}\left(x,y;t\right)$
is given by
\begin{equation}
K_{\D,\bm\ell}{}^{\bm\mu\bm\nu}(x,y;t) = \sum_{n} \psi_{n}^{\bm\mu}(x)\,
\psi_{n}^{\bm\nu}(y)^*\,e^{-t\left[E_n+M^2_{\D,\bm\ell}\right]}\,.
\end{equation}
For homogeneous spaces like spheres and hyperboloids (Euclidean AdS), the coincident heat kernel 
 is independent of the position $x$\,, hence we find for AdS,
\be
	K_{\D,\bm\ell}(t) = \text{Vol}_{\text{AdS}}\, k_{\D,\bm\ell}(t)\,,
	\qquad
	k_{\D,\bm\ell}(t):=
	K_{\D,\bm\ell}{}^{\bm\mu}{}_{\bm\mu}(x,x;t)\,,
\ee 
where 
the volume of AdS, $ \text{Vol}_{\text{AdS}}$\,,
is a divergent quantity which 
requires a regularization.
Besides this AdS IR divergence, we also have 
the usual UV divergences coming from the loop integrals.
In the representation \eqref{hk}, it arises
from the small $t$ integral region, which corresponds 
to the short-distance heat propagation. 
This UV divergence can be regularized by 
a Mellin transform, namely
the zeta function $\zeta_{\D,\bm\ell}$\,, as 
\begin{equation}
	\G^{\sst (1)}_{\D,\bm\ell}(\L)=-\frac12\,\G(z)\,\zeta_{\D,\bm\ell}(z)\,,
	\qquad
\zeta_{\D,\bm\ell}\left(z\right) = \int_0^\infty \frac{dt}t\,\frac{t^{z}}{\Gamma(z)}\, K_{\D,\bm\ell}(t)\,.
\label{spec zeta}
\end{equation}
Collecting the contributions from the entire field content, we obtain
\be
	\G^{\sst (1)}_{\rm\sst AdS}(z)
	=-\frac12\,\G(z)\,\zeta_{\cH}(z)\,,
	\qquad
	\zeta_{\cH}(z)=\sum_{\D,\bm\ell}\,N_{\D,\bm\ell}\,\zeta_{\D,\bm\ell}(z)\,,
	\label{zeta AdS}
\ee
where $\cH$ is the physical Hilbert space of the theory \eqref{decomp H}.
The dimensionless parameter $z$ plays the role of UV cut-off:
\be
	\frac1z=\log\!\left(\L_{\rm\sst UV}\,R\right),
\ee
with the UV cut-off scale  $\L_{\rm\sst UV}$
and  the AdS radius $R$\,,
hence
the renormalized vacuum energy  of the theory is given by
\be
	\G^{\sst (1)\,\rm ren}_{\sst\rm AdS}=-\frac{\zeta_{\cH}(0)}2\,
	\log\!\left(\mu\,R\right)
	-\frac{\zeta_{\cH}'(0)}2\,,
\ee
whereas the coefficient of the UV divergence is given by
\be
	\G^{\sst(1)\,\rm div}_{\sst\rm AdS}=-\frac{\zeta_\cH(0)}2\,
	\log\frac{\L_{\rm\sst UV}}\mu\,.
\ee
Here, $\mu$ is the renormalization scale.

To summarize, the one-loop vacuum energy
of the AdS theory is
given by the function $\zeta_{\cH}(z)$ \eqref{zeta AdS}, 
which is determined by the multiplicity of the spectrum $N_{\D,\bm\ell}$ and the spectral zeta function $\zeta_{\D,\bm\ell}$ \eqref{spec zeta}. 
The calculation of the latter has been solved in great detail using 
group theoretic properties of AdS spaces \cite{Camporesi:1990wm,Camporesi:1993mz,Camporesi:1994ga,Camporesi:1995fb,Camporesi}, as we shall shortly review. 
About the multiplicity $N^{\cH}_{\D,\bm\ell}$, we extract the spectrum $\{\varphi^I_{\D,\bm\ell}\}$ of AdS fields  by identifying it with the spectrum  $\{\cO^I_{\D,\bm\ell}\}$ of conformal primaries in the dual CFT in the $\mathsf{N}\to\infty$ limit.
This is in any case a necessary condition for the validity of the given AdS/CFT duality.

Even then the problem of arriving at the total one-loop vacuum energy is still non-trivial since, as we saw, the general expression for the multiplicity of conformal primaries in a free CFT rapidly becomes very complicated as we start including contributions from higher and higher powers of singletons $\bm\phi$\,. 
We will shortly see this very explicitly for the case of AdS$_4$.

\section{AdS$_4$ with $S^{3}$ Boundary}
\label{sec: AdS4}

The first case we study is the computation of the one-loop vacuum energy in $AdS_{4}$ with $S^{3}$ boundary. This was also the first case that was considered in the higher spin/CFT duality computations for matching the one-loop free energy for Vasiliev higher spin theories with their CFT duals, the 
$U(N)$ and $O(N)$ vector models \cite{Giombi:2013fka}. While the vector models have a relatively simple spectrum of conformal primaries, as the spectrum involves only the square of singletons, 
the $SU(N)$ adjoint model has a far richer spectrum as now arbitrarily high powers of singletons may be taken. In this respect, the case of AdS$_{4}$ is particularly simple because it cannot have any mixed symmetry tensor representations.

\subsection{Zeta Functions of  AdS$_4$ Fields}

 Fields in AdS$_4$ are labeled by their $so(2,3)$ quantum numbers $\left(\Delta,s\right)$. The spectral zeta function for a field labeled by these quantum numbers is given by \cite{Camporesi:1994ga}
\begin{equation}\label{Zeta}
\zeta_{\D,s}\!\left(z\right)={1\over 3}\int_0^\infty du\,u\,\tanh\pi u\, D_{\D}\!\left(z,u\right) S_s\!\left(u\right),
\end{equation}
where $D_{\D}\!\left(z,u\right)$ and $S_s\!\left(u\right)$ are
\begin{equation}
D_{\D}\!\left(z,u\right) = {1\over\left [u^2+\left(\D-{3\over 2}\right)^2\right]^z},\quad S_s\!\left(u\right)=\frac{2\,s+1}2
\left[u+\left(\frac{2\,s+1}2\right)^2\right].
\end{equation}
We shall provide two strategies for evaluating the zeta function of matrix model CFTs. First, we shall work explicitly with the spectra of conformal primaries obtained from considering powers of singletons and sum over the contributions of each such field. We shall find that this method quickly becomes prohibitive for two reasons. Firstly, as already encountered in the case of vector models \cite{Giombi:2013fka,Giombi:2014iua}, these sums are naively divergent and need to be regularized. Secondly, the spectrum of conformal primaries becomes increasingly complicated as higher powers of singletons are taken. This leads us to consider an alternative method for computing the zeta function, using the character (or generalized partition function)
of the dual CFT.

\subsection{Vacuum Energy from Infinite Series}
\label{sec: summation}

We note that expressions for $\zeta_{\D,s}\!\left(0\right)$ and $\zeta_{\D,s}'\!\left(0\right)$ obtained from evaluating \eqref{Zeta} are already known  \cite{Camporesi:1993mz}. They are given by 
\begin{equation}\label{zeta}
\zeta_{\D,s}\!\left(0\right)= \frac{2s+1}{24}
\left[\left( \Delta- \frac{3}{2}\right)^4 
- \left(\frac{2\,s+1}{2}\right)^2 \left( 2 \left( \Delta- \frac{3}{2} \right)^2 + \frac{1}{6}\right) - \frac{7}{240} \right], 
\end{equation}
and 
\begin{equation}\label{zeta'}
\begin{split}
\zeta_{\D,s}'\!\left(0\right)= \frac{2\,s+1}3\,&
\Bigg[\frac{1}{8}\left( \Delta- \frac{3}{2}\right)^4 + \frac{1}{48} 
\left( \Delta- \frac{3}{2}\right)^2 + c_3 + \left(\frac{2\,s+1}{2}\right)^2 c_1 \\
&+ \int_0^{\Delta- \frac{3}{2}} dx \left( \left(\frac{2\,s+1}{2} \right)^2 - x^2 
\right)x\,\psi\!\left(x+\frac{1}{2}\right)\Bigg].
\end{split}
\end{equation}
The constants $c_n$, where $n=1,\,3$,  and the digamma function $\psi(x)$ are defined as
\begin{equation}
c_n = \int_0^\infty du\,\frac{2\,u^n\,\ln u}{e^{2 \pi u}+1}, \quad
\psi(x) = \int_0^{\infty} dt \left( \frac{e^{-t}}{t} - \frac{e^{-x t}}{1-e^{-t}}\right).
\end{equation}
These expressions were used to evaluate $\zeta_{\D,s}\!\left(0\right)$ and $\zeta_{\D,s}'\!\left(0\right)$ for Rac$^{\otimes 2}$/$T_{\rm cyc}^{\sst(2)}({\rm Rac})$, the case relevant to the $U(N)$/$O(N)$ vector models in \cite{Giombi:2013fka,Giombi:2014iua}. 

For the vacuum-energy of the AdS theory under consideration,
we need to use the tensor products $T^{\sst (n)}_{\rm cyc}$\,.
However, as we have seen in Section \ref{sec: decomp ex},
the decomposition rule becomes quickly complicated as $n$ increases.
Hence, in the following, we will conduct the calculations first for
the `toy models'  Rac$^{\otimes 3}$ and Rac$^{\otimes 4}$
in order to see how the methods adopted in \cite{Giombi:2013fka,Giombi:2014iua}
can be extended to the $SU(N)$ adjoint model. 

First, to compute the zeta function that includes the contribution of all the AdS fields corresponding to the conformal primaries contained in Rac$^{\otimes 3}$ we use the decomposition rule \eqref{rac3_decom}, and regulate the infinite sum over spins using zeta function regularisation. Firstly, it is easy to see that $\zeta_{{\rm Rac}^{\otimes 3}}(0)$ vanishes:
\begin{equation}
\begin{split}
\zeta_{{\rm Rac}^{\otimes 3}}(0) &= \lim_{\alpha \to 0} \,\sum_{s=0}^{\infty} \left( \Delta - \frac{3}{2} \right)^{\!\!-\alpha} \zeta_{\Delta,s}(0)\\ &= \lim_{\alpha \rightarrow 0}\, \sum_{s=0}^{\infty} s^{-\alpha}\, \zeta_{s+{3\over 2},s}(0) +\lim_{\alpha \rightarrow 0}\, \sum_{s=0}^{\infty} \left(s+1\right)^{-\alpha} \,\zeta_{s+{5\over 2},s}(0)\\&= -\frac{7}{13824} + \frac{7}{13824} = 0\,.
\end{split}
\end{equation}
Hence the logarithmically divergent part of the vacuum energy vanishes. We now consider the finite part of the one-loop free energy, contained in $\zeta_{{\rm Rac}^{\otimes 3}}'(0)$. The nontrivial contribution to this comes solely from the digamma dependent part,
\begin{equation}
\begin{split}
\zeta_{{\rm Rac}^{\otimes 3}}'(0)=
\lim_{\a\to0}\, \sum_{s=0}^{\infty}\,&\int_0^\infty dt 
\left[ s^{-\alpha}\, (s+1)
\int_0^s dx
+(s+1)^{-\alpha}\,s\int_0^{s+1} dx\right]\times \\
& \times \frac23
\left(\frac{2\,s+1}{2}\right)\left[\left(\frac{2\,s+1}{2}\right)^2 x-x^3\right] \left( \frac{e^{-t}}{t} - \frac{e^{-(x+\frac{1}{2})\,t}}{1-e^{-t}}\right).
\end{split}
\end{equation}
After carrying out the integration over $x$ and the summation over $s$, we obtain
\begin{equation}
\begin{split}
\zeta_{{\rm Rac}^{\otimes 3}}'(0) = &\int_0^\infty dt \, \frac{-2\, e^{-\frac{7 t}{2}}-6\, e^{-\frac{5\, t}{2}}-6\, e^{-\frac{3\, t}{2}}-2\, e^{-\frac{t}{2}}}{\left(1-e^{-t}\right)^4 t^4}\\ &+  \int_0^\infty dt \,\frac{-12\, e^{-\frac{7\, t}{2}}-24\, e^{-\frac{5\, t}{2}}-12\, e^{-\frac{3\, t}{2}}}{\left(1-e^{-t}\right)^5 t^3} \\ &+  \int_0^\infty dt \, \frac{e^{-\frac{11\, t}{2}}-23 e^{-\frac{9\, t}{2}}-362\, e^{-\frac{7 t}{2}}-362\, e^{-\frac{5\, t}{2}}-23\, e^{-\frac{3\, t}{2}}+e^{-\frac{t}{2}}}{12 \left(1-e^{-t}\right)^6 t^2}\\ &+  \int_0^\infty dt  \, \frac{5\, e^{-\frac{7\, t}{2}}+18\, e^{-\frac{5\, t}{2}}+5 e^{-\frac{3\, t}{2}}}{2 \left(1-e^{-t}\right)^5 t}\,. 
\end{split}
\end{equation}
By expressing these integrals in terms of the Lerch transcendent, we find that
\begin{equation}
	\zeta_{{\rm Rac}^{\otimes 3}}'(0) = -\frac{\ln2 }{128} -\frac{19\, \zeta(3)}{3840\, \pi ^2} +\frac{25\, \zeta (5)}{256 \,\pi ^4} -\frac{63\, \zeta (7)}{512 \,\pi ^6}\,. \label{rac3_result}
\end{equation}
We next turn to the case of Rac$^{\otimes 4}$. In this case we will require the decomposition rule \eqref{rac4_decom}. Using this decomposition and \eqref{zeta}, we see that
\ba
\zeta_{{\rm Rac}^{\otimes 4}}(0)\eq\lim_{\alpha \rightarrow 0}\, \sum_{s=0}^{\infty}\, \sum_{n=1}^{\infty} \left( s+n+\frac{1}{2}\right)^{\!\!-\alpha} \frac{2\,n\, (2\,s+1)+3\, (-1)^n+(2\,s+1)^2}4\, \zeta_{s+n+2,s}(0) \nn
&& +\lim_{\alpha \rightarrow 0}\, \sum_{s=0}^{\infty} \left( s+\frac{1}{2}\right)^{\!\!-\alpha}\frac{(s+1)\,(s+2)}2\, \zeta_{s+2,s}(0)\,,
\label{zeta_rac4}
\ea
where
\begin{equation}
\zeta_{s+n+2,s}(0) = \frac{2\,s+1}{24} \left[\left(s+n+\frac{1}{2} \right)^4-\left(s+\frac{1}{2}\right)^2 \left(2\left(s+n+\frac{1}{2}\right)^2+\frac{1}{6} \right) - \frac{7}{240} \right].
\end{equation}
To do the double summation we introduce a new parameter
 $m \equiv n+ s$. After this replacement, we find that the first line of \eqref{zeta_rac4} gives
\begin{equation}
\lim_{\alpha \rightarrow 0} \sum_{m=1}^{\infty} \sum_{n=1}^{m} \left(m+\frac{1}{2}\right)^{-\alpha} \frac{2\,n\,[2\, (m-n)+1]+3\,(-1)^n+[2\,(m-n)+1]^2}4\, \zeta_{m+2,m-n}(0)\,,
\end{equation}
which is equal to $-\frac{7963}{232243200}$\,.
By a similar calculation,  one can show that the second line gives $+\frac{7963}{232243200}$. Hence, the two contributions cancel each other and we again find that the UV divergent term $\zeta_{\rm Rac^{\otimes 4}}(0)$ vanishes. The finite part of the vacuum energy 
is again given by the digamma piece of \eqref{zeta'}, and we get
\ba
&& \zeta_{\rm Rac^{\otimes 4}}'(0) \nn 
&&=\,\sum_{s=0}^{\infty} \int_0^\infty dt \int_0^{s+\frac{1}{2}} dx \left(s+\frac{1}{2}\right)^{1-\alpha}\frac{(s+1)(s+2)}3\left( \left(s+\frac{1}{2}\right)^2 x-x^3\right) \left( \frac{e^{-t}}{t} - \frac{e^{-(x+\frac{1}{2})t}}{1-e^{-t}}\right)\nn 
                           &&\ +\, \sum_{m=1}^{\infty} \sum_{n=1}^{m} \int_0^\infty dt \int_0^{m+\frac{1}{2}} dx \left(m+\frac{1}{2}\right)^{-\alpha}\,
                           \frac{2\,n\,[2(m-n)+1]+3(-1)^n+[2(m-n)+1]^2}{12}
                            \nn
                           &&\hspace{90pt} \times [2(m-n)+1] \left( \left(m-n+\frac{1}{2}\right)^2 x-x^3\right) \left( \frac{e^{-t}}{t} - \frac{e^{-(x+\frac{1}{2})t}}{1-e^{-t}}\right).
\ea
We first carry out the integration over $x$ and the summation over $n$. Next, the summation over $m$ is carried out by Hurwitz zeta function regularization and we finally obtain
\ba
&&\zeta_{\rm Rac^{\otimes 4}}'(0)= \int_0^\infty dt \,
\frac{2\, e^{-8\, t} + 2\, e^{-7\, t} - 6\, e^{-6\, t} - 6\, e^{-5\, t} + 6\, e^{-4 \,t} + 
 6\, e^{-3 \,t} - 2 \,e^{-2\, t} - 2\, e^{-t}}{\left(1-e^{-t}\right)^9 t^4} \nn
                        &&\quad + \int_0^\infty dt \,
                           \frac{-e^{-8\, t}-19 \,e^{-7\, t}-17\, e^{-6\, t}+37\, e^{-5\, t}+37\, e^{-4\, t}-17\, e^{-3\, t}-19\, e^{-2\, t}-e^{-t}}{\left(1-e^{-t}\right)^9 t^3}  \nn
                           &&\quad + \int_0^\infty dt \, \frac{e^{-8\, t}+73 \,e^{-7\, t}+501\, e^{-6\, t}+429 \,e^{-5\, t}-429\, e^{-4\, t}-501\, e^{-3\, t}-73\, e^{-2\, t}-e^{-t}}{6 \left(1-e^{-t}\right)^9 t^2}  \nn
                           &&\quad + \int_0^\infty dt \, \frac{-40\, e^{-6\, t}-184\, e^{-5\, t}-184\, e^{-4\, t}-40\, e^{-3\, t}}{\left(1-e^{-t}\right)^9 t}\,.
\ea
To obtain an analytic expression we again express the above in terms of the Lerch transcendent, and finally obtain
\begin{equation}
\zeta_{{\rm Rac}^{\otimes 4}}'(0) = -\frac{29\, \zeta (3)}{2520\, \pi ^2}-\frac{\zeta (5)}{60\, \pi ^4}+\frac{\zeta (7)}{8 \,\pi ^6}-\frac{\zeta (9)}{8\, \pi ^8}\,. \label{rac4_result}
\end{equation}
This completes the explicit evaluation of the zeta functions for the conformal primaries contained in Rac$^{\otimes 3}$ and Rac$^{\otimes 4}$. 
Even though we managed to get the results,
it is hard to continue this method to higher order of singletons as
the decomposition rules becomes quickly very complicated.
Moreover if we consider the realistic matrix models 
instead of the toy models of Rac$^{\otimes n}$, even
the order four level becomes highly non-trivial  and it makes the computation more prohibitive. 
We now turn to an alternative approach to the same problem, which turns out to be of greater utility as it allows the powers of the singleton to increase.

\subsection{Zeta Function from Character}

We have seen above that while it is in principle possible to carry out a brute-force evaluation of the summation of the vacuum energies for higher powers of singletons, it is in practice quite prohibitive both due to increasing complexity of the spectrum and the careful regularization required at several steps in the calculation. We now introduce an alternative approach to the problem, based on the fact that the spectrum of conformal primaries is encoded in the 
corresponding character, namely  the  
generalized
partition function of the CFT. In particular, we would aim to write the zeta function corresponding to a given conformal primary in terms of the corresponding $so(2,3)$ character, and then formally carry out the sum over characters and multiplicities to write an expression for the zeta function for all the bulk fields in one go. As a starting point, we will rewrite the character \eqref{so23 ch} of $\cV(\D,s)$ in the form
\begin{equation}
	\chi_{\D,s}(\b,\a)=\frac{e^{-(\D-\frac32)\,\beta}\,
	\sin\left(\frac{2\,s+1}2\,\a\right)}{\eta(\beta,\phi)}\,,
\end{equation}
where $q=e^{-\beta}$ and $x=e^{i\,\a}$ are used in \eqref{so23 ch} and 
\begin{equation}
	\eta(\beta,\a)=8\,\sinh\tfrac{\beta}{2}\,\sin\tfrac\a2
	\left(\sinh^{2}\tfrac\beta2+\sin^{2}\tfrac\a2\right).
\end{equation}
In order to relate $\zeta_{\D,s}$ to $\chi_{\D,s}$\,, we take the inverse Laplace transform of $D_{\D}$ as
\begin{equation}
	D_{\D}(z,u)=\frac{\sqrt{\pi}}{\Gamma(z)}\,
	\int_{0}^{\infty}d\beta\,e^{-\beta\,(\D-\frac32)}
	\left(\frac\beta{2\,u}\right)^{z-\frac12}\,J_{z-\frac12}(u\,\beta)\,,
	\label{D}
\end{equation}
where we assumed $\D\ge\tfrac32$\,. Next, we recast $S_{s}$ as 
\begin{equation}
	S_{s}(u)
	=\frac{d}{d\a}
	\left(u^{2}-\frac{d^{2}}{d\a^{2}}\right)\sin\!\left[(s+\tfrac12)\a\right]
	\Big|_{\a=0}\,.
	\label{S}
\end{equation}
Using \eqref{D} and \eqref{S}, the zeta function $\zeta_{\D,s}$ is related to the character $\chi_{\D,s}$ as
\begin{equation}\label{zetachar}
	\zeta_{\D,s}(z)=\frac{1}{\Gamma(z)}
	\int_{0}^{\infty}d\beta
	\left(\mu(z,\beta)+\nu(z,\beta)\,\frac{\partial^{2}}{\partial\a^{2}}\right)
	\chi_{\D,s}(\beta,\a)\,\Big|_{\a=0}\,,
\end{equation}
where $\mu(z,\beta)$ and $\nu(z,\beta)$ are independent of $\D$ and $s$. Their expressions are given by
\begin{equation}
\begin{split}
\mu(z,\b)
	&= \frac13\,
	\sinh\tfrac\beta2\left[
	\left(\sinh^{2}\tfrac\beta2-6\right) f_{1}(z,\beta)
	+4\,\sinh^{2}\tfrac\beta2\,f_{3}(z,\beta)\right]\\
	\nu(z,\beta)&=-4\,\sinh^{3}\tfrac\beta2\,f_{1}(z,\beta),
\end{split}
\end{equation}
with $f_{n}(z,\beta)$ defined as
\begin{equation}
	f_{n}(z,\beta)=\sqrt{\pi}\int_{0}^{\infty}du\,u^{n}\,\tanh\pi u
	\left(\frac{\beta}{2\,u}\right)^{z-\frac12}\,
	J_{z-\frac12}(\beta\,u).
	\label{f int}
\end{equation}
Notice that $\mu(z,\b)$ and $\nu(z,\b)$ are simply fixed functions
hence the relation \eqref{zetachar} can be 
extended from $\cV(\D,s)$ to any Hilbert space $\cH$\,.
In other words, it defines a universal relation between the spectral zeta function
and the character.
However, the identification of the functions $\mu(z,\b)$ and $\nu(z,\b)$ are
non-trivial.
When $z=0$, the functions $\mu(z,\beta)$ and $\nu(z,\beta)$ have the simple form.
\begin{equation}
	\mu(0,\beta)=\frac{2}{\beta}\,\frac{\cosh\frac\beta2}{\sinh\tfrac\beta2}\,,
	\qquad
	\nu(0,\beta)
	=\frac{1}{\beta}\,\sinh\beta\,.
\end{equation}
Let us notice that if we use $\mu(0,\beta)$ and $\nu(0,\beta)$ 
instead of $\mu(z,\beta)$ and $\nu(z,\beta)$\,,
then the $\beta$ integral will be divergent: the $z$-dependence
was introduced precisely to regularize this UV divergence. 
Since we have freedom to choose convenient 
regularization scheme, we suggest a modification of zeta function \eqref{zetachar} to 
\be
		\tilde\zeta_{\cH}(z)=
	\int_{0}^{\infty}d\beta\,
	\frac{\beta^{2z-1}}{\Gamma(2z)}\,
	\frac{\cosh\frac\beta2}{\sinh\frac\beta2}\left(1
	+\sinh^2\tfrac\beta2\,
	\partial_\a^2\right)
	\chi_{\cH}(\beta,\a)\,\Big|_{\a=0},
	\label{AdS4 zeta}
\end{equation}
which makes use of slightly different $z$-dependence 
which nevertheless does regularize the integral.
Explicit comparison with the standard zeta function shows that 
their value at $z=0$ coincide with each other implying that the log divergence of vacuum energy coincide.
Moreover, they have the same  first derivatives --- proportional to the finite part of the vacuum energy ---
for the character which is even in $\b$\,.
This is the case for  Rac 
itself as well as any tensor product of it.
Therefore, for our purpose, the formula \eqref{AdS4 zeta}
is fully equivalent to the standard zeta function
up to the $z^2$ order which is physically irrelevant.
The details of the comparison can be found in Appendix \ref{sec: AdS4 zeta}.

\subsection{A Few Tests}

\subsubsection{Vector Models}

Before analyzing the matrix models, we revisit the vector models as testing examples of the new method with $\tilde\zeta_{\cH}$\,. 
The spectrum of non-minimal Vasiliev theory is given by
the character
\be
	\chi_{\textrm{non-min}}(\b,\a)=\chi_{\rm Rac}{}^2(\b,\a)\,,
	\label{non min}
\ee
where the character of Rac   is given in \eqref{Di Rac} in $(q=e^{-\b}, x=e^{i\,\a})$\,,
which can be re-expressed in $(\b,\a)$ as
\be
	\chi_{\rm Rac}(\b,\a)=\frac{\cosh\frac\b2}{\cosh\b-\cos\a}\,.
\ee
By plugging \eqref{non min} to \eqref{AdS4 zeta}, we find that 
the integrand itself vanishes, hence
\be
	\tilde\zeta_{\textrm{non-min}}(z)=0\,.
\ee
As a result, both of the log divergent and finite parts are zero. 

In the case of the minimal Vasiliev theory dual to free scalar $O(N)$ vector model, 
the character is given by
\be
		\chi_{\textrm{min}}(\b,\a)=
		\frac{\chi_{\rm Rac}^2(\b,\a)+\chi_{\rm Rac}(2\b,2\a)}2\,.
		\label{char min}
\ee
Since the $\chi_{\rm Rac}^2=\chi_{\textrm{non-min}}$ gives vanishing zeta function, the only contribution comes from 
the second term $\chi_{\rm Rac}(2\b,2\a)/2$\,.  After some manipulation, one can show that
\be
	\tilde\zeta_{\rm min}(z)
	=4^{-z}\,\tilde\zeta_{\rm Rac}(z)
	=\tilde\zeta_{\rm Rac}(z)+\cO(z^{2})\,,
\ee
where $\tilde\zeta_{\rm Rac}(z)$ reads
\be
	\tilde\zeta_{\rm Rac}(z)
	=
	\zeta(2z-2,\tfrac12)+\frac14\,\zeta(2z,\tfrac12)
	=-\left[\frac{\ln2}4-\frac{3\,\zeta(3)}{8\,\pi^{2}}\right]z+\cO(z^{2})\,.
\ee
Interestingly, $\tilde\zeta_{\rm Rac}$ 
reproduces the vacuum energy of the conformal scalar on the boundary.\footnote{See \cite{Beccaria:2016tqy} for more discussion on this occurrence between the partition functions in different dimensions.}
Therefore, the examples of vector models shows the agreement.

\subsubsection{Toy Models: Rac$^{\otimes 3}$ and Rac$^{\otimes 4}$}

Let us move to the toy model  examples of Rac$^{\otimes n}$\,.
The corresponding vacuum energies have been calculated
in Section \ref{sec: summation},
and in below, we shall check whether the new method can reproduce
the result correctly.
 The character for Rac$^{\otimes n}$ is simply
 \be
	\chi_{{\rm Rac}^{\otimes n}}(\b,\a)=\chi_{\rm Rac}{}^n(\b,\a)\,,
 \ee
 and the corresponding  spectral zeta function 
 leads to the integral,
\be
	\tilde{\zeta}_{{\rm Rac}^{\otimes n}}(z) 
	=
	-\frac{n-2}2\,\int_{0}^{\infty}d\beta\,
	\frac{\b^{2z-1}}{\Gamma(2z)}\,
	e^{-\frac n2\,\b}\,
	\frac{(1+e^{-\b})^{n+1}}{(1-e^{-\b})^{2n+1}}\,.
\ee
For specific values $n=3$ and $n=4$, 
 the integrals can be evaluated by Taylor expanding the 
 integrand in $e^{-\b}$ and eventually resumming. 
 In the end, we obtain
\ba
	\tilde{\zeta}_{\rm Rac^{\otimes 3}}(z)
	\eq  \frac{\zeta(2\,z, \frac{3}{2})}{128} -\frac{19\,\zeta(2\,z-2, \frac{3}{2}) }{1440}
	-\frac{5\,\zeta(2\,z-4, \frac{3}{2})}{72}- \frac{\zeta(2\,z-6, \frac{3}{2})}{90} 
	 \nn
	\eq  z \left( -\frac{\ln 2}{128} -\frac{19\,  \zeta (3)}{3840\, \pi ^2}+\frac{25  \,\zeta (5)}{256\, \pi ^4}-\frac{63 \, \zeta (7)}{512\, \pi ^6}\right) +\cO(z^{2})\,,\\
\tilde{\zeta}_{\rm Rac^{\otimes 4}}(z) 
           \eq    \frac{29\,\zeta (2\,z-2)}{1260} -\frac{\zeta (2\,z-4)}{90}-\frac{\zeta (2\,z-6)}{90}-\frac{\zeta (2\,z-8)}{1260}
          \nn
           \eq z \left(-\frac{29\, \zeta(3)}{2520\, \pi^2} -\frac{ \zeta(5)}{60\,\pi^4} 
            +\frac{\zeta(7)}{8\,\pi^6} -\frac{\zeta(9)}{8\,\pi^8} \right)
            +\cO(z^{2})\,.
\ea
By comparing these with the results  (\ref{rac3_result}) and (\ref{rac4_result})
obtained in Section \ref{sec: summation} by summation over the 
field content,
we find the these results exactly matches to those.
One can see that compared to 
the calculations in Section \ref{sec: summation}, 
the new integral representation $\tilde\zeta_{\cH}(z)$
requires a considerably shorter calculation.

\subsection{Vacuum Energy for the AdS Dual of $SU(N)$ Adjoint Model}

Now, let us turn to the vacuum energy of the AdS theory dual to the $SU(N)$
adjoint matrix model. 

\subsubsection{Vacuum Energies for the first Few Regge Trajectories}
\label{sec: ads4 traj}

To begin with,
we calculate the spectral zeta function 
for the first few Regge trajectories,
that is, for the first few orders in
the power of singleton $\bm\phi$\,.
In the end, the total vacuum energy is the sum of the ones of 
the order from two to infinity.

\paragraph{Order Two}
At the order two, the field content coincides with that of the minimal Vasiliev theory hence
their vacuum energies coincides with that of Rac:
\be 
	\tilde\zeta_{\rm cyc^2}{}'(0) =-\frac{\ln2}4+\frac{3\,\zeta(3)}{8\,\pi^{2}}
	\simeq -0.127614\,.
\ee
\paragraph{Order Three}
At the order three, the $SU(N)$ adjoint model
corresponds to the cyclic character,
\be
	\chi_{\rm cyc^3}(\b,\a)=	\frac{\chi_{\rm Rac}{}^3(\b,\a)+
	2\,\chi_{\rm Rac}(3\,\b,3\,\a)}3\,,
	\label{char 3}
\ee
and gives the zeta function,
\ba
\tilde\zeta_{\rm cyc^3}(z) \eq \frac{\zeta \left(2 z,\frac{3}{2}\right)}{1152}
- \frac{19 \, \zeta \left(2 z-2,\frac{3}{2}\right)}{17280}
 -\frac{5 \, \zeta \left(2 z-4,\frac{3}{2}\right)}{216}
 -\frac{\zeta \left(2 z-6,\frac{3}{2}\right)}{270} \nn
&& +\, 2^{2 z}\,3^{-2z-1} \left[\zeta (2 z)+\zeta (2 z-2)\right]
-3^{-2 z-1}\left[ \zeta (2 z)+4\,\zeta (2 z-2)\right]\\
&&+\,
3^{-2 z-3}\,\frac{57}{32} \left[\zeta \left(2 z,\tfrac{1}{6}\right)+\zeta \left(2 z,\tfrac{5}{6}\right)-8\,\zeta \left(2 z-1,\tfrac{1}{6}\right)+8\,\zeta \left(2 z-1,\tfrac{5}{6}\right)\right]\nn
&& 
+\,3^{-2 z-1} \left[ \zeta \left(2 z-2,\tfrac{1}{6}\right)
+\zeta \left(2 z-2,\tfrac{5}{6}\right)
-3\,\zeta \left(2 z-3,\tfrac{1}{6}\right)
+3\,\zeta \left(2 z-3,\tfrac{5}{6}\right)\right]. \nonumber
\ea
It is free from the UV divergence because $\tilde\zeta_{\rm cyc^3}(0)=0$\,.
However,  
the finite part does not vanish but gives
\ba
	\tilde\zeta_{\rm cyc^3}{}'(0) \eq -\frac{43\,\ln2}{128}
 	+\frac{1487\,\zeta (3)}{3840\,\pi ^2}+\frac{25\,\zeta (5)}{768 \,\pi ^4}
 	-\frac{21\,\zeta (7)}{512\, \pi ^6}+\frac{4\, \pi}{27\sqrt{3}}
 	-\frac{19\,\psi ^{(1)}\!\left(\frac{1}{3}\right)}{72 \sqrt{3}\,\pi }
 	 	+\frac{\psi ^{(3)}\!\left(\frac{1}{3}\right)}{96 \sqrt{3}\, \pi ^3} \nn
	&\simeq& -0.311588\,,
\ea
where $\psi^{(n)}$ is the polygamma function of order $n$\,.
The order three trajectory
has more than two times the vacuum energy than the order two.
Let us also note that the particular property of the order two does not seem to continue to the order three.

\paragraph{Order Four}
At  the level order, the $SU(N)$ adjoint model character is given by
\be
	\chi_{\rm cyc^4}(\b,\a)=	\frac{\chi_{\rm Rac}{}^4(\b,\a)+
	\chi_{\rm Rac}{}^2(2\,\b,2\,\a)+
	2\,\chi_{\rm Rac}(4\,\b,4\,\a)}4\,,
	\label{char 4}
\ee
and the corresponding zeta function reads
\ba
&& \tilde\zeta_{\rm cyc^4}(z) =
\frac{5\, \zeta (2 z)}{16}-\frac{47\, \zeta (2 z-2)}{2520}
-\frac{23\, \zeta (2 z-4)}{720}-\frac{\zeta (2 z-6)}{90}
-\frac{\zeta (2 z-8)}{1260}
\nn
&&\quad +\,2^{-2z}\left[\frac{3\,\zeta (2 z)}2
+\frac{4\,\zeta (2 z-2)}{3}  
+\frac{2\,\zeta (2 z-4)}{3} \right]\\
&&\quad +\,2^{-4z}\left[
\frac{5\, \zeta\! \left(2 z-1,\frac{3}{4}\right)}{6}
-\frac{5\,\zeta\! \left(2 z-1,\frac{1}{4}\right)}{6} 
+\frac{8\,\zeta\! \left(2 z-3,\frac{3}{4}\right)}{3} 
-\frac{8\, \zeta\! \left(2 z-3,\frac{1}{4}\right)}{3}
-\frac{\zeta (2 z)}2\right]. \nonumber
\ea
Again, one can show that the vacuum energy is finite, $\tilde\zeta_{\rm cyc^4}(0)=0$\,,
and the finite part is given by
\ba
\tilde\zeta_{\rm cyc^4}{}'(0)\eq
-\frac{25\,\ln2}{64}
-\frac{4573\,\zeta (3)}{20160\, \pi ^2}
+\frac{457\,\zeta (5)}{1920\, \pi ^4}+\frac{\zeta (7)}{32\, \pi ^6}
-\frac{\zeta (9)}{32 \,\pi ^8}
+\frac{\pi}{32}-\frac{5\,\psi^{(1)}\!\left(\frac{1}{4}\right)}{96\, \pi }+\frac{\psi^{(3)}\!\left(\frac{1}{4}\right)}{384\,\pi ^3}  
\nn
&\simeq& -0.353518\,.
\ea
The vacuum energy of the fourth order trajectory
is increased again compared to the order three,
but only by a small amount.

\paragraph{Higher Orders}

\begin{figure}[h]
\centering
  \includegraphics[width=0.7\linewidth]{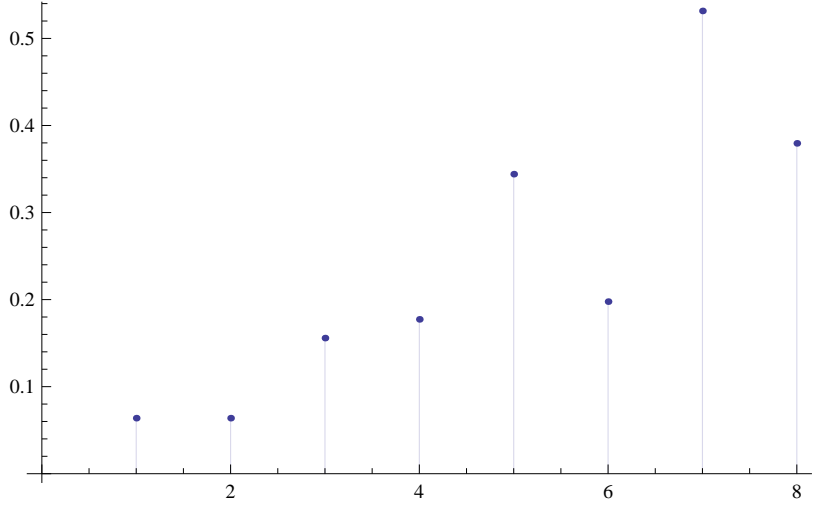}
  \caption{Vacuum energies of AdS${}_4$ fields in different orders}
  \label{fig1}
\end{figure}

In order to see a pattern, we can proceed with a few more orders:
the vacuum energies of the first 8 order Regge trajectories are plotted in  Fig.\,\ref{fig1}.
One can see that the energies exhibit a rough linear growth.
Assuming that the pattern continues, we can conclude that the total vacuum energy --- the sum of all vacuum energies in different trajectories --- 
will be divergent.
Therefore, we need to perform a regularization yet another time.
If we had analytic expressions for the vacuum energies for different orders,
we may consider various possible regularizations, but unfortunately 
we could not find the analytic formula.

\subsubsection{Vacuum Energies in Different Slices}
\label{sec: ads4 slice}

In the previous section, we have calculated the vacuum energy
of the AdS theory for the first few Regge trajectories,
which show a growing pattern.
In this section, we will consider the vacuum energy 
as a difference series.
As we have explained in Section \ref{sec: single trace},
in the limit $\mathsf{N}\to\infty$\,,
the characters for the matrix models can be partly resummed 
ending up with a logarithmic function with Euler totient function as \eqref{cyc}.
By expressing it as
\be
	\chi_{\rm adj}=	-\chi_{\rm Rac}+\sum_{k=1}^{\infty}\,\frac{\varphi(k)}{k}\,
	\chi_{{\rm log},k}\,,
	\label{cyc log}
\ee
with
\be 
	\chi_{{\rm log},k}(\b,\a)=
	-\log\!\left[1-\chi_{\rm Rac}(k\,\b,k\,\a)\right],
\ee
we can focus on the spectral zeta function 
corresponding to $\chi_{{\rm log},k}$\,,
\ba
	\tilde\zeta_{{\rm log},k}(z)\eq -
	\int_{0}^{\infty} d\b\,\frac{\b^{2z-1}}{\Gamma(2z)}\,\Bigg[
	\frac{\cosh\frac\b2}{\sinh\frac\b2}\,
	\log\!\left(1-\frac{\cosh \frac{k\b}2}{2\,\sinh^{2}\frac{k\b}2}\right)\nn
	&&\qquad -\,\frac{k^{2}}2\,
	\sinh\b
	\left(\frac1{1+\cosh\frac{k \b}2-\cosh k\b}+\frac1{\cosh k\b-1}\right)\Bigg]\,.
	\label{AdS4 log}
\ea
Working with a fixed value of $k$ corresponds
to considering the full vacuum energy as a series in $k$\,.
To distinguish with the previous organization of the spectrum of primaries into Regge trajectories,
we refer to all the primaries appearing with a fixed $k$ as belonging to the same `slice'.
The evaluation of the above integral is technically prohibitive,
but one can easily check that the above two functions vanish when $z=0$\,,
implying that the vacuum energy is free from the logarithmic UV divergence.
The renormalized vacuum energy corresponding to the first derivative of $\tilde \zeta_{{\rm log},k}$ is not easily accessible by analytic method.
We can nevertheless proceed numerically (see Appendix \ref{sec: AdS4 zeta}), but the validity of such result is not fully clear.

\bigskip

In the above, we were attacking the computation of the vacuum energy
of the AdS theory dual to $SU(N)$ adjoint model as two different series:
one as a series of the order of fields (considered in Section \ref{sec: ads4 traj}), 
and the other as a series with a fixed Euler totient function (considered in Section \ref{sec: ads4 slice}).
In both case, we could obtain the contributions for a few low orders, but
an analytic expression for whole sequence was not available. 
At this point, one may wonder
whether we can consider the
full character where different $k$ contributions are summed up.
In a sense, this idea seems to be in the continuation of our 
reasoning:
whenever we face a series which is divergent, we consider
directly the character which originates the series.
However at this time, the situation is different.
Clearly, the full character for the adjoint model
has an infinite number of singularities at \mt{\beta = \frac{\beta_c}{k}},
where $\beta_c$ is the singular point for the $k=1$ part. 
Moreover, around \mt{\b=0}, the singularities corresponding higher $k$ values
are accumulated making the character highly non-analytic around 
\mt{\b=0} point. 
Hence, if we consider the full character function, 
it seems to be impossible to control the divergence arising from the small $\b$ region
because its non-analyticity is severe. 
Therefore, the only well-defined ways to address the adjoint models
would be what we have been considered here:
as a series in the order of trajectories or as a series in the order of `slices'.

\section{AdS$_5$ with $S^4$ Boundary}
\label{sec: AdS5}

We now extend the approach of writing the bulk zeta function in terms of the character
--- that is, the generalized partition function --- of 
the boundary CFT to the case of AdS$_5$. This is the first instance where fields of mixed symmetry make an appearance. Since the case of completely symmetric fields is well understood \cite{Camporesi:1994ga} and has already been applied to this context in \cite{Giombi:2014iua}, we omit that discussion and consider directly
the most general case of mixed symmetry fields. The expressions obtained in the next section for the zeta function have previously been obtained in Appendix C of \cite{Beccaria:2014xda}. Nonetheless, it is useful to review this expression as it will prove important for the subsequent analysis.

\subsection{Zeta Functions of  AdS$_5$ Fields}

We start from an AdS$_5$ field that labelled by the $so(2,4)$ quantum numbers $(\Delta,(\ell_1,\ell_2))$. 
Furthermore, we assume this to be a long representation of the conformal algebra, as characters of (semi-)short representations can be written as sums and differences of characters of long representations. It was observed in \cite{Camporesi:1990wm,Camporesi:1993mz, Camporesi:1995fb,Camporesi,Camporesi:1994ga} that for many classes of fields --- including symmetric transverse traceless tensors, spinors, and $p$-forms --- the coincident heat kernel for the Laplacian on AdS$_{2n+1}$ is simply obtained by an analytic continuation\footnote{Subtleties related to even dimensional spheres and hyperboloids are discussed at length in \cite{Camporesi:1990wm,Camporesi:1995fb,Camporesi,Camporesi:1994ga}.} of the  corresponding quantity on $S^{2n+1}$. Inspired by these lessons, we will compute the heat kernel over generic mixed symmetry fields on AdS$_5$ by relating it to a heat kernel over $S^5$ via analytic continuation. We now describe how this is done, using the following results from harmonic analysis on spheres.  
Let $\cS$ be the space of fields on a five-sphere $S^5= SO(6)/SO(5)$,
then the space $\cS$ carries a UIR of $SO(5)$\,.
The eigenvalues of the Laplacian acting on $\cS$ 
are determined by the quadratic Casimirs $Q_\cR$ of the UIRs $\cR$ of $SO(6)$
whose restriction to $SO(5)$ include the representation $\cS$\,. 
By imposing irreducibility conditions (such as transversality or tracelessness) on the fields, one can further constrain the permitted set of $\cR$s. Then by definition, the traced and  integrated heat kernel is given by
\begin{equation}
K_\cS(t)=\sum_{\cR|\cS}d_\cR\, e^{-t\, Q_\cR},
\end{equation}
where $\cR|\cS$ means that all the UIR $\cR$ of $SO(6)$ whose restriction to $SO(5)$ 
include $\cS$\,.
The degeneracy of the eigenvalue,  $d_{\cR}$, is given by the dimension of the representation $\cR$\,. 
Using the homogeneity of the sphere, we conclude that the corresponding coincident heat kernel is given by
\begin{equation}
k_\cS(t)=\sum_{\cR|\cS}{d_{\cR}\over {\rm Vol}_{S^5}}\,e^{-t\, Q_{\cR}}.
\label{kcoins5}
\end{equation}
More concretely, we consider the space of  the fields on $S^5$ which transforms in
\be
	\cS=\left(\ell_1,\ell_2\right)\qquad [\ell_1\ge \ell_2\ge0]\,,
\ee 
representation of $SO(5)$ and satisfy irreducibility conditions.
Then, the $SO(6)$ UIRs which can branch into $\cS$ are 
\be
	\cR=\left(\ell_0,\ell_1,\pm \ell_2\right) \qquad [\ell_0\ge \ell_1]\,.
\ee
Hence, for a given $\cS$, namely $(\ell_1,\ell_2)$, we have one free integer parameter $\ell_0$ for $\cR$
and the choice of the sign in $\pm\ell_2$\,.
The dimension $d_{\cR}$ of the above represention is
\begin{equation}
d_{\cR}=d_{\left(\ell_0,\ell_1,\pm\ell_2\right)}={\left(\ell_0+2\right)^2-\left(\ell_1+1\right)^2\over 2^2-1^2}\,{\left(\ell_0+2\right)^2-\ell_2^2\over 2^2-0^2}\,{\left(\ell_1+1\right)^2-\ell_2^2\over 1^2-0^2}\,,
\end{equation}
and it does not depend on the sign in $\pm\ell_2$\,. Moreover the quadratic Casimirs also do not 
depend on this sign: $Q_{(\ell_0,\ell_1,+\ell_2)}=Q_{(\ell_0,\ell_1,-\ell_2)}$\,,
hece the coincident heat kernel $k_{(\ell_1,\ell_2)}(t)$ receives the same contribution twice,
once from $(\ell_0,\ell_1,+\ell_2)$  and the other time from  $(\ell_0,\ell_1,-\ell_2)$.
By imposing a duality condition on $\cS$\,, one can also restrict $\cR$ to one of these two.

The corresponding computation for AdS$_5$ is more subtle because $AdS_5=SO(1,5)/SO(5)$, and $SO(1,5)$ being non-compact admits infinite dimensional unitary irreducible representations.
It has been explicitly demonstrated for many classes of fields 
in \cite{Camporesi:1990wm,Camporesi:1993mz,Camporesi:1995fb,Camporesi,Camporesi:1994ga} that the coincident heat kernel \eqref{kcoins5} on $S^5$ may be analytically continued to that on $AdS_5$ 
 via 
 \be
 	\ell_0\mapsto i\,u-2 \qquad [0<u<\infty]\,.
\ee 
Moreover
the sum over $\ell_0$ gets mapped to an integral over $u$
and $t$ becomes $-t$\,,  and we obtain
\begin{equation}
k_{\left(\ell_1,\ell_2\right)}(t)=\int_0^\infty d u\, 
\mu\!\left(u;\ell_1,\ell_2\right) e^{t\,Q_{(i\,u-2,\ell_1,\ell_2)}},
\end{equation}
where the measure,
\begin{equation}
\mu\!\left(u;\ell_1,\ell_2\right)={d_{(i\,u -2,\ell_1,\ell_2)}\over {\rm Vol}_{S^5}}
=
\frac{\left(u^2+(\ell_1+1)^2\right)\left(u^2+\ell_2^2\right)
\left(\ell_1+\ell_2+1\right)\left(\ell_1-\ell_2+1\right)}{12\,{\rm Vol}_{S^5}},
\end{equation}
is known as the Plancherel measure and intuitively corresponds to the degeneracy of the eigenvalue $Q_{(i\,u-2,\ell_1,\ell_2)}$. It was observed in \cite{Lal:2012ax} that to reproduce the thermal partition function of a conformal primary $\cD(\Delta,(\ell_1,\ell_2))$\,, one has to consider quadratic fluctuations of fields carrying the representation $\left(i\,u-2,\ell_1,\ell_2\right)$ of $SO(6)$\,, but 
the eigenvalues of the kinetic operator should be shifted such that 
the coincident heat kernel of the kinetic operator is replaced by
\begin{equation}
k_{\Delta,(\ell_1,\ell_2)}(t)=\int_0^\infty du\, \mu\!\left(u;\ell_1,\ell_2\right)
e^{-t\left[u^2+(\D-2)^2\right]}\,.
\end{equation}
On going through these replacements and taking the Mellin transform, we arrive at the following expression for the zeta function corresponding to the coincident heat kernel.
\begin{equation}
\zeta_{\Delta,(\ell_1,\ell_2)}(z)=\int_0^\infty \frac{dt}t\,\frac{t^{z}}{\G(z)}\,
K_{\Delta,(\ell_1,\ell_2)}(t)
                   =\frac{{\rm Vol}_{AdS_5}}{3\,\pi^3}\int_0^\infty du\,D_{\D}(z,u)\,S_{(\ell_1,\ell_2)}(u)\,,
                   \label{zetaads5}
\end{equation}
where $D_{\D}$ and $S_{(\ell_1,\ell_2)}$ are given by
\ba
	D_{\D}(z,u)\eq\frac{1}{\left[u^2+\left(\Delta-2\right)^2\right]^z}\,,
	\label{D fn}\\
	S_{(\ell_1,\ell_2)}(u)
	\eq\frac{\ell_1+\ell_2+1}2\,\frac{\ell_1-\ell_2+1}2
	\left(u^2+(\ell_1+1)^2\right)\left(u^2+\ell_2^2\right).
	\label{S fn}
\ea
For the case of symmetric tensors, i.e. $\ell_2=0$\,, this expression agrees with \cite{Giombi:2014iua}. 
For the later analysis, it is convenient
to work also with the $su(2)\oplus su(2)$ label $[j_+,j_-]$ 
together with the $so(4)$ one $(\ell_1,\ell_2)$\,. 
The relation between $[j_+,j_-]$ and $(\ell_1,\ell_2)$ is given by
\eqref{ell j}, and the function $S_{[j_+,j_-]}=S_{(\ell_1,\ell_2)}$ reads
\be
	S_{[j_+,j_-]}(u)
	=\frac{2\,j_++1}2\,\frac{2\,j_-+1}2
	\left(u^2+(j_++j_-+1)^2\right)\left(u^2+(j_+-j_-)^2\right).
\ee
Finally, the volume of AdS$_5$ can be regularized to
\be
	{\rm Vol}_{AdS_5}=\pi^2\,\log(\mu\,R)\,,
\ee
as in \cite{Diaz:2007an}. Hence, 
the IR regularized vacuum energy in AdS$_5$ is proportional to
the logarithm of the AdS radius $R$
and the holographic renormalization scale $\mu$\,.
Hereafter, we suppress the dependence of $\mu$ as it 
always appears with $R$\,.

\subsection{Zeta Function from Character}

Next we will show how this zeta function may be written in terms of the character \eqref{so24 ch} of the representation $\cV(\Delta,[j_+,j_-])$\,.
The latter is given in the variables $(\b,\a_+,\a_-)$\,, with $q=e^{-\b}$ and $x_{\pm}=e^{i\,\a_\pm}$\,, by
\begin{equation}
\chi_{\Delta,[j_+,j_-]}(\beta,\alpha_+,\alpha_-)
={e^{-\left(\Delta-2\right)\beta}\,
\sin\!\left(\frac{2\,j_++1}2\,\alpha_+\right)
\sin\!\left(\frac{2\,j_-+1}2\,\alpha_-\right)\over \eta(\beta,\alpha_+,\alpha_-)}\,,
\label{char jj}
\end{equation}
where
\be
\eta(\beta,\alpha_+,\alpha_-)=4\,\sin{\alpha_+\over 2}\,\sin{\alpha_-\over 2}\left(\cosh\beta - \cos\!\left({\alpha_++\alpha_-\over 2}\right)\right)\left(\cosh\beta - \cos\!\left({\alpha_+-\alpha_-\over 2}\right)\right).
\ee
In the formula \eqref{zetaads5} for the zeta function,
we replace the factor $D_{\D}$ \eqref{D fn} by
\begin{equation}
D_{\D}(z,u)=
{\sqrt{\pi}\over\Gamma\left(z\right)}\int_0^\infty d\beta\,
e^{-\left(\Delta-2\right)\b}\,\left(\beta\over 2\,u\right)^{z-{1\over 2}}J_{z-{1\over 2}}\!\left(u\,\beta\right),
\label{D int}
\end{equation}
and the factor $S_{[j_+,j_-]}$ \eqref{S fn} by
\be
\begin{split}
	S_{[j_+,j_-]}(u)=&\ \partial_{\a_+}\,\partial_{\a_-}
	\left(u^2-(\partial_{\a_+}+\partial_{\a_-})^2\right)
	\left(u^2-(\partial_{\a_+}-\partial_{\a_-})^2\right)\times\\
	&\times
	\sin\!\left(\frac{2\,j_++1}2\,\alpha_+\right)
\sin\!\left(\frac{2\,j_-+1}2\,\alpha_-\right)
\bigg|_{\a_\pm=0}\,.
\end{split}
\label{S differ}
\ee
The left hand sides of \eqref{D int} and \eqref{S differ}
involve the depedence in $\D$ and $j_\pm$ only through $e^{-(\D-2)\,\b}$ and 
$\sin\!\left(\frac{2\,j_++1}2\,\alpha_+\right)\sin\!\left(\frac{2\,j_-+1}2\,\alpha_-\right)$ which 
are nothing but the  numerator of the character $\chi_{\D,[j_+,j_-]}$ \eqref{char jj}.
Therefore, similarly to AdS$_4$ case, this observation allows to relate
the zeta function $\zeta_{\D,[j_+,j_-]}$ to the character $\chi_{\D,[j_+,j_-]}$\,.
An important advantage from the AdS$_4$ case ---  
where $\tanh \pi u$ term \eqref{f int} complicates the integral ---
 is that the original $u$-integral in \eqref{zetaads5}
can be exactly evaluated in AdS$_5$ as
\begin{equation}
	\sqrt{\pi}\int_{0}^{\infty}du\,u^{2n}\,
	\left(\frac{\beta}{2\,u}\right)^{z-\frac12}
	J_{z-\frac12}(\beta\,u)=
	\frac{\sqrt{\pi}}2\,\frac{\Gamma(n+\frac{1}2)}{\Gamma(z-n)}\left(\frac\b2\right)^{2(z-1-n)}
	\quad [z>n]\,,
\end{equation} 
after  exchanging  the order of integrals in $\b$ and $u$\,.
Finally, we obtain the zeta function as 
the sum of three pieces:
\begin{equation}
\zeta_{\cH}\!\left(z\right):=
\zeta_{\cH|2}\!\left(z\right)+\zeta_{\cH|1}\!\left(z\right)+\zeta_{\cH|0}\!\left(z\right)\,,
\label{zeta sum}
\end{equation}
where $\zeta_{\cH|n}$ are the Mellin transforms,
\be
\frac{\Gamma(z)\,\zeta_{\cH|n}\!\left(z\right)}{\log R}=\int_0^\infty 
d\beta\,\frac{\big(\frac\beta2\big)^{2(z-1-n)}}{\Gamma\!\left(z-n\right)}\,
f_{\cH|n}(\b)\,,
\label{zeta f}
\ee
of the functions $f_{\cH|n}(\b)$ given by
\ba
f_{\cH|2}\!\left(\b\right)\eq{1\over 8}\,\partial_{\alpha_+}\,\partial_{\alpha_-}
\Big[\eta(\b,\a_+,\a_-)\,\chi_\cH(\b,\a_+,\a_-)\Big]_{\a_\pm=0}=\frac{\sinh^4\tfrac{\b}2}2\,\chi_\cH(\b,0,0)\,,\\
f_{\cH|1}\!\left(\b\right)\eq-{1\over 6}\,
\partial_{\alpha_+}\,\partial_{\alpha_-}\!\left(\partial_{\alpha_+}^2+\partial_{\alpha_-}^2\right)
\Big[\eta(\b,\a_+,\a_-)\,\chi_\cH(\b,\a_+,\a_-)\Big]_{\a_\pm=0}\nn
\eq\sinh^2\tfrac\b2\,
\left[\frac{\sinh^2\tfrac\b2}3-1
-\sinh^2\tfrac\b2\left(\partial_{\a_1}^2+\partial_{\a_2}^2\right)\right]
\chi_\cH(\b,\a_1,\a_2)\Big|_{\a_i=0},
\ea
and 
\be\begin{split}
&f_{\cH|0}\!\left(\b\right)={1\over 6}\,
\partial_{\alpha_+}\,\partial_{\alpha_-}\!\left(\partial_{\alpha_+}^2-\partial_{\alpha_-}^2\right)^{\!2}
\Big[\eta(\b,\a_+,\a_-)\,\chi_\cH(\b,\a_+,\a_-)\Big]_{\a_\pm=0}\\
&=\bigg[1
+\frac{\sinh^2\frac\b2\,(3-\sinh^2\frac\b2)}3\left(\partial_{\a_1}^2+\partial_{\a_2}^2\right)\\
&\qquad\qquad
-\frac{\sinh^4\tfrac\b2}3 \left(\partial_{\a_1}^4
-12\,\partial_{\a_1}^2\partial_{\a_2}^2+\partial_{\a_2}^4\right)\bigg]
\chi_\cH(\b,\a_1,\a_2)\Big|_{\a_i=0}\,.
\end{split}\ee
Note that 
we have related the  zeta function of a set of fields given by a Hilbert space $\cH$\,,
\begin{equation}
\zeta_{\cH}(z) = \sum_{\Delta,j_+,j_-}N^{\cH}_{\Delta,[j_+,j_-]}\,\zeta_{\D,[j_+,j_-]}(z)\,,
\end{equation}
to the corresponding character,
\begin{equation}
\chi_{\cH}(\b,\a_+,\a_-) = \sum_{\Delta,j_+,j_-}N^{\cH}_{\Delta,[j_+,j_-]}\,\chi_{\D,[j_+,j_-]}(\b,\a_+,\a_-) \,,
\end{equation}
where $N^\cH_{\D,[j_+,j_-]}$
is the multiplicity of $\cV(\D,[j_+,j_-])$ representation in the space $\cH$\,.
The existence of such relation is due to the
fact that the formulas  are linear
 and do not involve any explicit dependence on $\D$ or $j_\pm$\,.

\bigskip

Let us emphasize that in AdS$_5$
it was not necessary  to change the regularization scheme 
from the ordinary zeta function $\zeta_{\cH}$ to a deformed one $\tilde\zeta_{\cH}$
 as in AdS$_4$\,. Hence, 
what we shall
 compute in the following
 are  the standard zeta functions.
This should  be the case also for other odd dimensional AdS spaces
where the absence of $\tanh\pi u$ term makes possible to 
evaluate the $u$ integral.

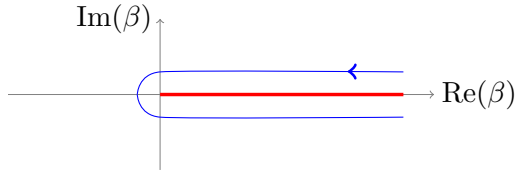
\begin{figure}[h]
\centering
\begin{tikzpicture}
\draw [help lines,->] (-2,0) -- (3.6,0);
\draw [help lines,->] (0,-1) -- (0,1);
\node at (4.2,0){Re$(\b)$};
\node at (-0.6,1) {Im$(\b)$};
\draw [line width =1.3pt, red] (0,0) to (3.2,0);
\draw [blue,
   decoration={markings,
  mark=at position 0.2 with {\arrow[line width=1.2pt]{>}}
  }
  ,postaction={decorate}]
  plot [smooth, tension=0.3]
  coordinates {(3.2,0.3) (0,0.3) (-0.3,0)};
  \draw [blue, postaction={decorate}]
  plot [smooth, tension=0.3]
  coordinates {(-0.3,0) (0,-0.3)  (3.2,-0.3)};
\end{tikzpicture}
\caption{Integration contour for the zeta function}
\label{fig: ct1}
\end{figure}

 Another important property of AdS$_5$ zeta function,
 which should hold in  other odd dimensions,
 is the presence of gamma functions in the right hand sides
of the formula \eqref{zeta f}.
Thanks to this property, one can easily show that the UV divergence of the vacuum energy
--- corresponding to $\zeta_{\cH}(0)$ --- is 
universally absent, as is a well-known fact of odd dimensions:
\be
	\int_0^\infty d\b\,\frac{\left(\frac{\b}2\right)^{2(z-1-n)}}{\G(z-n)}\,
	f_{\cH|n}(\b)
	=-2\,\gamma_{\cH|n}+\mathcal{O}(z)\,.
	\label{f integral}
\ee
Moreover the finite part of the vacuum energy will be 
entirely captured by the divergence arising from the neighborhood of $\b=0$\,.
If the function $f_{\cH|n}$ does not have any singularity 
around the positive real axis of $\b$ except for the pole at $\b=0$, then
the integral \eqref{f integral} with a sufficient large ${\rm Re}(z)$
can be recast into to the contour integral,
\be
    \frac i{2\,\sin (2\pi z)}\,\oint_C d\b\,\frac{\left(-\frac{\b}2\right)^{2(z-1-n)}}{\G(z-n)}\,
	f_{\cH|n}(\b)\,,
	\label{f contour}
\ee 
where the contour encircles the branch cut generated by $\b^{2(z-1-n)}$
in the counter-clockwise direction (see Fig.\,\ref{fig: ct1}). 

Differently from the form \eqref{f integral}, the above contour integral 
is well-defined for any value of $z$\,,  hence we can directly put $z=0$\,.
With this evaluation, the integrand becomes free from branch cut
and the contour can be shrunken to a small circle around $\b=0$\,.
In the end, the finite part $\g_{\cH|n}$ will be given by the residue, 
\be 
	\g_{\cH|n}
	=-(-4)^{n}\,n!\oint \frac{d\b}{2\,\pi\,i}\,
	\frac{f_{\cH|n}(\b)}{\b^{2(n+1)}}\,.
	\label{g H n}
\ee
Practically, this amounts to
expanding the functions $f_{\cH|2}, f_{\cH|1}$
and $f_{\cH|0}$ around $\b=0$ and picking up the term proportional
to $\b^5, \b^3$ and $\b$, respectively.
Note that the residue vanishes 
for any even  functions $f_{\cH|n}(\b)$\,,
which is guarantied when the character $\chi_{\cH}$ itself is even in $\beta$\,.
In the end, the one-loop  vacuum energy of the AdS$_5$  theory with the spectrum $\cH$   is given by the sum of three coefficients as
\be
	\Gamma^{\sst (1)\, \rm ren}_{\cH}=\log R \left(\gamma_{\cH|2}
	+\gamma_{\cH|1}+\gamma_{\cH|0}\right).
	\label{fin G g}
\ee
To summarize,
the one-loop vacuum energy can be obtained from
\begin{description}
  \item[Prescription 1] Extracting $\b^{2n+1}$ coefficients
    of the functions $f_{\cH|n}$
    \label{pres 1}
    
     \item[Prescription 2]  The integral  \eqref{f integral} with an analytic continuation on $z$
    \label{pres 2}
    
    \item[Prescription 3] The contour integral \eqref{f contour}
    \label{pres 3}
\end{description}
These three prescriptions are equivalent
as far as the functions $f_{\cH|n}$
free from any singularity around
positive real $\b$ axis except for the pole
at $\b=0$\,.
This is the case for any one particle state in AdS$_5$\,,
as well as for the spectrum of Vasiliev's theory.
However, we will see that it is no more
true for the AdS dual to a matrix model CFT.
We shall come back to this issue 
after considering a few examples:
 the vector models as the first example,
then the sample calculations for the spectra of the second
and third Regge trajectories (that correspond
to the CFT operators involving three and four fields $\bm\phi$, respectively).
For a better illustration, we will consider both the exact evaluation of the $\b$ integral keeping the $z$ dependence and the residue results \eqref{g H n} and \eqref{fin G g}.

\subsection{Test with Vector Models}

In order to test the result in the previous section,
we compute the zeta function of the non-minimal/minimal Vasiliev theory, dual to 
free $U(N)$/$O(N)$ model CFT.

Let us first consider the non-minimal Vasiliev theory,
whose spectrum corresponds to the tensor product of two Rac representation.
The character of Rac obtained in \eqref{char Rac5} can be written 
in terms of $(\b,\a_1,\a_2)$ variables as 
\be
	\chi_{\rm Rac}(\b,\a_1,\a_2)=\frac{\sinh\b}{2\left(\cosh\b-\cos\a_1\right)
	\left(\cosh\b-\cos\a_2\right)}\,,
\ee
hence the character of the non-minimal model is
\be
	\chi_{\textrm{non-min}}(\b,\a_1,\a_2)=\left[\chi_{\rm Rac}(\b,\a_1,\a_2)\right]^2
	=\frac{\sinh^2\b}{4\left(\cosh\b-\cos\a_1\right)^2
	\left(\cosh\b-\cos\a_2\right)^2}\,.
\ee
One can first notice that the above character is even in $\beta$
hence the residues \eqref{g H n} vanish implying that the one-loop vacuum energy vanishes.
We may nevertheless proceed to evaluate the zeta function 
for more concrete understanding.
For the evaluation of the zeta function, we need to first
calculate the corresponding $f_{\textrm{non-min}|n}(\b)$ functions.
They are given by
\be
	f_{\textrm{non-min}|2}(\b)= \frac{\coth^2\frac \b2}{32}\,, \quad
	f_{\textrm{non-min}|1}(\b)= \frac{\coth^2\frac \b2}{48}-
	\frac{\coth^2\frac \b2}{24\,\sinh^2\frac\b2}\,, \quad
	f_{\textrm{non-min}|0}(\b)= 0\,.
\ee
With these we can obtain $\zeta_{\textrm{non-min}|n}(z)$'s by
performing Mellin transforms as \eqref{zeta f}.
However, the integral is divergent for large $\b$ 
for a $z$ which regularizes the small $\b$ divergence.
Actually this large $\b$ divergence is due to the contribution
of $\cD(2,0)$ which is the lightest scalar field in AdS$_5$  in the spectrum.
The latter has already ill-defined zeta function at the level of \eqref{zetaads5}.
One can regularize this divergence by increasing its mass by infinitesimally small amount as in \cite{Giombi:2014iua}.
This can be realized by replacing the $\D$ value from 2 to $2+\e$\,, 
and it amounts to
inserting a $e^{-\e\,\b}$ term in the $\b$ integral.
In this scheme, we get
\ba
	&& \frac{\G(z)\,\zeta_{\textrm{non-min}}(z)}{\log R}\nn
	&&=\,
	\lim_{\e\to0} \int_0^\infty d\b\,e^{-\e\,\b}\left[
	\frac{\big(\frac\b2\big)^{2(z-3)}}{\G(z-2)}\,\frac{\coth^2\frac \b2}{32}
	+\frac{\big(\frac\b2\big)^{2(z-2)}}{\G(z-1)}
	\left( \frac{\coth^2\frac \b2}{48}-
	\frac{\coth^2\frac \b2}{24\,\sinh^2\frac\b2}\right)\right]\nn
	&& = \, \frac{9\, \G(z-\frac{5}{2}) \, \zeta(2z-6) - 8\, \G(z-\frac{3}{2}) \, \zeta(2z-6) + 2 \, \G(z-\frac{3}{2}) \, \zeta(2z-4)}{72\,\sqrt{\pi }}
	\label{ads5ractest}
\ea
where we used the integration method of Appendix \ref{sec: AdS5 zeta}. 
In the $z\to0$ limit, the ratios of gamma functions in the right hand side
are finite whereas the zeta functions vanish: $\zeta(-6)=\zeta(-4)=0$\,. Hence,
we can verify that the one-loop vacuum energy vanishes. Alternatively, we can use Laurent expansion of $f_{\text{non-min}|n}$ with the variable $q \equiv e^{-\b}$. By this, we can check that the quantities $\gamma_{\text{non-min}|n}$ defined in \eqref{g H n} individually vanish for $n=0,1,2$ (since the functions are even in $\b$). This result again verify the vanishing of the one-loop vacuum energy.

We next consider the minimal Vasiliev theory with
only even spins. The corresponding character is given 
again in terms of $\chi_{\rm Rac}$ as \eqref{char min}. 
Since the $\chi_{\rm Rac}^2=\chi_{\textrm{non-min}}$ gives trivial vacuum energy,
 the only non-trivial contribution may come from 
the second term $\chi_{\rm Rac}(2\b,2\a_1,2\a_2)/2=
\chi_{\rm min}(\b,\a_1,\a_2)-\chi_{\textrm{non-min}}(\b,\a_1,\a_2)/2=:\chi_{\rm R}(\b,\a_1,\a_2)
$\,. After some manipulation, one get
\be
	f_{\rm R|2}(\b)=\frac{\cosh\b\,\sinh^4\frac\b2}{16\,\sinh^3\b}\,,
	\quad
	f_{\rm R|1}(\b)=\frac{\cosh\b\,(\sinh^2\frac\b2-2)\,\sinh^6\frac\b2}{6\,\sinh^5\b}\,,
	\quad
	f_{\rm R|0}(\b)=0\,.
\ee
The corresponding zeta function is given by
\ba
	&&\frac{\G(z)\,\zeta_{\textrm{R}}(z)}{\log R}= \frac{\G(z)}{\log R}
	\left(\zeta_{\textrm{min}}(z)-\frac12\,\zeta_{\textrm{non-min}}(z)\right)\nn
	&&= \lim_{\e\to0} \int_0^\infty d\b\,e^{-\e\,\b}\left[
	\frac{\big(\frac\b2\big)^{2(z-3)}}{\G(z-2)}\,
	\frac{\cosh\b\,\sinh^4\frac\b2}{16\,\sinh^3\b}
	+\frac{\big(\frac\b2\big)^{2(z-2)}}{\G(z-1)}\,
	\frac{\cosh\b\,(\sinh^2\frac\b2-2)\,\sinh^6\frac\b2}{6\,\sinh^5\b}
	\right]\nn
	&&=-\frac{2^{1-4z} \, \Gamma(2z-5) \left[ (2^{2z} -256)  \, \zeta(2z-7) + (2^{2z} - 64) \, \zeta(2z-5)\right]}{\Gamma(z-2)}\nn
	&&\ \ \ -\frac{2^{-4z} \, \Gamma(2z-3) \left[ (2^{2z} -256)  \, \zeta(2z-7) + 
	3\,(2^{1+2z}-128)\, \zeta(2z-5) + (2^{2z}-16) \, \zeta(2z-3)\right]}{3\,\Gamma(z-1)} \nn
	&&= -\frac{1}{45}+ \mathcal{O}(z)\,.
	\label{min Vas 5}
\ea
We therefore correctly reproduce the result of \cite{Giombi:2014iua} for the one-loop vacuum energy of the minimal Vasiliev theory
\be
	\G^{\sst (1)\,\rm ren}_{\rm min} =-\frac{1}{2} \,\zeta_{\rm R}'(0) = \frac{\log R}{90}\,.
	\label{min AdS5}
\ee
Again, instead of computing the zeta function explicitly, one can directly
identify the vacuum energy from the residue calculations.
By Laurent expanding the $f_{{\rm R}|n}(\b)$ we get
\be
	\gamma_{\rm R|2}=\frac{13}{1920},
	\qquad
	\gamma_{\rm R|1}=\frac{5}{1152},
	\qquad 
	\gamma_{\rm R|0}=0\,,
\ee
whose sum again gives 1/90.
Therefore, the examples of vector models show the agreement.
We will now compare the above result with $\Gamma^{\sst (1)\,\rm ren}_{\sst\rm Rac}$\,, the one-loop vacuum energy associated to the Rac representation in AdS$_5$. This does not represent a propagating degree of freedom  in the bulk. Nonetheless, we can formally define and evaluate a one-loop determinant corresponding to this field. 
We will find that the answer correctly reproduces the $a$-anomaly of the conformal scalar on the boundary. See \cite{Beccaria:2016tqy} for related discussions.
We also remark here that this result will be useful for computing the one-loop vacuum energy of the 
AdS dual of the free $SU(N)$ adjoint scalar field theory as well. Again, we first calculate the functions $f_{{\rm Rac}|n}$ and get
\be
	f_{\rm Rac|2}(\b)=\frac{\sinh\b}{16}\,,
	\qquad
	f_{\rm Rac|1}(\b)=\frac{\sinh\b}{24}\,,
	\qquad
	f_{\rm Rac|0}(\b)=0\,.
	\label{f Rac}
\ee
With these, the zeta function is given by the Mellin tranform,
\ba
	\frac{\G(z)\,\zeta_{\textrm{Rac}}(z)}{\log R}
	\eq\lim_{\e\to0} \int_0^\infty d\b\,e^{-\e\,\b}\left[
	\frac{\big(\frac\b2\big)^{2(z-3)}}{\G(z-2)}\,
	\frac{\sinh\b}{16}
	+\frac{\big(\frac\b2\big)^{2(z-2)}}{\G(z-1)}\,
	\frac{\sinh\b}{24}
	\right]\nn
	\eq  -\frac{(-1)^{2z}\left[1+(-1)^{2z}\right] (z-1)\,\Gamma(z-\frac{5}{2})}{48\, \sqrt{\pi}} = -\frac{1}{45} + \mathcal{O}(z)\,,
\ea
where we have again regularized the divergence coming from large $\b$ region
by introducing a $e^{-\e\,\b}$ damping factor. This divergence arises because the conformal weight is smaller than 2. By expanding the zeta function around $z=0$\,,
one can show that it gives the same value as the vacuum energy of the conformal scalar on the boundary. Once again, the finite part of the vacuum energy can be directly extracted 
by expanding \eqref{f Rac} around $\b=0$\,, and we immediately get
\be
	\gamma_{\rm Rac|2}=-\frac{1}{60},
	\qquad
	\gamma_{\rm Rac|1}=\frac{1}{36},
	\qquad 
	\gamma_{\rm Rac|0}=0\,.
\ee
By summing the above three contributions, we get 
\be
	\G^{\sst (1)\,\rm ren}_{\rm Rac} = \frac{\log R}{90}\,.
	\label{Vac Rac}
\ee
Interestingly again, as mentioned at the outset, the IR log divergence of $\zeta_{\rm Rac}$ gives the UV log divergence of the vacuum energy 
of the conformal scalar on the boundary.

\subsection{Vacuum Energy for the AdS Dual of $SU(N)$ Adjoint Model}
We now turn to the main result of the paper, one-loop vacuum energy computation of the AdS$_5$ theory which is dual to the free $SU(N)$ adjoint scalar CFT on $S^4$. We first compute the vacuum energy of the fields in the 
first few Regge trajectories
using the method of residues, following the prescription \eqref{g H n}. We present a trend of the vacuum energy growth by the power of the fields $\bm\phi$\,, exhibited in Fig.\,\ref{fig}. Next, we will take the limit $N \to\infty$, and compute the one-loop vacuum energy for the corresponding theory. This indicates a non-trivial shift in the relation between the bulk dimensionless coupling $g$ and the boundary parameter $N$.

\subsubsection{Vacuum Energies for a Few Low Orders}

We first carry out the determination of the one-loop vacuum energy of the fields in  the 
first few Regge trajectories using the method of residues. We have explicitly evaluated and exhibited the contributions of terms up to order 4 in the fields $\bm\phi$ and results up to order 32 are then displayed graphically. While the overall pattern for the one-loop vacuum energy is chaotic, and indeed non-monotonic as well, we do observe the trend that the vacuum energies corresponding to $n=2,4,8,16,32$
are exactly $1,2,4,8,16$ times of the $n=1$ case, respectively. We start with the order two contribution.

\paragraph{Order Two}
The order two spectrum coincides with that of minimal Vasiliev theory,
hence the vacuum energy is equal to that of Rac:
\be
	\G^{\sst (1)\,\rm ren}_{\rm cyc^2}=	\frac{\log R}{90}
	\simeq 0.0111111\,\log R\,.
\ee
\paragraph{Order Three}
The order three cyclic character is given analogously to \eqref{char 3}.
From that, we first calculate $f_{{\rm cyc^3}|n}(\b)$'s
and their residues,
\be 
	\g_{\rm cyc^3|2}=\frac{7627}{475200}\,,
	\quad
	\g_{\rm cyc^3|1}=\frac{46483}{3991680}\,,
	\qquad
		\g_{\rm cyc^3|0}=-\frac{3071}{561330}\,.
\ee
By summing the above, we get
\be 
	\G^{\sst (1)\,\rm ren}_{\rm cyc^3}=	\frac{362911}{16329600}\,\log R=
	\simeq 0.0222241\,\log R\,,
\ee
which is roughly twice of the order two contribution.

\paragraph{Order Four}
The order four cyclic character is given analogously to \eqref{char 4},
and we obtain similarly the residues,
\be 
	\g_{\rm cyc^4|2}=\frac{253}{15360}\,,
	\quad
	\g_{\rm cyc^4|1}=\frac{125}{9216}\,,
	\qquad
		\g_{\rm cyc^4|0}=-\frac{1}{128}\,.
\ee
whose sum simplifies ending up with
\be 
	\G^{\sst (1)\,\rm ren}_{\rm cyc^3}=	\frac{\log R}{45}
	\simeq 0.0222222\,\log R\,.
\ee
Interestingly, the order four contribution is exactly twice of 
the contribution of Rac or the order two.
We can see that the vacuum energies do not increase monotonically
as the order four part is slightly smaller than the order three one.

\paragraph{Higher Orders}

\begin{figure}[h]
\centering
  \includegraphics[width=0.8\linewidth]{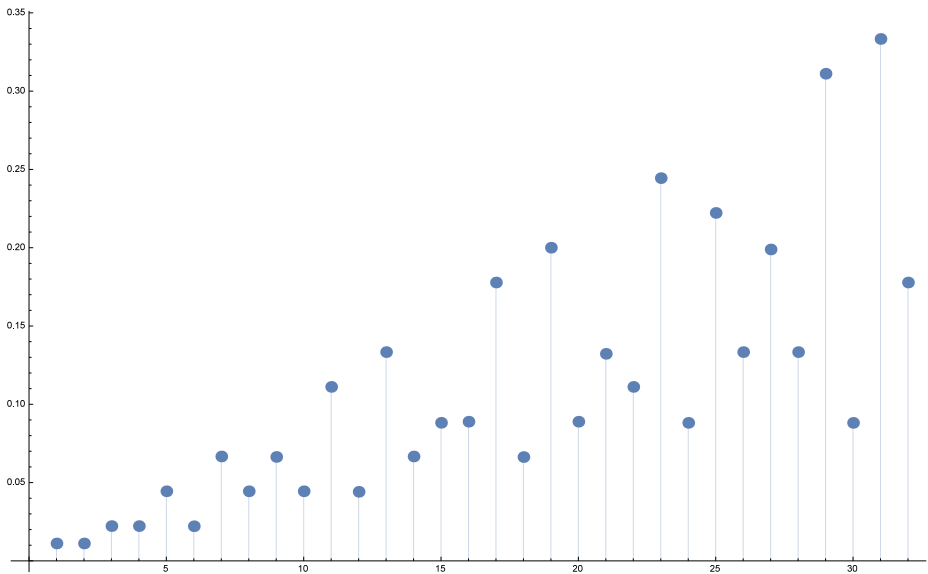}
  \caption{Vacuum energies of AdS${}_5$ fields in different orders}
  \label{fig}
\end{figure}

\begin{figure}[h]
\centering
  \includegraphics[width=0.8\linewidth]{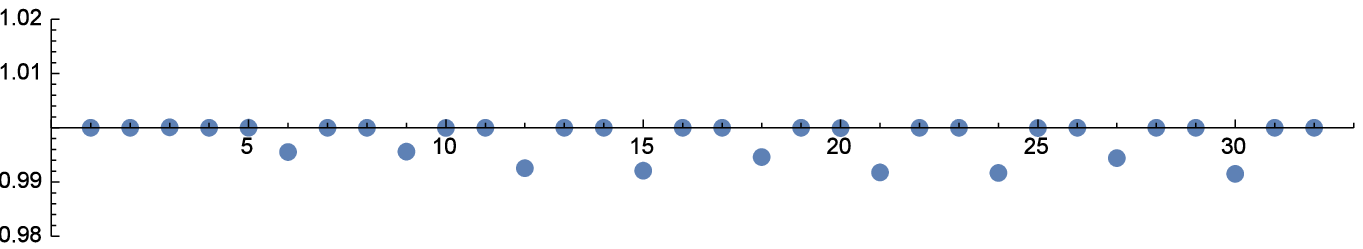}
  \caption{Plot of $\G^{\sst (1)\,\rm ren}_{\sst{\rm cyc}^n}/(\varphi(n)\,\G^{\sst (1)\,\rm ren}_{\sst\rm Rac})$ in AdS$_5$ from $n=1$ to 32. }
  \label{figReg}
\end{figure}

In order to have a better idea, we can proceed to calculate the higher order contributions to
the vacuum energy. Fig.\,\ref{fig} shows the values of the vacuum energies
for the fields corresponding to $\chi_{{\rm cyc}^n}$ up to the order $n=32$.
As we have commented previously, the energies show an approximate linear growth with some chaotic oscillations.
Additionally,
 the vacuum energies corresponding to $n=2,4,8,16,32$
are exactly $1,2,4,8,16$ times of the $n=1$ case, respectively. 
One can notice that the first eight values of the vacuum energies
show very similar pattern as the AdS$_4$ case (Fig.\,\ref{fig1}).
In fact, both vacuum energies follow the pattern of the Euler totient function very closely.
See Fig.\,\ref{figReg}. 
Apart from the $n=2^m$  cases, the values plotted in Fig\,\ref{figReg}
are not exactly 1 but have very small fluctuations. 
It would be interesting to investigate this pattern further.

\subsubsection{Vacuum Energy in Different Slices}
When the rank of the gauge group becomes infinite, 
the character encoding the set of all single trace operators in the CFT
can be simplified to obtain an expression in terms of the Euler totient function, given in the second equality of \eqref{cyc}, reproduced here for convenience:
\be
	\chi_{\rm adj}=	-\chi_{\rm Rac}+\sum_{k=1}^{\infty}\,\frac{\varphi(k)}{k}\,
	\chi_{{\rm log},k}\,.
	\label{cyc log 5}
\ee
The $\chi_{{\rm log},k}$ are given in the AdS$_5$ case by
\be 
	\chi_{{\rm log},k}(\b,\a_1,\a_2)=
	-\log\!\left[1-\chi_{\rm Rac}(k\,\b,k\,\a_1,k\,\a_2)\right].
	\label{log k}
\ee
The first term in \eqref{cyc log 5} subtracts 
the single field contribution from the rest, and its contribution to the vacuum energy
has been calculated in \eqref{Vac Rac}. We will therefore focus on 
the logarithmic term \eqref{log k} which is new. As we already argued in the AdS$_4$ case, carrying out this computation while summing over all $k$ leads to an infinite number of singular points in the $\beta$ place, clustering around $\beta=0$, making the partition function highly non-analytic. Therefore to carry out this computation in a well-defined way, we need to work at a fixed $k$, compute the one-loop vacuum energy contribution, and then sum over all different $k$ contributions 
with the weight $\varphi(k)/k$\,.
For these reasons, from now on let us focus on $\chi_{{\rm log},k}$ \eqref{log k}
and calculate the corresponding contribution to the vacuum energy
$\G^{\sst (1)\,\rm ren}_{{\rm log},k}$\,.
For that, we need to first identify
the functions $f_{{\rm log},k|n}(\b)$\,. 
After some compuations, we get
\ba
	f_{{\rm log},k|2}(\b)
	\eq \frac{\sinh ^4\frac{\b}{2}}{2}\, 
	\log\!\left[1-\frac{\sinh (k\b)}{8\,\sinh^4\frac{k\b}{2}}\right],
    \nn
	f_{{\rm log},k|1}(\b)
	\eq \frac{ \sinh ^2\frac{\b}{2}\left(\sinh ^2\frac{\b}{2}-3\right)}{6}\,
	\log\!\left[1-\frac{\sinh (k\b)}{8\,\sinh^4\frac{k\b}{2}}\right]
	+ \frac{k^2 \sinh ^4\frac{\b}{2}\, 
	\coth\frac{k\b}2}{\sinh (k\b) -8\,\sinh^4\frac{k\b}2}\,,
	 \nn
	f_{{\rm log},k|0}(\b)
	\eq
	\log\!\left[1-\frac{\sinh (k\b)}{8\,\sinh^4\frac{k\b}{2}}\right]
	-\frac23\,k^2\left[(k^2-1)\,\sinh ^2\tfrac{\b}{2}+3\right]	
	 \frac{\sinh ^2\tfrac{\b}{2}\,\coth\frac{k\b}{2}}{ 
	\sinh (k\b) -8\,\sinh^4\frac{k\b}2}\nn
	&&-\,2\,k^4
	\left(
	 \frac{\sinh ^2\tfrac{\b}{2}\,\coth\frac{k\b}{2}}{ 
	\sinh (k\b) -8\,\sinh^4\frac{k\b}2}\right)^2.
	\label{AdS5 log}
\ea
The next step is the identification of $\g_{{\rm log},k|n}$ with \eqref{g H n}.
Note that the issue of different prescriptions enters here.
As one can see, the function  $f_{{\rm log},k|n}(\b)$ do have 
additional branch cuts on the positive real axis of $\b$\,.
In fact, such singularities precisely
coincide with the Hagedorn phase transition which appears 
at the thermal AdS partition function,
where we interpret the integration variable $\b$ as 
the inverse temperature.
In our work, we are focusing on the AdS space with sphere boundary hence
there is no notion of temperature a priori. 
Nevertheless, the technical simplification of the vacuum energy calls for the use of the full character.
This
can be physically interpreted as the generalized partition function, which does see the Hagedorn transition at the specific point $\b = \b_{\rm H}$.
We emphasize again the $\b$ does not carry a 
thermal/geometric meaning on the boundary/bulk of the AdS space,
but simply enters to address properly the spectrum of the theory in consideration.
As we have shown before, the one-loop vacuum energy is given through the integral
of such partition function and the singularity arising at $\b_{\rm H}$
introduces an ambiguity of the prescription.
In other words, as we have seen from the computation of the vacuum energy of a fixed Regge trajectory,
the vacuum energy increases as the order of $\bm\phi$ in the dual CFT operator increases.
Hence, the full one-loop vacuum energy will be given as a divergent series.
This divergence, arising while summing over infinitely many trajectories, 
is not automatically regularized by the introduction of the UV regulator $z$
as in the Vasiliev's model case,
but requires a new regularization prescription.
This shows a clear difference between Vasiliev's theory (dual to vector models)
and stringy AdS theory (dual to matrix models).

The necessity of introducing a new regularization method is translated here 
to the choice of prescription among three possibilities proposed in Section \ref{pres 1}.
Let us examine the three prescriptions, one by one, for the AdS theory
dual to the $SU(N)$ adjoint model.

\subsubsection*{Prescription 1}

The prescription 1 is the simplest option and has well defined meaning even 
for the functions $f_{{\rm log},k|n}(\b)$:
it is sufficient to expand the
functions $f_{{\rm log},k|n}(\b)$ around the $\b=0$ point.
Considering first $f_{{\rm log},k|2}(\b)$, one can show that
the $\b^5$ coefficient is absent:
\be
	f_{{\rm log},k|2}(\b)
	=
	\frac{-3\,\log\b+\log(-\frac 2{k^3})}{32}\,\b^4
	+\frac{-3\,\log\b+\log(-\frac 2{k^3})}{192}\,\b^6
	-\frac{k^3}{64}\,\b^7+\cO\!\left(\b^8\right),
\ee
implying $\g_{{\rm log},k|2}=0$\,.
Similarly, we obtain the expansions of the remaining functions as
\ba
 	f_{{\rm log},k|1}(\b)\eq
	\frac{3\, \log \b-\log(-\frac{2}{k^3})+1}8\,\b^2 -\frac{k^2-2}{96}\,\b^4 +\frac{k^3}{8}\,\b^5 +\cO\!\left(\b^6\right),\\
	f_{{\rm log},k|0}(\b)\eq
	 -3 \log \b+\log (-\tfrac{2}{k^3})-\frac{3}{2}+\frac{k^2-1}{12}\, \b^2 
	 -\frac{3\,k^2}2\, \b^3 +\cO\!\left(\b^4\right),
\ea
and can verify that they miss the $\b^3$ and $\b$ coefficients, respectively.
Hence, $\g_{{\rm log},k|1}=\g_{{\rm log},k|0}=0$\,,
and in this prescrition, we are led to conclude
that the one-loop vacuum energy for 
a fixed slice vanishes.

\subsubsection*{Prescription 2}

The prescription 2 is the integral along the postive real axis of $\b$\,.
This integral is convergent only for large enough $z$ hence requires an analytic continuation on $z$\,.
In the following, we sketch 
how we can carry out the computation of 
such integral for $\zeta_{{\rm log},k|2}(z)$.
We first recast the integral into
\ba
&& \frac{\G(z)\, \zeta_{\textrm{log},k|2}(z)}{\log R} =  -\frac{(2k)^{5-2z}\, \Gamma (2 z-5)}{16\,\Gamma (z-2)}\times \nn 
&& \qquad \times
\left[\Psi_k(p_1,2z-5)+\Psi_k(p_2,2z-5)+\Psi_k(p_3,2z-5)-
3\,\Psi_k(1,2z-5)\right],
\label{Log k=2}
\ea
by introducing 
\be
    \Psi_k(p,z)=6 \, \tilde\Phi(p,z,0)-4\,\tilde\Phi(p,z,-\tfrac{1}{k})
    -4\,\tilde\Phi(p,z,\tfrac{1}{k})
    +\tilde\Phi(p,z,-\tfrac{2}{k})+\tilde\Phi(p,z,\tfrac{2}{k})\,.
    \label{f phi}
\ee
Here $\tilde\Phi(p,z,a)$
is the function defined for the region $|p|<1$ by
\be
    \tilde\Phi(p,z,a)
    =\int_0^\infty d\b\,\frac{\b^{z-1}}{\G(z)}\,\log(1-p\,e^{-\b})\,e^{-a\,\b}\,,
    \label{tilde phi}
\ee
and for the other region by analytic continuations. 
The $p_1$, $p_2$ and $p_3$ are defined through the factorization\footnote{The argument of the log function which appears in all of the $f_{\text{log,k}|n}$ is $1-{\sinh\left(k\beta\right)\over 8\sinh^4{k\beta\over 2}}=
{1 - 4\,q^{k} + 2\,q^{2k} -q^{3k}\over\left(1-q^k\right)^3}$, where \mt{q=e^{-\beta}}. Factorizing the numerator as in eq.\,\eqref{q eq} enables us to replace $\log\left[1-{\sinh\left(k\beta\right)\over 8\sinh^4{k\beta\over 2}}\right]$ by $\log\left(1-p_1\, q^k \right)+\log\left(1-p_2\, q^k\right)+\log\left(1-p_3\, q^k\right)-3 \log\left(1- q^k\right)$, from which we obtain decompositions like eq.\,\eqref{Log k=2}. See Appendix \ref{sec: AdS5 zeta} for details.},
\be 
    1 - 4\,q + 2\,q^2 -q^3 = \left(1-p_1\,q\right)\left(1-p_2\,q\right)\left(1-p_3\,q\right)\,.
    \label{q eq}
\ee
For ${\rm Re}(z)>1$ and $p\neq1$\,, the integral \eqref{tilde phi} is equivalent to the contour integral,
\be
    \tilde\Phi(p,z,a)
    =\frac{i}{2\,\sin\pi z}\,\oint d\b\,\frac{(-\b)^{z-1}}{\G(z)}\,\log(1-p\,e^{-\b})\,e^{-a\,\b}\,,
    \label{tilde phi ct}
\ee
with the contour depicted in Fig.\,\ref{fig: ct1}.
The above is again well-defined for any value of $z$\,,
hence one can immediately put $z=-5$ and consider the residue at the origin.
This way, we obtain
\be 
    \Psi_k(p,-5)
    =\frac{120}{k^4}\,\frac{p}{1-p}\,.
    \label{ff eq}
\ee
Note that the above result can be applied to $p=p_1, p_2$ and $p_3$ but not to $p=1$\,,
as one can check that \eqref{ff eq} diverges in the latter case.
This means that Prescription 2 does not give a
finite value for $\zeta_{\log,k|2}'(0)$ even after an analytic continuation on $z$\,.
To proceed, we need to properly extract a finite part from $\Psi_k(1,-5)$. 
The shift of the branch point at the origin to $\b=-\e$ corresponds to
the replacement of $\Psi_k(1,-5)$ by
\be 
    \Psi_k(e^{-\e},-5)
    =\frac{120}{k^4}\left(\frac1{\e}-\frac12+\cO(\e)\right).
    \label{f reg}
\ee
By taking only the finite part $-1/2$, the total contribution for \eqref{Log k=2}
is proportional to
\be
    \frac{p_1}{1-p_1}+\frac{p_2}{1-p_2}+\frac{p_3}{1-p_3}-3\left(-\frac12\right).
\ee
Using \eqref{q eq} one can show that the above four terms exactly cancel each others
implying $\zeta_{{\rm log},k|2}'(0)=0$\,.
Let us remark that the regularization \eqref{f reg} is equivalent 
to taking only the $\b^5$ term from the integrand of \eqref{tilde phi} for the residue
and ignoring the presence of the branch cut.
Therefore, Prescription 2 with the regularization \eqref{f reg} 
is equivalent to Prescription 1.
For $\zeta_{{\rm log},k|1}$ and $\zeta_{{\rm log},k|0}$\,,
one can do similar analysis and show
their first derivatives vanish at $z=0$.
See Appendix \ref{sec: AdS5 zeta} for more details.

\medskip

Let us focus on the result obtained in Prescription 1 and 2
and provide some interpretation. 
As all three contributions $\g_{\log,k|2}$,
$\g_{\log,k|1}$ and $\g_{\log,k|0}$
vanish. Hence, we can conclude
that the $k$-th contribution to the vacuum energy
vanishes according to  \eqref{fin G g}:
\be
	\G^{\sst (1)\,\rm ren}_{{\rm log},k}=0\,.
\ee
Because each $k$ contribution vanishes, the total vacuum energy
is
\be
	\G^{\sst (1)\,\rm ren}_{\rm AdS_5}
	=-\G^{\sst (1)\,\rm ren}_{\rm Rac}+\sum_{k=1}^\infty
	\frac{\varphi(k)}{k}\,\G^{\sst (1)\,\rm ren}_{{\rm log},k}
	=-\G^{\sst (1)\,\rm ren}_{\rm Rac}\,.
\ee
Therefore,  we conclude that the expansion of free energy near AdS$_5$ vacuum is
\be
\Gamma^{\sst (1)\,\rm ren}_{\rm AdS_5} = \left(g \, \cL_0 - \frac{1}{90}\right) \log R + 
\mathcal{O}(g^{-1})\,,
\ee
where $\cL_0=S_0/{\rm Vol_{AdS_5}}$ is the classical Lagrangian
evaluated on the $AdS_5$ vacuum solution (see \eqref{S expansion}),
and $g$ is the dimensionless coupling constant defined in \eqref{coup g}. From the AdS/CFT correspondence, the free energy $F_{\rm CFT_4}$ of boundary scalar field theory should be identified to the $\Gamma_{\rm AdS_5} $. The free energy of conformal $SU(N)$ matrix scalar on the $S^4$ has a logarithmic divergence corresponding to the conformal $a$-anomaly \cite{Duff:1977ay,Christensen:1978md},
\be
	F_{\rm CFT_4} = \frac{N^2 - 1}{90}\,\log \L+\cO(\L^0)\,.
\ee
Using the correspondence between IR and UV divergences respectively in $AdS_5$ and $S^4$\,,
we get
\be
	g \, \cL_0 - \frac{1}{90}=\frac{N^2 - 1}{90}\,.
\ee
As in the vector model cases \cite{Giombi:2013fka,Giombi:2014iua,Giombi:2014yra}, 
this formula suggests the  relations,
\be 
	g = N^2\,,\qquad
	\cL_0 = \frac{1}{90}\,.
\ee
It may be worth to note that 
 the second equation 
is compatible with the vector model cases,
assuming that the matter sectors do not contribute
to the Lagrangian value $\cL_0$.
This is a reasonable assumption from ordinary field theory 
point of view 
because only gravity can have a non-trivial background value.

\subsubsection*{Prescription 3}

\begin{figure}[h]
\centering
  \includegraphics[width=0.5\linewidth]{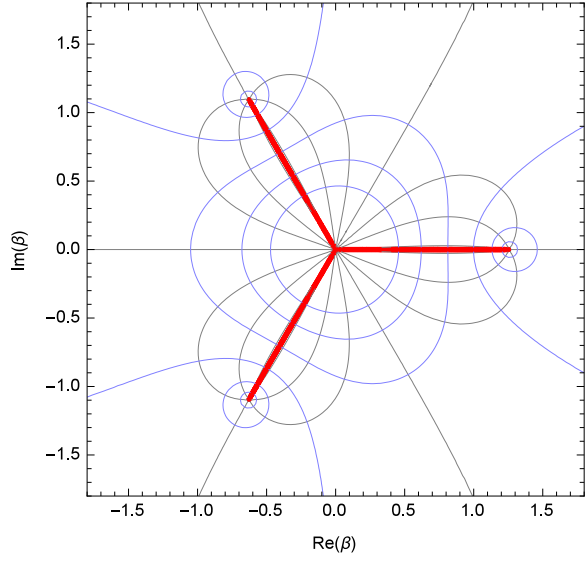}
  \caption{The blue/grey lines are 
  the contour plots for the real/imaginary part of the function \eqref{Log fn}. The red line corresponds to the branch cut.}
  \label{fig: Log}
\end{figure}

Finally in the prescription 3, we have to consider a contour which encircles all the singularities of the integrand. 
In particular, the logarithm function,
\be
	\log\!\left[1-\frac{\sinh (k\b)}{8\,\sinh^4\frac{k\b}{2}}\right]\,,
	\label{Log fn}
\ee
generates the branch cuts which is depicted in Fig.\,\ref{fig: Log}.
With the evaluation \mt{z=0}\,, the branch cut 
of $(-\b)^{2(z-1-n)}$ along the positive real axis of $\b$ disappears,
and the remaining singularities are composed of 
the branch cuts in Fig.\,\ref{fig: Log} and the poles at the end points of the cuts.
\begin{figure}[h]
\centering
\begin{tikzpicture}
\draw [help lines,->] (-3.6,0) -- (3.6,0);
\draw [help lines,->] (0,-3) -- (0,3);
\node at (3.6,-0.4){Re$(\b)$};
\node at (-0.6,3) {Im$(\b)$};
\draw [line width =1.3pt, red] (0,0) to (2.5,0);
\draw [line width =1.3pt, red] (0,0) to (120:2.5);
\draw [line width =1.3pt, red] (0,0) to (-120:2.5);
\draw [red,fill] (0,0) circle [radius=0.05];
\draw [red,fill] (2.5,0) circle [radius=0.05];
\draw [red,fill] (120:2.5) circle [radius=0.05];
\draw [red,fill] (-120:2.5) circle [radius=0.05];
\draw [blue, decoration={ markings,
  mark=at position 0.30 with {\arrow[line width=1.2pt]{>}},
  mark=at position 0.65 with {\arrow[line width=1.2pt]{>}},
  mark=at position 0.97 with {\arrow[line width=1.2pt]{>}}
  },
  postaction={decorate}] plot [smooth cycle] coordinates {(-10:2.5) (0:3) (10:2.5) 
  (60:0.7) 
  (110:2.5) (120:3) (130:2.5)
  (180:0.7) 
  (-130:2.5) (-120:3) (-110:2.5) 
  (-60:0.7)};
\end{tikzpicture}
\caption{Integration contour for $\zeta'_{\log,k|n}(0)$}
\label{fig: ct2}
\end{figure}
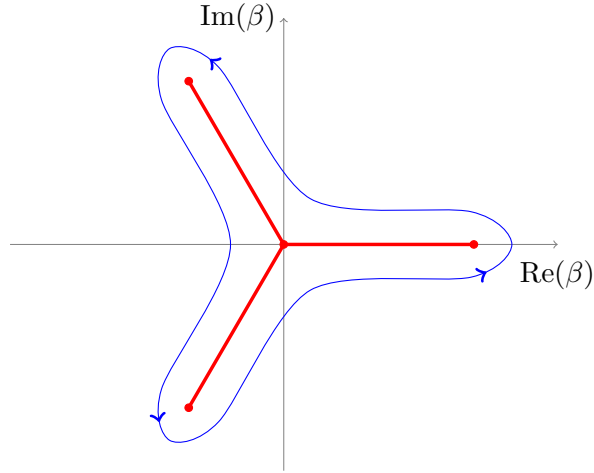

\begin{figure}[h]
\centering
\begin{tikzpicture}
\draw [help lines,->] (-3.6,0) -- (3.6,0);
\draw [help lines,->] (0,-3) -- (0,3);
\node at (3.6,-0.4){Re$(\b)$};
\node at (-0.6,3) {Im$(\b)$};
\draw [line width =1.3pt, red] (1,0) to (2.5,0);
\draw [line width =1.3pt, red] (120:1) to (120:2.5);
\draw [line width =1.3pt, red] (-120:1) to (-120:2.5);
\draw [red,fill] (0,0) circle [radius=0.05];
\draw [red,fill] (2.5,0) circle [radius=0.05];
\draw [red,fill] (1,0) circle [radius=0.05];
\draw [red,fill] (120:2.5) circle [radius=0.05];
\draw [red,fill] (120:1) circle [radius=0.05];
\draw [red,fill] (-120:2.5) circle [radius=0.05];
\draw [red,fill] (-120:1) circle [radius=0.05];
\draw [blue, decoration={ markings,
  mark=at position 1 with {\arrow[line width=1.2pt]{>}}
  },
  postaction={decorate}] plot [smooth cycle, tension =1.4] coordinates {(0:2.8) (13:1.8) 
  (0:0.7) (-13:1.8)};
  \draw [blue, decoration={ markings,
  mark=at position 1 with {\arrow[line width=1.2pt]{>}}
  },
  postaction={decorate}] plot [smooth cycle, tension =1.25] coordinates {(120:2.8) (120+13:1.8) 
  (120:0.7) (120-13:1.8)};
  \draw [blue, decoration={ markings,
  mark=at position 1 with {\arrow[line width=1.2pt]{>}}
  },
  postaction={decorate}] plot [smooth cycle, tension =1.4] coordinates {(-120:2.8) (-120+13:1.8) 
  (-120:0.7) (-120-13:1.8)};
\draw [blue, decoration={ markings,
  mark=at position 1 with {\arrow[line width=1.2pt]{>}},
  },
  postaction={decorate}] plot [smooth cycle, tension =1.6] coordinates 
  {(60:0.4) (180:0.4) (-60:0.4)};
\end{tikzpicture}
\caption{Integration contours for shifted branch points}
\label{fig: ct3}
\end{figure}
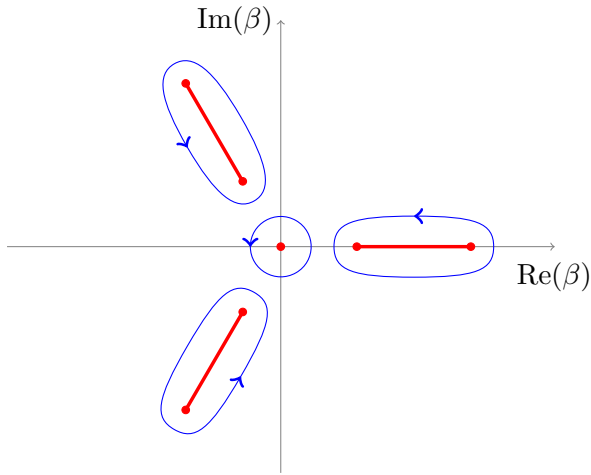
Hence we can consider the contour of Fig.\,\ref{fig: ct2}.
Such integral will give a finite and non-trivial result for a fixed slice $k$\,.
Once the integral is evaluated, we 
can sum them for all $k$ according to 
\eqref{cyc log 5} to get
the full vacuum energy.
However, in this paper we do not evaluate 
them as we could not find 
any analytic measure to do so.
Instead,
in order to see
the relation of Prescrition 3
to the others, let us 
consider a deformation of the branch points at $\b=0$ while keeping all the poles untouched.
The original integral can now be split into four pieces as in Fig.\,\ref{fig: ct3}.
After evaluating the four contour integrals separately, 
we can eventually take the limit that the shifted branch points
tend back to the origin.
Each of four contributions diverges in the latter limit,
while their sum is finite.
In this set-up, Prescription 1 would
correspond to taking the finite part of 
the contour around the origin but discarding all other contributions.


\section{Conclusion}
\label{sec: conclusion}

Let us briefly summarize the objectives and  the results obtained in this paper.
The physics we attempted to explore is the holography of the free $SU(N)$ adjoint
scalar CFT in the $N\to \infty$ limit. We assume this theory to admit an AdS dual, 
an extension of Vasiliev's theory with infinitely many higher spin massive multiplets.
This theory can be considered as a toy model for more realistic and/or
 interesting stringy models.
The field content of the AdS theory 
is to match with the spectrum of single trace operators in the CFT.
This  is a necessary condition for formulating a meaningful AdS/CFT duality 
in all known cases.
The main aspect of this holography that we investigated
is the one-loop correction of the AdS theory
which ought to correspond to the first  $1/N$ correction of the CFT.
Since the CFT is free, the latter correction vanishes implying 
the triviality of the AdS one-loop effect.

The one-loop triviality of AdS theory is a highly non-trivial property
as an infinite number of contributions from each field contents
should sum up to cancel precisely.
In order to test this  property,
we first analyzed the single trace operator content of the 
 CFT by
decomposing tensor product of the singleton representation
which accounts the degrees of freedom of free conformal scalar. 
Explicit decompositions were  carried out for the three and four  tensor products
in the three dimensional CFT,
and in principle the analysis is extendable to other dimensions and higher powers as well. 
We found that the decomposition rules increase in complexity as higher powers of singletons are considered, but in every case explicit closed form expressions can be obtained.

Next, we turn to the simplest one-loop effect, the  vacuum energy of the AdS theory.
Using the explicit results obtained for the spectrum of single trace operators,
we compute the vacuum energies of the AdS fields
in a first few Regge trajectories. However, this method quickly becomes prohibitive both due to increasing complexity of the spectrum and the careful regularization required at various steps in the calculation. 

We therefore developed a new formalism for writing down the zeta function, exploiting the fact that the 
character of the singleton encodes the spectrum of single trace operators.
This formalism greatly simplifies computations of the one-loop vacuum energies. 
In the case of AdS$_4$, it allowed us to calculate the vacuum energies for the first 8 trajectory fields. 
In the case of AdS$_5$, the integral involved becomes even simpler, hence
we could calculate the vacuum energies for the first 32 trajectories.
In both cases, the calculation of vacuum energy for any fixed trajectory
can be done in principle, but requires more computing powers 
for higher trajectories.
An analytic expression of the vacuum energy for an arbitrary level
is nevertheless unavailable. It is partly due to the fact
that the corresponding character involves 
the number theoretic Euler totient function.

In order to avoid the difficulty related to the Euler totient function,
we consider the different summation (or slice) for the full vacuum energy which is valid only in the $N\to\infty$ limit.
At this point, it might be worth to remark that
the integral and series that we are considering
--- $\sum_n \int d\b$ or $\sum_k \int d\b$ ---
is all the time `two-dimensional'. This is to be compared 
with the two-dimensional fundamental domain of torus appearing in the string loop:
\be
	\sum_{\D,\bm\ell}\, N_{\D,\bm\ell}\
	\parbox{63pt}{
	\begin{tikzpicture}
	\draw [semithick, blue] (0,0) circle [radius=1];
	\draw [semithick] (0,0) circle [radius=0.4];
	\node at (0,-0.6) {$\st \D,\bm\ell$};
	\end{tikzpicture}}
	=\sum_{n\,{\rm or}\,k}\int d\b\left(\,\cdots\right)
	=\,
	\parbox{60pt}{
	\begin{tikzpicture}
	\draw [semithick, blue] (0,0) circle [radius=1];
	\draw [line width=11, lightgray] (0,0) circle [radius=0.5];
	\draw [gray] (0,0) circle [radius=0.3];
	\draw [gray] (0,0) circle [radius=0.7];
    \draw[dotted,xscale=2] (-0.15,0)
    arc [radius=0.1, start angle=0, end angle=180];
     \draw[xscale=2] (-0.35,0)
    arc [radius=0.1, start angle=180, end angle=360]; ;
	\end{tikzpicture}}\,.
\ee
With the new slice,
the AdS$_4$ vacuum energy for a given slice is
given by an integral, which could be evaluated only by a numerical method (see Appendix \ref{sec: AdS4 zeta}).
Turning to AdS$_5$ case, 
we note first that the derivation of the vacuum energy
from the character is no more unique for 
AdS duals of matrix model CFTs.
We considered three prescriptions which
become equivalent for a single particle state in AdS$_5$,
but differ from each other for the matrix model case.

In the first prescription,
we find that the vacuum energy exactly vanishes for each slice. 
This result can be equally obtained in the second prescription
upon introducing an addition regularization.
Hence,
the one-loop vacuum energy of the AdS$_5$ theory dual to the free $SU(N)$ adjoint scalar CFT on $S^4$ coincides with minus of the free energy of a boundary scalar.
This is somewhat analogous to the duality between
the minimal Vasiliev theory and the free $O(N)$ vector model,
hence suggests that the loop-expansion parameter in the AdS$_5$ theory should be identified to $N^2$ rather than $N^2-1$\,, the dimension of $SU(N)$.

 In the last prescription, 
 the vacuum energy for a fixed slice
 is given by a contour integral
 surrounding the singularities.
 The latter contains
 not only poles 
 but also branch cuts,
 which are generated by the logarithm
 arising as a result of summing over trajectories.
 These singularities are precisely where
 the Hagedorn phase transition takes place.
 The appearance of such singularities
 and the possibility of their relevance to
 one-loop vacuum energy is interesting 
 but also intriguing,
 because the parameter $\beta$ is not a physical temperature for the CFT in our case.
 Rather it is just a parameter which is being used to count the number of the spectrum of the theory,
 and in this sense one might not expect the Hagedorn transition to play a role in the vacuum energy.
 In this work, for the technical reason, we have not explore sufficiently the contributions coming from
 the Hagedorn-related singularities. 
 We hope to revisit this issue in near feature. 
 One interesting direction in this respect would
 be the computation of the vacuum energy in the thermal AdS$_{5}$ with $S^3\times S^1$ boundary,
 where the Hagedorn transition has a physical meaning.
 We hope to report soon about the latter computation.

Finally, the results contained in this paper can be extended in many directions.
The most interesting one would be to apply these tests to explore the duality of tensionless string theory to free CFTs,
e.g. the duality between the Type IIB string theory on $AdS_5\times S^5$ and the $\mathcal{N}=4$  Super Yang-Mills (SYM) with the gauge group $SU(N)$. The planar ($N\rightarrow \infty$) and free  ($g^2_{YM}N\rightarrow 0$) limit of this theory should correspond in the bulk to the tensionless limit of Type IIB string theory. While this phase of string theory is still fairly poorly understood, one may use the boundary CFT data available to us as a means of getting some insight into the bulk physics \cite{HaggiMani:2000ru,Aharony:2003sx,Sundborg:1999ue,Sundborg:2000wp,wittenHS}. In particular, one may expect to identify the field content of the bulk theory from the operator spectrum of the dual CFT, and carry out the checks we have done in this paper.
Let us also note that in \cite{Basar:2014hda,Cherman:2015rpc}, related issues have been
discussed from the boundary theory point of view: the vacuum energy of large $N$ gauge
theories on $S^3\times S^1$ was shown to vanish with a particular choice of branch.

Additionally, for technical reasons we mainly focused on free field theories whose operator spectrum is completely classified by taking tensor products of the singleton representation. However, our methods allow us in principle to compute the one-loop vacuum energy for the bulk dual of any theory whose planar operator spectrum is known. These include the Chern-Simons with Matter theories of \cite{Aharony:2011jz,Giombi:2011kc} and the interacting theories obtained by flowing to Wilson-Fisher fixed points by turning on double-trace deformations for the free theory, much as in the case of vector models. These, and related questions, are work in progress and we hope to report on them soon.

\acknowledgments

We are grateful to
Dongsu Bak,
Kangsin Choi,
Justin David,
Dongmin Gang,
Rajesh Gopakumar,
Seungho Gwak,
Jaewon Kim,
Jihun Kim,
Ruslan Metsaev,
Karapet Mkrtchyan,
Jeong-hyuck Park,
Soo-Jong Rey,
Evgeny Skvortsov
and
Arkady Tseytlin
for useful discussions.
The work of JB is supported by the National Research Foundation of Korea grant number NRF-2015R1D1A1A01059940. JB thank the organizers of the workshop ``Current Topics in String Theory'' for valuable discussions related to this work. 
The work of EJ was supported in part by the National Research Foundation of Korea through the grant NRF-2014R1A6A3A04056670 and the Russian Science Foundation grant 14-42-00047 associated with Lebedev Institute.
The work of SL is supported by the Marie--Sklodowska Curie Individual Fellowship 2014. SL would also like to thank the Indian Institute of Science, Bangalore for hospitality during the course of this work. 

\appendix

\section{Oscillator Analysis for Tensor Products}
\label{app: oscil}

In order to find out the explicit formulas for $N^{(\e_{1},\ldots,\e_{k})}_{(n_{a},n_{b})}$\,,
let us recast the branching \eqref{branching} in terms of the character:
\be
	\chi_{(n_{a},n_{b})}(\S)
	=\sum_{\e_{1}=\pm1}\cdots \sum_{\e_{p}=\pm1}
	N^{(\e_{1},\ldots,\e_{k})}_{(n_{a},n_{b})}\,
	\chi_{(\e_{1},\ldots,\e_{k})}(\S)\,,
\ee
where $\chi_{(n_{a},n_{b})}$ is the $O(k)$ character in the $(n_{a},n_{b})$ representation.
$\S=\s_1^{\d_{1}}\cdots \s_{k}^{\d_{k}}$ (with $\d_{i}=0,1$)
is an element of $\mathbb Z_{2}^{\otimes k}\subset O(k)$\,,
and $\chi_{(\e_{1},\ldots,\e_{k})}$ is its character
in the  $(\e_{1},\ldots,\e_{k})$ representation:
\be
	\chi_{(\e_{1},\ldots,\e_{k})}(\s_1^{\d_{1}}\cdots \s_{k}^{\d_{k}})
	=\chi_{\e_{1}}(\s_{1}^{\d_{1}})\cdots \chi_{\e_{k}}(\s_{k}^{\d_{k}})\,,
	\qquad
	\chi_{\e}(\s^{\d})=\e^{\d}\,.
\ee
The multiplicities can be obtained using these properties as
\ba
	N^{(\e_{1},\ldots,\e_{k})}_{(n_{a},n_{b})}
	\eq\frac1{\big|\mathbb Z_{2}^{\otimes k}\big|}
	\sum_{\S\in\mathbb Z_{2}^{\otimes k}}
	\chi_{(\e_{1},\ldots,\e_{k})}(\S^{-1})\,\chi_{(n_{a},n_{b})}(\S)\nn
	\eq\frac1{2^{k}}
	\sum_{\d_{1}=0,1}\cdots\sum_{\d_{k}=0,1}
	\e_{1}^{\d_{1}}\cdots \e_{k}^{\d_{k}}\,
	\chi_{(n_{a},n_{b})}(\s_1^{\d_{1}}\cdots \s_{k}^{\d_{k}})\,.
\ea
From the above formula, one can see that the multiplicity (that is the tensor product decomposition) depends on the number of Di's and Rac's but not their order:
\be
	N^{(\e_{1},\ldots,\e_{k})}_{(n_{a},n_{b})}
	=N^{[l,k-l]}_{(n_{a},n_{b})}\,,
	\qquad
	k-2l=\e_{1}+\cdots+\e_{k}\,,
\ee
here $l$ and $k-l$ are respectively the number of Di's and Rac's.
The multiplicity $N^{[l,k-l]}_{(n_{a},n_{b})}$ is simply given by
the $O(k)$ character as
\ba
	N^{[l,k-l]}_{(n_{a},n_{b})}
	\eq\frac1{2^{k}}
	\sum_{\d_{1}=0,1}\cdots\sum_{\d_{k}=0,1}
	(-1)^{\d_{1}+\cdots+\d_{l}}
	\chi_{(n_{a},n_{b})}(\s_1^{\d_{1}}\cdots \s_{k}^{\d_{k}})\nn
	\eq
	\frac1{2^{k}}
	\sum_{j=0}^{k}\,c^{[l,k-l]}_{k}\ \chi_{(n_{a},n_{b})}(\s_{1}\cdots \s_{j})\,,
\ea
where the coefficients $c^{[l,k-l]}$ satisfy
\be
	\sum_{j=0}^{k}\,c^{[l,k-l]}_{j} a^{k-j}\,b^{j}
	=(a-b)^{l}\,(a+b)^{k-l}\,.
	\label{c coef}
\ee
Now it is turn to evaluate the character $\chi_{(n_{a},n_{b})}$
for the element $\s_{1}\cdots \s_{j}$\,.
The $O(k)$ character in $(n_{a},n_{b})$ representation is given by
\be
	\chi_{(n_{a},n_{b})}
	=(h_{n_{a}}-h_{n_{a}-2})(h_{n_{b}}-h_{n_{b}-4})
	-(h_{n_{a}+1}-h_{n_{a}-3})(h_{n_{b}-1}-h_{n_{b}-3})\,,
\ee
where $h_{n}$ is the homogeneous symmetric polynomials of
the eigenvalues of the representation matrix.
For the element $\S_{j}=\s_{1}\cdots\s_{j}$\,, it is given by
\be
	\sum_{n=0}^{\infty}h_{n}(\S_{j})\,z^{n}
	=\frac1{\det(1-z\,\S_{j})}=\frac1{(1+z)^{j}(1-z)^{k-j}}\,.
\ee
Defining the multiplicities generating function
\be
	N^{[l,k-l]}(z,w)=\sum_{n_{a}=-1}^{\infty}\sum_{n_{b}=0}^{\infty}
	N^{[l,k-l]}_{(n_{a},n_{b})}\,z^{n_{a}}\,w^{n_{b}}\,,
	\label{N zw}
\ee
we obtain
\ba
	N^{[l,k-l]}(z,w)=\frac{n(z,w)}{2^{k}}\sum_{j=0}^{k}\frac{c^{[l,k-l]}_{j}}
	{[(1+z)(1+w)]^{j}[(1-z)(1-w)]^{k-j}}\,.
\ea
with
\be
	n(z,w)=(1-z^{2})(1-w^{4})-(z^{-1}-z^{3})
	(w-w^{3})
	=\frac{(1-z^{2})(1-w^{2})(z-w)(1-z\,w)}{z}\,.
\ee
Finally, using \eqref{c coef}, we get a simple form,
\be	
	N^{[l,k-l]}(z,w)=n(z,w)\left(\frac{z+w}{(1-z^{2})(1-w^{2})}\right)^{l}
	\left(\frac{1+z\,w}{(1-z^{2})(1-w^{2})}\right)^{k-l}\,.
\ee
As we shall see below, this generating function 
of multiplicities has a simple relation to the character.

Defining the generating function  $N_{\cH}(q,x)$
of the multiplicities as 
\be
	N_{\cH}(q,x)=\sum_{\Delta,s} N^{\cH}_{\cV(\Delta,s)}\,q^\Delta x^s\,,
\ee
it has a simple relation 
to the one in \eqref{N zw} for $\cH={\rm Di}^{\otimes l}\otimes {\rm Rac}^{\otimes(k-l)}$  as
\be
	N_{\cH}(q,x)=q^{\frac{k}2}\,N^{[l,k-l]}\left(\sqrt{q\,x},\sqrt{q\,x^{-1}}\right).
 \ee

Let us first consider the cyclic tensor product of Di\,$\oplus$\,Rac.
\be
	{\rm Cyc}^{\otimes p}({\rm Di}\oplus{\rm Rac})
	=\bigoplus_{n_{a}=0}^{\infty}\bigoplus_{n_{b}=0}^{n_{a}}
	N^{{\rm cyc}(p)}_{(n_{a},n_{b})}\,
	D\Big(\frac{n_{a}+n_{b}+p}2,\frac{n_{a}-n_{b}}2\Big)\,. 
	\label{DiRac cyc p}
\ee
Again, we consider the branching of $O(p)$ to its cyclic subgroup $\mathbb Z_{p}$\,:
\be
	\p^{\sst O(p)}_{(n_{a},n_{b})}\,\Big|_{\mathbb Z_{p}}=
	N^{{\rm cyc}(p)}_{(n_{a},n_{b})}\,
	\pi^{\sst \mathbb Z_{p}}_{\rm cyc}\oplus\cdots,
\ee
where the multiplicity of the cyclic singlet can be obtained as
\be
	N^{{\rm cyc}(p)}_{(n_{a},n_{b})}=\frac1{p}\sum_{l=0}^{p-1}\,
	\chi_{(n_{a},n_{b})}(C^{l})\,,
\ee
where $C$ is the cyclic permutation.
The character with cyclic group element can be evaluated as
\be
	\sum_{n=0}^{\infty}h_{n}(C^{l})\,z^{n}=
	\frac1{\det(1-z\,C^{l})}
	=\frac1{\left(1-z^{\frac{p}{\gcd(l,p)}}\right)^{\gcd(l,p)}}\,.
\ee
Finally, the multiplicities are generated by
\ba
	N^{{\rm cyc}(p)}(z,w)\eq
	\sum_{n_{a}=-1}^{\infty}\sum_{n_{b}=0}^{\infty}
	N^{{\rm cyc}(p)}_{(n_{a},n_{b})}\,z^{n_{a}}\,w^{n_{b}}\nn
	\eq
	\frac{n(z,w)}{p}\sum_{l=0}^{p-1}
		\frac1{\left[\left(1-z^{\frac{p}{\gcd(l,p)}}\right)
		\left(1-w^{\frac{p}{\gcd(l,p)}}\right)\right]^{\gcd(l,p)}}\,.
\ea

\section{AdS$_4$ Zeta Function}
\label{sec: AdS4 zeta}

\subsection{Modified Zeta Function Regularization}

As we are only interested in the zeta function up to order $z$ in the small $z$ expansion, it is sufficient to check that the $\zeta_{\D,s}(z)$ and $\tilde\zeta_{\D,s}(z)$ agree up to order $z^2$. To carry out this check, it is more convenient to work with $y=\cos\a$ instead of $\a$ itself.
The character of $\cV(\D,s)$ may be rewritten as
\begin{equation}
	\chi_{\D,s}(\beta,y)=
	\frac{e^{-(\D-\frac32)\,\beta}}{4\,\sinh\frac\beta2}\,
	\frac{V_{2s}(y)}{\cosh\beta-y}\,,
\end{equation}
where $V_{n}(y)$ is shorthand for $U_{n}(\sqrt{(y+1)/2})$, and $U_{n}$ is the Chebyshev polynomial of the second kind. We need only first two Taylor coefficients of $V_{n}$.
\begin{equation}
	V_{n}(1)=n+1\,,
	\qquad 
	V'_{n}(1)=\frac{n(n+1)(n+2)}{12}\,.
\end{equation}
With these inputs, $\tilde\zeta_{\D,s}$  reduces to the following expressions:
\be
	\tilde\zeta_{\D,s}(z,0)=
	\frac{2s+1}{6}\,\zeta_{4}(2z,\D-\tfrac32)
	-\frac{s(s+1)(2s+1)}{6}\,\zeta_{2}(2z,\D-\tfrac32)\,
\ee
with
\begin{equation}
	\zeta_{n}(z,a)=
	\frac{(n-1)!}{2^{n}}\int_{0}^{\infty}d\beta\,e^{-a\,\beta}\,
	\frac{\beta^{z-1}}{\Gamma(z)}\,
	\frac{\cosh\frac\beta2}{\sinh^{n}\frac\beta2}\,.
\end{equation}
The explicit evaluation of $\zeta_{n}(z,a)$ for $n=2,4$ yields
\begin{equation}
\begin{split}
	\zeta_{2}(z,a)&=
	\zeta(z-1,a+\tfrac12)-a\,\zeta(z,a+\tfrac12)\\
	\zeta_{4}(z,a)&=
	\zeta(z-3,a+\tfrac32)-3\,a\,\zeta(z-2,a+\tfrac32)\\ &\quad 
	+\left(3\,a^{2}-\tfrac14\right) \zeta(z-1,a+\tfrac32)
	-a
	\left(a^{2}-\tfrac14\right) \zeta(z,a+\tfrac32)\,.
\end{split}
\end{equation}
From this, one can first calculate the constant part with $a=\D-\frac32$,
\begin{equation}
	\tilde \zeta_{a+\frac32,s}(0)=
	\frac{2s+1}{6}
	\left[	
	\frac{a^{4}}{4}-\frac{a^{2}}{8}-\frac{17}{960}
	-
	s(s+1)\left(\frac{a^{2}}2+\frac1{24}\right)\right],
\end{equation}
using the identity $\zeta(-n,x)=-B_{n+1}(x)/(n+1)$, where $B_{n}(x)$ is Bernouilli polynomial. This result agrees with \eqref{zeta}. Then we move to the first derivative part,
\begin{equation}
\begin{split}
	\tilde\zeta_{a+\frac32,s}'(0)=
	\frac{2s+1}{3}&\,\Big[
	\zeta'(-3,a+\tfrac32)-3a\,\zeta'(-2,a+\tfrac32)
	+\left(3a^{2}-\tfrac14\right) \zeta'(-1,a+\tfrac32)\\
	&-a\left(a^{2}-\tfrac14\right) \zeta'(0,a+\tfrac32)-
	s(s+1)\left[
	\zeta'(-1,a+\tfrac12)-a\,\zeta'(0,a+\tfrac12)\right]\Big]\,.
	\label{tzeta'}
\end{split}
\end{equation}
In order to show the coincidence with \eqref{zeta'}, we first compare them for the value $a=0$\,:
\begin{equation}
\begin{split}
	\tilde\zeta'_{\frac32,s}(0)&=\frac{2s+1}3\left[\zeta'(-3,\tfrac12)-\left(s+\tfrac12\right)^{2}\zeta'(-1,\tfrac12)\right],\\
	 \zeta'_{\frac32,s}(0)&=\frac{2s+1}3\left[c_{3}+\left(s+\tfrac12\right)^{2}c_{1}\right].
\end{split}
\end{equation}
One can check that $\zeta'(-3,\tfrac12)=-c_{3}$
and $\zeta'(-1,\tfrac12)=c_{1}$ so they match.
Then, we consider their $a$ derivatives, we get
\begin{equation}
\begin{split}
	\frac{\partial}{\partial a}\tilde\zeta_{a+\frac32,s}'(0)
	&=\frac{2s+1}{3}\,\Big[
	-\zeta(-2,a+\tfrac32)+3\,a\,\zeta(-1,a+\tfrac32)-(3a^{2}-\tfrac14)\,
	\zeta(0,a+\tfrac32)\\
	&\hspace{45pt}-\,a\left(a^{2}-\tfrac14\right)\psi(a+\tfrac32)
	+s(s+1)\,a\left[\psi(a+\tfrac12)+1\right]\Big]
	\\
	&=\frac{2s+1}{3}\,\Big[
	\tfrac{11}6\,a^{3}-\tfrac{5}{24}\,a+s(s+1)\,a+
		\left[\left(s+\tfrac12\right)^{2}-a^{2}\right]a\,\psi(a+\tfrac12)\Big]\,,
		\label{TZeta'}
\end{split}
\end{equation}
and 
\begin{equation}
\begin{split}
	\frac{\partial}{\partial a}\zeta'_{a+\frac32,s}(0) =
	\frac{2s+1}3\,
\Big[\tfrac{1}{2}\,a^3 + \tfrac{1}{24}\,a+
 \left[ \left(s+\tfrac{1}{2} \right)^2 - a^2
\right]a\,\psi(a+\tfrac{1}{2})\Big]\,,
\label{Zeta'}
\end{split}
\end{equation}
where we have used the following identities of Hurwitz zeta function,
\begin{equation}
\frac{\partial}{\partial a}\zeta(s,a)=-s\,\zeta(s+1,a)\,,
\qquad
\frac{\partial}{\partial a}\zeta'(0,a)=\psi(a)\,.
\end{equation}
Finally, comparing \eqref{TZeta'} with \eqref{Zeta'}, we 
get a non-trivial difference between two zeta functions:
\begin{equation}
    \tilde\zeta_{a+\frac32,s}(z)-\zeta_{a+\frac32,s}(z)
	=\tfrac13\left(s+\tfrac12\right)a^{2}
	\left[\tfrac16\,{a^{2}}-\left(s+\tfrac12\right)^{2}\right]
	z+\mathcal{O}(z^{2})\,.
	\label{diff zeta}
\end{equation}
The above difference can be also expressed from the character as
\ba
	&&\D\G^{\sst (1)\,\rm ren}_{\cH}=
	\tilde\G^{\sst (1)\,\rm ren}_{\cH}-\G^{\sst (1)\,\rm ren}_{\cH}\nn
	&&=\,
	\frac{1}{2\,\pi\,i}\oint d\b\,\frac{2\,\sinh^3\frac\b2}{\b^3}\left(
	\frac{8}{3\,\b^2}+\frac{2}{\sinh^2\frac\b2}-\frac13
	+4\,\partial_\a^2\right)\chi_{\cH}(\b,\a)\,\bigg|_{\a=0}\,,
	\label{diff chi}
\ea
where the contour encircles the $\b=0$ point.
Hence, one can simply adjust the 
formula \eqref{AdS4 zeta} by \eqref{diff chi}.
One can notice that the difference term vanishes
when the character $\chi_{\cH}$ is even function of $\b$\,.

\subsection{Numerical Approach to Small $\b$ Cut-Off Regularization}

\begin{figure}[h]
\centering
  \includegraphics[width=0.8\linewidth]{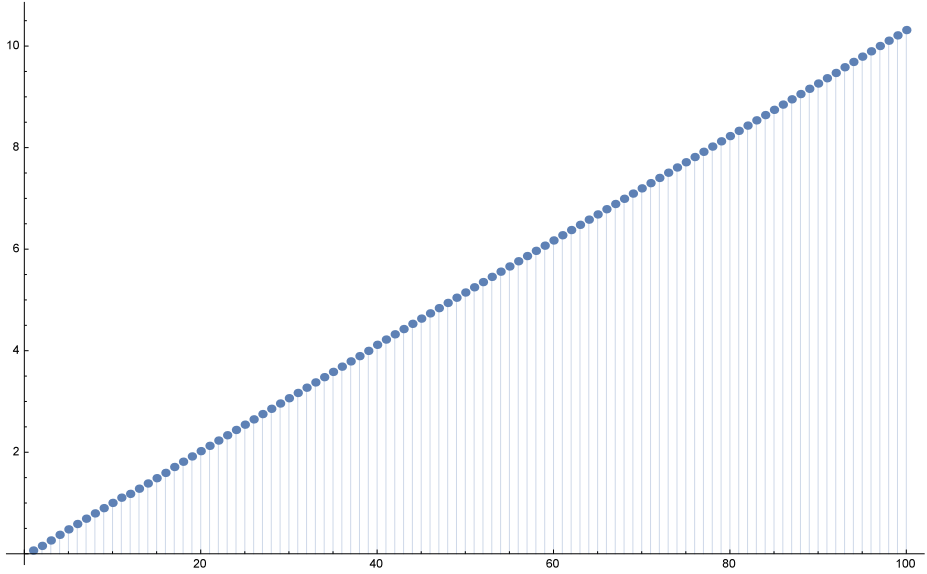}
  \caption{Vacuum energy of AdS${}_4$ theory in different slices}
  \label{fig2}
\end{figure}

In Section \ref{sec: AdS4}, we noted the difficulty of the analytic computation 
of $\tilde\zeta_{\sst \log,k}(z)$\,.
We can proceed numerically by change the regularization scheme 
from zeta function one to short $\b$ cut-off one:
\be 
	\int_0^\infty d\b\,\b^{2z}\,f(\b) \quad \to \quad \int_\e^\infty d\b\,f(\b)\,,
\ee
where we simply remove all the divergence arising in the limit $\e\to 0$\,.
The absence of $\tilde \zeta_{{\rm log},k}(0)$ can be verified in the new scheme
from the absence of $1/\b$ term of the integrand $f(\b)$\,. 
Thanks to this, the splitting of $f(\b)$ into the regular  and singular parts  is unambiguous. Truncating the integrand only to the regular part,
the $\e\to0$ limit of the integral become well-defined and can be evaluated numerically. 
The plot in Fig.\,\ref{fig2}
shows the numerical values of the $\G^{\sst (1)\,\rm ren}_{{\rm log},k}$ 
contributions 
for $k=1,2,\ldots,100$.
It shows an approximate linear growth
with a very small perturbations:
\be
	\G^{\sst (1)\,\rm ren}_{{\rm log},k}\simeq
	-0.049+0.108\,k\,,
	\label{lin}
\ee
which a linear fit of our numerical data set. 
The total vacuum energy is given by
\be
	\G^{\sst (1)\,\rm ren}_{\rm AdS_4}
	=-\G^{\sst (1)\,\rm ren}_{\rm Rac}+\sum_{k=1}^\infty
	\frac{\varphi(k)}{k}\,\G^{\sst (1)\,\rm ren}_{{\rm log},k}\,.
\ee
Since each of $\G^{\sst (1)\,\rm ren}_{{\rm log},k}$ is positive and growing almost linearly,
the total vacuum energy 
is again given by a divergent series, to which we do not have an analytic access.
We may nevertheless proceed the summation based on the numerical fitting \eqref{lin}
and the regularization of the Euler totient sum,
$\sum_{k=1}^\infty \varphi(k)\,k^{-z}=\zeta(z-1)/\zeta(z)$ but the result\footnote{If we plug in the approximation \eqref{lin} into the above, we are lead to
\be
	\sum_{k=1}^\infty
	\frac{\varphi(k)}{k}\left(-0.049+0.108\,k\right)
	=-0.049\,\frac{\zeta(0)}{\zeta(1)}+0.108\,\frac{\zeta(-1)}{\zeta(0)}
	\simeq 0.018\,.
\ee
Combining with the contribution $-\G^{\sst (1)\,\rm ren}_{\rm Rac}$\,,
we get the following value,
\be 
		\G^{\sst (1)\,\rm ren}_{\rm AdS_4}
	\simeq 0.018-0.064=-0.046\,.
\ee
}
obtained in this way would be hardly reliable because of the unjustifiable 
combination of numerical approximations
and analytic continuations.

\section{AdS$_5$ Zeta Function}
\label{sec: AdS5 zeta}

The zeta functions are given by $\beta$-integrals
whose $z$-dependence regularizes
the pole of the integrand at $\b=0$\,.
Upon analytic continuations in $z$, the zeta functions
give the vacuum energies by their first derivatives at $z=0$\,.
The typical way to evaluate these $\beta$-integrals
is by partial fraction decomposition of the integrand in $q = e^{-\beta}$
and by using the integral representation of (derivatives of) Lerch zeta function,
\be
 \int_0^\infty d\b\, \frac{\b^{z-1} \, q^{n+a}}{(1-p \, q)^{n+1}} = \frac{\G(z)}{n!} \,\partial_p^n\,
    \Phi(p,z,a)\,.
    \label{Phi rep}
\ee
and the identity,
\be \label{Phi deriv}
    \Phi(p,z-1,a)=\left(a+p\,\partial_p\right) \Phi(p,z,a)\,.
\ee
We will illustrate this method very soon for the case of the Vasiliev theory in AdS$_5$. Also eventually for the matrix model we will need the function $\tilde\Phi$, defined in \eqref{tilde phi}, given by
\be
\tilde\Phi\left(p,z,a\right)=\int_0^\infty d\beta\,{\beta^{z-1}\over\Gamma\left(z\right)}\log\left(1-p\,e^{-\beta}\right)e^{-a\beta},
\ee
which is related to the Lerch zeta function via
\begin{equation}
    \partial_p\tilde{\Phi}\left(p,z,a\right) = -\Phi\left(p,z,a+1\right).
\end{equation}
i.e. it is the $p$-primitive of Lerch function, up to shifts in the argument $a$. It is also useful to note that the function $\tilde{\Phi}$ has a series representation, given by
\begin{equation}\label{tildephiseries}
    \tilde\Phi\left(p,z,a\right)= -\sum_{m=1}^\infty {p^m\over m\left(m+a\right)^z}.
\end{equation}

\subsection{Zeta Function of the Vector Model}

In this section we will present the detailed computations for the vacuum energy for the non--minimal and minimal Vasiliev theories in AdS$_5$, outlined respectively in \eqref{ads5ractest} and \eqref{min Vas 5}. 
Let us consider first the zeta function for the non-minimal theory given by the first line of \eqref{ads5ractest}.
We first rewrite the integrand as a rational function of $q=e^{-\beta}$\,,
then do the partial fraction decomposition:
\ba
	&& 
	\frac{\G(z)\,\zeta_{\textrm{non-min}}(z)}
	{\log R}
	=\, \lim_{\e\to0} \int_0^\infty d\b\,e^{-\e\,\b}\left[ \frac{(1+q)^2 \, 2^{1-2 z} \, \beta ^{2 z-6}}{(1-q)^2 \, \G(z-2)} + \frac{(1+q)^2 \, (1-10\,q +q^2) \beta ^{2 z-4}}{4^z\times 3 \, (1-q)^4 \, \G(z-1)} \right] \nn
	&&=\, \lim_{\e\to0} 
\frac{2^{1-2 z} }{\G(z-2)} 
 \int_0^\infty d\b\,\beta ^{2 z-6}
 \left[q^{\e}-4\,\frac{q^\e}{1-q}+4\,\frac{q^{\e}}{(1-q)^2}\right] + \nn 
 &&\lim_{\e\to0} \frac{4^{-z}}{3\, \G(z-1)}
  \int_0^\infty d\b\,\beta ^{2 z-4}
  \left[q^{\e}+4\,\frac{q^{\e}}{1-q}-36\,\frac{q^{\e}}{(1-q)^2}
 +64\,\frac{q^{\e}}{(1-q)^3}-32\,\frac{q^{\e}}{(1-q)^4}\right].
\ea
Thus, \textit{via} the partial fraction decomposition above, the $\beta$ integrals have reduced integral representations for the Lerch zeta functions and their $p$ derivatives, as in \eqref{Phi rep}, evaluated at $p=1$. 
We next use \eqref{Phi deriv}, and the identity $\Phi(1,z,a) = \zeta(z,a)$ to show that
\ba
	&& 
	\frac{\G(z)\,\zeta_{\textrm{non-min}}(z)}
	{\log R}
	=\, \frac{2^{3-2 z} \,\G(2z-5) }{\G(z-2)} \bigg[ - \,  \zeta(2z-5,0) +\, \zeta(2z-6, -1) + \, \zeta(2z-5,-1) \bigg] \nn
 &&\quad +\,\frac{4^{1-z} \,\G(2z-3)}{3\, \Gamma (z-1)} \bigg[ -\frac{4}{3} \, \zeta(2 z-6,-2) + 4 \, \zeta(2 z-5,-2)+\frac{64}{3} \, \zeta(2 z-4,-2) \nn
 && \qquad -9 \, \zeta(2 z-4,-1)+16 \, \zeta(2 z-3,-2)-9 \, \zeta (2 z-3,-1)+\zeta(2 z-3,0)    \bigg] \nn
 &&= \frac{\G(z-\frac{5}{2}) \, \zeta(2z-6)}{8 \sqrt{\pi}} - \frac{\G(z-\frac{3}{2}) \, \zeta(2z-6)}{9 \sqrt{\pi}} + \frac{\G(z-\frac{3}{2})  \zeta(2z-4)}{36 \sqrt{\pi}},
\ea
which is nothing but the last line of \eqref{ads5ractest}.

We now turn to the case of the minimal Vasiliev theory, in which case we need to evaluate the second line of \eqref{min Vas 5}.
Following the method described above, we get
\ba
	&& \frac{\G(z)\,\zeta_{\textrm{R}}(z)}{\log R} 
	\nn 
	&&
	=\,\lim_{\e\to0} \int_0^\infty d\b\,e^{-\e\,\b}\,\bigg[ \frac{(1-q) \left(1+q^2\right)  \beta ^{2 z-6}}{4^{z}(1+q)^3 \, \G (z-2)} 
	+ \frac{(1-q) (1-10\,q+q^2) \left(1+q^2\right)  \beta^{2 z-4}}
	{2^{2 z+1}3 (1+q)^5 \, \G (z-1)} \bigg] \nn
    &&=\, \lim_{\e\to0} \frac{4^{-z} }{\G(z-2)}  \int_0^\infty d\b\,\beta ^{2 z-6} \left[-q^{\e}+\frac{4\, q^\e}{1+q}-\frac{6\, q^{\e}}{(1+q)^2}+\frac{4 \,q^{\e}}{(1+q)^3}\right] \nn 
    &&\quad + \, \frac{2^{-1-2z}}{3\, \G(z-1)} \int_0^\infty d\b\,\beta ^{2 z-4} \left[-q^{\e}+\frac{16 \, q^{\e}}{1+q}- \frac{66 \, q^{\e}}{(1+q)^2}+ \frac{124 \, q^{\e}}{(1+q)^3} - \frac{120 \, q^{\e}}{(1-q)^4}+ \frac{48 \, q^{\e}}{(1-q)^5}\right] \nn
	&&=-\frac{\Gamma(2z-5) \left[ (2^{2z} -256)  \, \zeta(2z-7) + (2^{2z} - 64) \, \zeta(2z-5)\right]}{2^{4z-1} \Gamma(z-2)}\nn 
	&&\quad -\,\frac{\Gamma(2z-3)(2^{2z} -256)  \, \zeta(2z-7)}{2^{4z}\times 3\,\Gamma(z-1)}\nn
	&&\quad -\,\frac{2^{-4z} \, \Gamma(2z-3) \left[ 3\,(2^{1+2z}-128)\, \zeta(2z-5) + (2^{2z}-16) \, \zeta(2z-3)\right]}{3\,\Gamma(z-1)}.
\ea
Again, using the the partial fraction expansion for the integrand, we could express all the $\beta$ integrals in terms of the Lerch function and its derivatives, and eventually the Hurwitz zeta function. On further simplification, we obtained the third line of \eqref{min Vas 5}.

\subsection{Zeta Function for the Matrix Model}

For the evaluation of the zeta function of the AdS theory dual to the free $SU(N)$ adjoint scalar CFT, 
we need to perform the integrals \eqref{zeta f} with $f_{\text{log},k\vert n}$ as defined in \eqref{AdS5 log} to arrive at the quantities $\zeta_{\text{log},k\vert n}$. We will show that these contributions to the zeta function vanish up to linear order in the small $z$ expansion.
Let us first focus on the integration of the $f_{\textrm{log},k\vert 2}$. The corresponding integral expression is given by
\ba
\frac{\G(z)\,\zeta_{\textrm{log},k|2}(z)}{\log R} \eq  \int_0^\infty d\beta \, \frac{\Big(\frac{\b}{2} \Big)^{2z-6}}{\G(z-2)} \frac{\sinh ^4\frac{\b}{2}}{2}\, 
	\log\!\left[1-\frac{\sinh (k\b)}{8\,\sinh^4\frac{k\b}{2}}\right] \nn
	\eq  \int_0^\infty d\beta \frac{2^{1-2 z} \, \beta ^{2 z-6}}{ \Gamma (z-2)}\,
	\frac{(1-q)^4 }{q^2}\,
	\log \left[\frac{1-4q^{k}+2q^{2k}-q^{3k}}{\left(1-q^k\right)^3}\right].
\label{zetaamat}
\ea
Using eq.\,\eqref{q eq} we may write the log function as 
\begin{equation}
\begin{split}
      \log \left[\frac{1-4q^{k}+2q^{2k}-q^{3k}}{\left(1-q^k\right)^3}\right]
      =\log\left[{\left(1-p_1\,q^k\right)\left(1-p_2\,q^k\right)\left(1-p_3\,q^k\right)\over\left(1-q^k\right)^3}\right],
\end{split}
\end{equation}
and express the zeta function as
\begin{equation}
    \begin{split}
    &\frac{\G(z)\,\zeta_{\textrm{log},k|2}(z)}{\log R} = \int_0^\infty d\beta\, \frac{2^{1-2 z} \, \beta ^{2 z-6}}{ \Gamma (z-2)}\,
	\frac{(1-q)^4 }{q^2} \log\left[{\left(1-p_1\,q^k\right)\left(1-p_2\,q^k\right)\left(1-p_3\,q^k\right)\over\left(1-q^k\right)^3}\right]\\
	&=\int_0^\infty d\beta \,\frac{2^{1-2 z}\, \beta ^{2 z-6}}{ \Gamma (z-2)}\left(q^{-2}-4q^{-1}+6-4q+q^2\right) \log\left[{\left(1-p_1\,q^k\right)\left(1-p_2\,q^k\right)\left(1-p_3\,q^k\right)\over\left(1-q^k\right)^3}\right].
    \end{split}
\end{equation}
On defining $\tilde{\beta}=k\beta$, and $\tilde{q}=e^{-\tilde{\beta}}$, we may rewrite the above expression as 
\begin{equation}
    \begin{split}
    \frac{\G(z)\,\zeta_{\textrm{log},k|2}(z)}{\log R} ={\left(2k\right)^{5-2z} \over 16\,\Gamma\!\left(z-2\right)}&\int_0^\infty d\tilde{\beta}\, \tilde{\beta} ^{2 z-6}\left(\tilde{q}^{-2/k}-4\tilde{q}^{-1/k}+6-4\tilde{q}^{1/k}+\tilde{q}^{2/k}\right)\times\\ &\times \log\left[{\left(1-p_1\,\tilde{q}\right)
    \left(1-p_2\,\tilde{q}\right)\left(1-p_3\,\tilde{q}\right)\over\left(1-\tilde{q}\right)^3}\right].
    \end{split}
\end{equation}
The above 
expression can be recast into \eqref{Log k=2}
using  \eqref{f phi} and \eqref{tilde phi},
and evaluated using the
contour integral \eqref{tilde phi ct}.

 While it is possible to evaluate the remaining contributions to the zeta function from $f_{\log,k|1}$
 and $f_{\log,k|0}$  in this manner, we shall provide an additional means to do so. Consider the following series expansion of the logarithmic function,
\be
\log \left[\frac{1-4q^{k}+2q^{2k}-q^{3k}}{\left(1-q^k\right)^3}\right] = -\sum_{\ell=1}^{\infty} \frac{1}{\ell}  \left(p_1^{\ell}+p_2^{\ell}+p_3^{\ell}-
3\right) q^{k \, \ell}\,, \label{log series}
\ee
where $p_i$'s are defined in \eqref{q eq}. 
By plugging this series expression into the integral \eqref{zetaamat}
and interchanging the order of $\ell$ summation 
and $\beta$ integration, 
the latter can be evaluated to give
\ba
&& \frac{\G(z)\, \zeta_{\textrm{log},k|2}(z)}{\log R} = 
-\frac{(2k)^{5-2z} \Gamma (2 z-5)}{16\,\Gamma (z-2)} \sum_{\ell=1}^{\infty} \frac{1}{\ell}
\left(p_1^{\ell}+p_2^{\ell}+p_3^{\ell}-3\right)\times\nn 
&&\times\,
\left[6 \, {\ell}^{5-2z} -4\left(\ell-\tfrac{1}{k}\right)^{\!5-2 z}-4 \left(\ell+\tfrac{1}{k}\right)^{\!5-2 z}  +\left(\ell-\tfrac{2}{k}\right)^{\!5-2 z}+\left(\ell+\tfrac{2}{k}\right)^{\!5-2 z}\right].
\label{mat k=2}
\ea
We further simplify the expression to \eqref{Log k=2} by
using the series form of $\tilde{\Phi}\left(p,z,a\right)$ given in \eqref{tildephiseries} to write
\begin{equation}
\begin{split}
    &\sum_{\ell=1}^{\infty} \frac{1}{\ell}p_1^{\ell}
\left[6{\ell}^{5-2z} -4\left(\ell-\tfrac{1}{k}\right)^{5-2 z}-4 \left(\ell+\tfrac{1}{k}\right)^{5-2 z}  +\left(\ell-\tfrac{2}{k}\right)^{5-2 z}+\left(\ell+\tfrac{2}{k}\right)^{5-2 z}\right]\\
    &=6\,\tilde\Phi\!\left(p_1,2z-5,0\right)
    -4\,\tilde\Phi\!\left(p_1,2z-5,\tfrac{1}{k}\right)
    -4\,\tilde\Phi\!\left(p_1,2z-5,-\tfrac{1}{k}\right)
    +\tilde\Phi\!\left(p_1,2z-5, \tfrac{2}{k}\right)\\&\qquad+\tilde\Phi\left(p_1,2z-5,-\tfrac{2}{k}\right)=\Psi_k\!\left(p_1,z\right),\label{series f}
\end{split}
\end{equation}
where $\Psi_k$ was defined in \eqref{f phi}. Hence we 
recover \eqref{Log k=2}. Note however that the series \eqref{series f} is not convergent
since one of $|p_i|$ is greater than 1. Nevertheless, once the series  is evaluated in its domain of convergence, it may be extended to $|p_i|>1$ as well through its representation in terms of $\tilde{\Phi}$.
The small $z$ expansion of this expression has already been carried out in the main text and it was shown that the expression vanishes up to linear order in $z$. That is,
\begin{equation}
    \zeta_{{\rm log},k|2}'(0)=0.
\end{equation}
In fact, there exists a shorter way to draw the same conclusion relying on an
ad-hoc regularization prescription. Since we are interested in the zeta function up to linear order in $z$, it is easy to see that
\begin{equation}
\begin{split}
    \frac{\zeta_{\textrm{log},k|2}(z)}{\log R} &= 
    -z\,{k^5\over 60} \left[\Psi\left(p_1,0\right)+\Psi\left(p_2,0\right)+\Psi\left(p_3,0\right)-3\Psi\left(1,0\right)\right]+\mathcal{O}\left(z^2\right)\\
    &= -z\,{k^5\over 60}\,\sum_{\ell=1}^\infty \left(p_1^{\ell}+p_2^{\ell}+p_3^{\ell}-3\right)+\mathcal{O}\left(z^2\right)\\& =  -z\,{k^5\over 60}\left(\sum_{i=1}^3 {p_i\over 1-p_i} -{3\over 2}\right) +\mathcal{O}\left(z^2\right) = \mathcal{O}\left(z^2\right). 
\end{split}
\end{equation}
In the above, the $z$ independent part in the summand gives a divergent series for some roots $p_i$ and for the $-3$ term. We have used the  
regularization prescriptions, 
\be
  \sum_{\ell=1}^\infty p_i^\ell\to \frac{p_i}{1-p_i}\,,
  \qquad
  \sum_{\ell=1}^\infty 1 \to \zeta\left(0\right)=-\frac12\,,
  \label{presc}
\ee 
to evaluate those sums, and subsequently also used the identity \eqref{q eq} to show that the order $z$ term vanishes, thus obtaining the same result.

For the sake of simplicity,
we shall use the above prescription 
to compute
$\zeta_{{\rm log},k|1}'(0)$
and $\zeta_{{\rm log},k|0}'(0)$
in the rest of this section.
We begin with the evaluation of $\zeta_{\textrm{log},k|1}$:
\be
\frac{\G(z)\, \zeta_{\textrm{log},k|1}(z)}{\log R}
= I_1(z)+I_2(z)\,,
\ee 
with 
\ba
    I_1(z) \eq  \int_0^\infty d\beta\,  \frac{\Big(\frac{\b}{2} \Big)^{2z-4}}{\G(z-1)}
\,\frac{ \sinh ^2\frac{\b}{2}\left(\sinh ^2\frac{\b}{2}-3\right)}{6}\,
	\log\!\left(1-\frac{\sinh (k\b)}{8\,\sinh^4\frac{k\b}{2}}\right),\\ 
	I_2(z) \eq  \int_0^\infty d\beta\,  \frac{\Big(\frac{\b}{2} \Big)^{2z-4}}{\G(z-1)}\, 
 \frac{k^2 \sinh ^4\frac{\b}{2}\, 
	\coth\frac{k\b}2}{\sinh (k\b) -8\,\sinh^4\frac{k\b}2}\,.
\ea
Rewriting the integrands in terms of $q$, they become
\ba
I_1(z)  \eq  \int_0^\infty d\beta\, \frac{ 2^{-2 z-1} \, \beta ^{2 z-4} }{3  \, \Gamma (z-1)}
\,\frac{(1-q)^2\,(1-14q+q^2) }{q^2}
\, \log \left(\frac{1-4q^k+2q^{2k}-q^{3k}}{\left(1 - q^k\right)^3}\right),
\label{I1}\\
I_2(z)\eq -  \int_0^\infty d\beta\,
\frac{k^2\,2^{1-2 z}  \, \beta ^{2 z-4}}{ \Gamma (z-1)}\,
\frac{(1-q)^4 \,q^{2 k-2} \, \left(1 + q^k\right) }{\left(1-q^k\right)^2 \left( 1-4q^k+2q^{2k}-q^{3k} \right)}\,. \label{I2}
\ea
Next, applying the series expansion \eqref{log series}, $I_1$ may be expressed as
\ba
&& I_1(z) = -\frac{(2k)^{3-2z}  \, \Gamma (2 z-3) \, }{48\, \Gamma (z-1)} \sum_{\ell=1}^{\infty} \frac{p_1^\ell+p_2^\ell+p_3^\ell-3 }{\ell}\times \nn 
&& \qquad \times \left[
30\,\ell^{3-2z}-
16\left(\ell-\tfrac{1}{k}\right)^{3-2 z}
-16\left(\ell+\tfrac{1}{k}\right)^{3-2 z} +\left(\ell-\tfrac{2}{k}\right)^{3-2 z}
+\left(\ell+\tfrac{2}{k}\right)^{3-2 z}\right]\nn
\eq  -\frac{(2k)^{3-2z}  \, \Gamma (2 z-3) \, }{48\, \Gamma (z-1)} \big[ \Psi^{(1)}(p_1, z) + \Psi^{(1)}(p_2, z) + \Psi^{(1)}(p_3, z)- 3 \Psi^{(1)}(1, z)\big]\,,
\ea
where the function $\Psi^{(1)}(p, z)$ is defined by
\ba
\Psi^{(1)}(p, z) \eq -30 \, \tilde{\Phi}(p_1,2z-3,0) - 16\, \tilde{\Phi}(p_1,2z-3,-\tfrac{1}{k}) - 16\, \tilde{\Phi}(p_1,2z-3,\tfrac{1}{k}) \nn 
&& \, +\, \tilde{\Phi}(p_1,2z-3,-\tfrac{2}{k})  + \tilde{\Phi}(p_1,2z-3,\tfrac{2}{k})\,.
\ea
The leading behavior of $I_1(z)$ is given by
\ba
I_1(z) \eq \frac{k^3}{36} \big[ \Psi^{(1)}(p_1, 0) + \Psi^{(1)}(p_2, 0) + \Psi^{(1)}(p_3, 0)- 3 \Psi^{(1)}(1, 0)\big]  \nn
\eq \sum_{\ell=1}^{\infty} k \left[ p_1^\ell+p_2^\ell+p_3^\ell-3\right] + \mathcal{O}(z).
\ea
Therefore, following the prescription \eqref{presc},
we obtain 
\ba
	I_1(0)= k \left( \frac{p_1 }{1-p_1}+\frac{p_2 }{1-p_2}+\frac{p_3 }{1-p_3}+\frac 32 \right)=0\,.
\ea
For the integrand of $I_2(z)$, we have the series expansion, 
\ba
\frac{q^{2k}\, (1+q^k)}{(1-q^k)^2 \, (1-4q^k+2q^{2k}-q^{3k})} \eq \sum_{\ell=0}^{\infty} \left( f_1 \, p_1^\ell+ f_3 \, p_2^\ell+ f_2 \, p_3^\ell-\ell \right) q^{k \ell}\,, \label{I2 expansion}
\ea
where
$p_i$'s are defined in the  \eqref{q eq} and $f_i$'s are three roots of the cubic equation $107 f^3 + 3 f - 2 = 0$.
Applying this expansion, \eqref{I2} becomes
\ba
I_2(z) \eq -\frac{ (2k)^{5-2z}\, \Gamma (2 z-3)}{16\,\Gamma (z-1)} \sum_{\ell=0}^{\infty} 
\left(f_1\,p_1^\ell+f_2\, p_2^\ell+f_3\, p_3^\ell+\ell \right) 
\times \nn 
&&\times 
\left(6 \, \ell^{3-2z} -4\, (\ell-\tfrac{1}{k})^{3-2 z}-4
\,(\ell+\tfrac{1}{k})^{3-2 z} 
+(\ell- \tfrac{2}{k})^{3-2 z} +(\ell+\tfrac{2}{k})^{3-2 z}\right)\nn 
\eq  -\frac{ (2k)^{5-2z}\, \Gamma (2 z-3)}{16\,\Gamma (z-1)} \bigg[  \Xi^{(1)}(p_1, 0) + \Xi^{(1)}(p_2, 0) + \Xi^{(1)}(p_3, 0) + 6\,\zeta(2z-4) \nn 
&& \qquad \qquad \qquad \qquad \qquad -\,3 \, \Phi^{+}(1,2z-4,\tfrac{1}{k}) + \tfrac{3}{k} \,  \Phi^{-}(1,2z-3,\tfrac{1}{k}) \bigg]\,,
\ea
where $\Phi^\pm$ is defined by
\be 
\Phi^{\pm}(p,z,a) = \Phi(p,z,a) \pm \Phi(p,z,-a)\,,
\ee 
and $\Xi^{(1)}$ by
\be
\Xi^{(1)}(p,z) = 6 \, \Phi(p,2z-3,0)  -4 \, \Phi^{+}(p,2z-3,\tfrac{1}{k})+ \Phi^{+}(p,2z-3,\tfrac{2}{k})\,.
\ee
Using an analogous prescription of \eqref{presc}, we get
$I_2(0)=0$\, at leading order of $z$.
Hence, we conclude that $\zeta'_{\textrm{log},k|1}(0)=0$\,. 

Finally, we consider the integral of $f_{\textrm{log},k|0}$
given by
\be 
    \frac{\G(z)\, \zeta_{\textrm{log},k|0}(z)}{\log R} 
    =J_1(z)+J_2(z)+J_3(z)\,,
\ee 
with
\ba
    J_1(z)  \eq  -\frac23\int_0^\infty d\beta \, \frac{\Big(\frac{\b}{2} \Big)^{2z-2}}{\G(z)}
	\,k^2\left[(k^2-1)\,\sinh ^2\tfrac{\b}{2}+3\right]	
	 \frac{\sinh ^2\tfrac{\b}{2}\,\coth\frac{k\b}{2}}{ 
	\sinh (k\b) -8\,\sinh^4\frac{k\b}2}\,,
	\label{J1} \\
    J_2(z)  \eq  -2\int_0^\infty d\beta\,  \frac{\Big(\frac{\b}{2} \Big)^{2z-2}}{\G(z)}
    \,k^4
	\left(
	 \frac{\sinh ^2\tfrac{\b}{2}\,\coth\frac{k\b}{2}}{ 
	\sinh (k\b) -8\,\sinh^4\frac{k\b}2}\right)^2\,,
	\label{J2}\\
	J_3(z)  \eq  \int_0^\infty d\beta \, \frac{\Big(\frac{\b}{2} \Big)^{2z-2}}{\G(z)}
	\log\!\left[1-\frac{\sinh (k\b)}{8\,\sinh^4\frac{k\b}{2}}\right]. \label{J3}
\ea
The integrands of the above can be expressed 
in terms of $q$ as
\ba
J_1(z)  \eq \int_0^\infty d\beta\,
\frac{k^2 \,4^{-z} \,
\beta ^{2 z-2}}{ \Gamma (z)}
\frac{(1-q)^2  \left[(k^2-1)\,q^2 - (2k^2-14)\,q+k^2-1\right]
q^{2 k-2}
\left(1+q^k\right)}{3 \left(1-q^k\right)^2 \left(1-4q^k+2q^{2k}-q^{3k}\right)},\nn
J_2(z) \eq  -\int_0^\infty d\beta\,
\frac{k^4 \, 2^{1-2 z} \,\beta ^{2 z-2}}{\Gamma (z)}\,
\frac{(1-q)^4 \, q^{4 k-2} \left(1+q^k\right)^2 
 }{\left(1-q^k\right)^4 \left(1-4q^k+2q^{2k}-q^{3k}\right)^2 },\nn
J_3(z) \eq \int_0^\infty d\beta \, \frac{4^{1-z} \beta ^{2 z-2} }{\Gamma (z)} \log\!
\left[ \frac{1-4q^k+2q^{2k}-q^{3k}}{\left(1-q^k\right)^3}\right].
\ea
For the integrand of $J_1$ and $J_3$, we use again the series expansion \eqref{log series} and \eqref{I2 expansion}, respectively. In case of $J_2$, we have the  expansion,
\ba
\frac{q^{4k}(1+q^k)^2}{(1-q^k)^4(1-4q^k+2q^{2k}-q^{2k})^2}  \eq \sum_{l=0}^{\infty}  \bigg[ l \left(c_1 \, p_1^{\ell+2}+e_2 \, p_2^{\ell+1}+ e_3 \, p_3^{\ell+1} \right) \nn
&&\quad   + d_1 \, p_1^\ell+d_3 \, p_2^\ell+d_2 \, p_3^\ell+\frac{\ell^3 -\ell +1}{6}  \bigg] q^{k \ell}\,.
\ea
Here the constant $c_1$ is the real solution of the cubic equation $ 11449 x^3-6527 x^2+957 x-4 = 0$. The $e_2$ and $e_3$ are imaginary solutions of the cubic equation $ 11449 x^3-1391 x^2+283 x-4 = 0$. And, the constants $d_1, d_2$ and $d_3$ are the three solutions of the cubic equation $ 1225043 x^3+1225043 x^2+409177 x+45401 = 0$.
Applying above series expansions, the function $J_1$ is given by
\ba
&& J_1(z)= - \sum_{\ell=1}^{\infty} \int_0^\infty d\beta \frac{\left(k^2 (1-e^{-\b})^2 \left[(k^2-1)\, e^{-2\b} - (2k^2-14)\, e^{-\b}+k^2-1\right] 4^{-z} \beta ^{2 z-2}\right)}{3  \, \Gamma (z)} \nn
&&\qquad \qquad \qquad \qquad \qquad \qquad \qquad \qquad \qquad \times \left(f_1 \, p_1^\ell+f_2 \, p_2^\ell+f_3 \, p_3^\ell+\ell\right) e^{2\b} e^{-k \, \b \, \ell}\nn
&&=\, - \frac{4^{-z} k^{3-2z} \G(2z-1)}{3 \, \G(z)} \bigg[ f_1 \, \Xi^{(2)}(p_1, z) + f_2 \, \Xi^{(2)}(p_2, z) + f_3 \, \Xi^{(2)}(p_3, z) + 6\, (k^2 - 5)\, \zeta(2z-2) \nn
&& \qquad \qquad \qquad \qquad  \quad  +\,(k^2 - 1) \Phi^{+}(1,2z-2,\tfrac{2}{k})-4\, (k^2-4) \, \Phi^{+}(1,2z-2,\tfrac{1}{k}) \nn
&& \qquad \qquad \qquad \qquad  \quad  + \,4\, (k - \tfrac{4}{k}) \, \Phi^{-}(1,2z-1,\tfrac{1}{k}) - 2 \, (k-\tfrac{1}{k}) \, \Phi^-(1,2z-1, \tfrac{2}{k}) \bigg]\,.
\ea
Here the definition of $\Xi^{(2)}$ is
\be
\Xi^{(2)}(p,z) = 6\, (k^2-5)\, \Phi(p,2z-1,0) + (k^2-1)\, \Phi^{+}(p, 2z-1, \tfrac{2}{k})  + (16-4k^2) \, \Phi^{+}(p, 2z-1, \tfrac{1}{k})\,. 
\ee
For the function $J_2$, we get
\ba
&& J_2(z) = -\sum_{\ell=1}^{\infty} \int_0^\infty d\beta \frac{k^4 \, (1-e^{-\b})^4 \, 2^{1-2 z} \, \beta ^{2 z-2}}{\Gamma (z)} \nn
&& \ \times \,\bigg( l \left(c_1 \, p_1^{\ell+2}+e_2 \, p_2^{\ell+1}+e_3 \, p_3^{\ell+1}\right)+d_1 \, p_1^\ell+d_3 \, p_2^\ell+d_2 \, p_3^\ell+\frac{\ell^3 -\ell +1}{6}\bigg) e^{2\b} e^{-k \, \b \, \ell} \nn
&&= \,\frac{\G(2z-1) \, 4^{-z}\, k^{5-2z}}{3\, \G(z)} \bigg[ d_1 \, \Xi^{(3)} (p_1, z) + d_3 \, \Xi^{(3)} (p_2, z) + d_2 \, \Xi^{(3)} (p_3, z) + c_1 \, p_1^2 \, \Xi^{(4)}(p_1,z) \nn
&& \ +\, e_2 \, p_2 \, \Xi^{(4)}(p_2,z) + e_3 \, p_3 \, \Xi^{(4)}(p_3,z)  + 6\, \zeta(2z-1) - 6 \, \zeta(2z-2) + 6\, \zeta(2z-4) \nn 
&& \ -\,4 \, \Phi^{+} (1,2 z-4,\tfrac{1}{k} )  +\Phi^{+} (1,2 z-4,\tfrac{2}{k})  + \tfrac{12}{k}  \, \Phi^{-} (1,2 z-3,\tfrac{1}{k}) -\tfrac{6}{k} \, \Phi^{-} (1,2 z-3,\tfrac{2}{k}) \nn
&& \ -\,4 \, \left(1+\tfrac{3}{k^2} \right)\, \Phi^{+}(1,2 z-2,\tfrac{1}{k}) + \left(1+\tfrac{12}{k^2} \right) \, \Phi^{+} (1,2 z-2,\tfrac{2}{k}) \nn
&& \ -\,4 \, \left(1+\tfrac{1}{k}+\tfrac{1}{k^3} \right) \, \Phi^{+} (1,2 z-1,\tfrac{1}{k}) +\left(1-\tfrac{2}{k}-\tfrac{8}{k^3} \right) \, \Phi^{+}(1,2 z-1,\frac{2}{k}) \nn
&& \ +\,\tfrac{4}{k} \, \Phi^{-}(1,2 z-1,\tfrac{1}{k}) -\tfrac{2}{k} \, \Phi^{+}(1,2 z-1,\tfrac{2}{k}) + \tfrac{4}{k^3} \, \Phi^{-}(1,2 z-1,\tfrac{1}{k})  - \tfrac{8}{k^3}\, \Phi^{+}(1,2 z-1,\tfrac{2}{k})\bigg]\,. \nn
\ea
In this expression, $\Xi^{(3)}(p,z)$ and $\Xi^{(4)}(p,z)$ are defined by
\ba
\Xi^{(3)}(p,z) \eq 36 \, \Phi(p,2z-1,0)+ 6 \, \Phi^{+}(p,2z-1,\tfrac{2}{k}) -24 \, \Phi^{+}(p,2z-1,\tfrac{1}{k})\,, \nn
\Xi^{(4)}(p,z) \eq \tfrac{6}{k \, p} \bigg[ 6\, k\, p \, \Phi(p,2z-2,0)  + k\, \Phi^{+}(p,2z-2,\tfrac{2}{k}) -4\, k\, \Phi^{+}(p,2z-2,\tfrac{1}{k})  \nn
&& \ - 2 \, \Phi^{-}(p,2z-1,\tfrac{2}{k}) + 4\, \Phi^{-}(p,2z-1,\tfrac{1}{k}) \bigg]\,.
\ea
Using the prescription \eqref{presc}, it is straightforward to check 
that $J_1(z) + J_2(z)$ vanishes at leading order in $z$. Finally, for the function $J_3$, we get
\ba
&& J_3(z) = -\sum_{\ell=1}^{\infty} \int_0^\infty d\beta \frac{4^{1-z} \beta ^{2 z-2}}{\Gamma (z)}  \frac{ \left(p_1^\ell+p_2^\ell+p_3^\ell-3\right)}{\ell} e^{-k \, \b \, \ell} \nn
\eq -\frac{\left(4^{1-z} \beta ^{2 z-2} k^{1-2z}\right)}{\Gamma (z)} \bigg[ \Phi(p_1, 1-2z, 0) + \Phi(p_2, 1-2z, 0) + \Phi(p_3, 1-2z, 0) - 3 \, \Phi(1, 1-2z, 0)\bigg]\,. \nn
\ea
Again through the prescription of \eqref{presc}, we get
\ba
J_3(z)\eq \sum_{\ell=1}^{\infty} 2 k \left(p_1^\ell+p_2^\ell+p_3^\ell-3\right) + \mathcal{O}(z) \nn
\eq -\bigg(\frac{2\,p_1 }{p_1-1}+\frac{2 \,p_2 }{p_2-1}+\frac{2\, p_3 }{p_3-1}-3 \bigg)k + \mathcal{O}(z) = \mathcal{O}(z)\,.
\ea
This completes the computation
of the zeta function up to
quadratic order in $z$ within 
Prescription 2.
We find that the linear term in $z$ is absent hence the corresponding
vacuum energy vanishes.

\bibliographystyle{JHEP}
\bibliography{matrix}

\end{document}